\documentclass[12pt]{article}

\usepackage{pifont}% http://ctan.org/pkg/pifont
\newcommand{\cmark}{\ding{51}}%
\newcommand{\xmark}{\ding{55}}%
\usepackage[utf8]{inputenc}
\usepackage{graphicx}
\usepackage{fancyref}
\usepackage{amsmath}
\usepackage{latexsym}
\usepackage{xcolor}
\usepackage[]{units}
\usepackage{float}
\usepackage{physics}
\usepackage{mathrsfs}
\usepackage{amssymb}
\usepackage{comment}
\usepackage{todonotes}
\usepackage{enumerate}
\usepackage{parskip}
\usepackage[margin=1in]{geometry}
\usepackage{mathtools}
\usepackage{bm}
\usepackage{slashed}
\usepackage{tensor}
\usepackage{caption}
\usepackage{subcaption}
\usepackage{cite}
\RequirePackage[colorlinks=true, urlcolor=black, menucolor=black, linkcolor=purple, citecolor=blue]{hyperref}

\numberwithin{equation}{section}

\newcounter{todocounter}

\title{\textbf{SU(6) Gauge-Higgs Grand Unification: Minimal Viable Models and Flavor}}
\author{\href{mailto:andrei.angelescu@mpi-hd.mpg.de}{Andrei Angelescu},
\href{mailto:andreas.bally@mpi-hd.mpg.de}{Andreas Bally},
\href{mailto:florian.goertz@mpi-hd.mpg.de}{Florian Goertz},
\href{mailto:sascha.weber@mpi-hd.mpg.de}{and Sascha Weber}\\\\ \it\small Max-Planck-Institut f{\"u}r Kernphysik,\\ \it\small
Saupfercheckweg 1, 69117 Heidelberg, Germany\\\\

}
\date{}
\begin{document}
\fontseries{mx}\selectfont	
\maketitle
\begin{abstract}
Gauge-Higgs grand unification theories are models of gauge-Higgs unification that extend the electroweak group into a simple group that includes the color symmetry. The minimal option is a gauge-Higgs grand unification based on the SU(6) gauge group, mirroring SU(5) grand unification in 4D while providing a solution to the hierarchy problem. We explore different minimal and realistic novel incarnations of SU(6) gauge-Higgs grand unification. We submit the setup to the various flavor hierarchies observed in nature and, utilizing the power of the fifth dimension, identify an embedding that provides a compelling model of quarks and leptons that naturally explains the mass hierarchies and the CKM/PMNS structure. We perform a detailed study of quark-- and lepton--flavor constraints (which are intimately related due to the GUT nature) together with an analysis of the Higgs potential which arises at the loop level. Electroweak precision constraints on the model are discussed and the rich scalar sector is analyzed. Future flavor constraints from upcoming experiments will provide a stringent test for this class of models, while a scalar singlet and leptoquark provide unique targets for current and future collider experiments to probe this solution to various open questions in nature.
\end{abstract}

\newpage
\setcounter{footnote}{0}

\tableofcontents

\newpage

%%%%%%%%%%%%%%%%%%%%%%%%%%%%%%%%%%%%%%%%%%%%%%%
%%%%%%%%%%%%%%%%%%%%%%%%%%%%%%%%%%%%%%%%%%%%%%%
\section{Introduction}\label{SEC-01}
Although the 2012 discovery of the Higgs boson \cite{ATLAS:2012yve,CMS:2012qbp} provided the last piece of the puzzle to finalize the standard model (SM) of particle physics as a successful theory surpassing the electroweak scale, the origin of electroweak symmetry breaking (EWSB) remains still unknown. Most intriguing in that context is the instability of the Higgs mass with respect to a higher UV scale, the hierarchy problem, which has motivated model building for the last decades. One elegant solution is the idea of gauge-Higgs unification (GHU) in which the Higgs boson is part of a higher dimensional gauge field \cite{HOSOTANI1983193,HOSOTANI1983309,MANTON1979141,FAIRLIE197997}. By higher dimensional gauge invariance the Higgs boson does not have a mass at the classical level. In such a setting it rather gets a radiative mass at one-loop due to symmetry breaking effects proportional to the size of the extra dimension: the Hosotani mechanism. These models are particularly well suited in a warped extra dimension \cite{Randall:1999ee} since this can dynamically explain the large hierarchy between the Planck scale and the electroweak vacuum expectation value (vev) \cite{Goldberger:1999uk}. GHU requires at least the propagation of the gauge bosons in the warped bulk  \cite{Davoudiasl:1999tf,Pomarol:1999ad} and it is plausible (and phenomenologically beneficial) to also allow the propagation of fermion fields in the bulk \cite{Grossman:1999ra,Gherghetta:2000qt}. 

Equally puzzling is the charge quantization that is observed in the SM and which may indicate the unification of the SM gauge group $G_{\textrm{SM}}$=SU(3)\textsubscript{c} $\times$ SU(2)\textsubscript{L} $\times$  U(1)\textsubscript{Y} into a larger gauge group of a grand unified theory (GUT) \cite{Georgi:1974sy,Fritzsch:1974nn,Pati:1973uk}, which is also hinted at by the approximate meeting of the gauge couplings at the unification scale $M_U$. The SM is particularly well suited for an SU(5) or SO(10) GUT, as its fermions fill complete representations for both of these gauge groups. The combination of both these ideas, that is, gauge-Higgs unification of the electroweak group together with the inclusion of the color group, is called a gauge-Higgs grand unified theory (GHGUT). Previously, gauge-Higgs unification of the electroweak gauge group SU(2)\textsubscript{L} $\times$  U(1)\textsubscript{Y}, thus omitting the strong interactions, was done in \cite{Contino:2003ve} and a custodial model was presented in \cite{Agashe:2004rs} based on an SO(5)$\times$ U(1) gauge group. The latter allows to have a relatively low scale of symmetry breaking, without violating electroweak precisision tests (EWPT). These models however generally feature light top partners that mix with the top and provide its large mass, while keeping the contribution to the Higgs mass moderate \cite{Contino:2006qr,Matsedonskyi:2012ym,Pomarol:2012qf,Carmona:2014iwa,Blasi:2019jqc,Blasi:2020ktl}. The absence of any signal at the LHC increases the required fine-tuning in these models. 

Models of GHGUT, thus including the strong interactions, are based on two main gauge groups: SU(6) and SO(11). The latter gauge group also enjoys the benefit of a custodial symmetry and was studied in \cite{Hosotani:2015hoa,Furui:2016owe}. However the presence of light exotics led the authors to go to six dimensions \cite{Hosotani:2017edv,Hosotani:2017hmu}. Phenomenology of these models was studied in \cite{ENGLERT2020135261,Englert:2020eep}. There has been a renewed interest in SO(5) $\times$ U(1)\textsubscript{X} $\times$ SU(3)\textsubscript{c} gauge-Higgs unification that is inspired by an SO(11)  GHGUT in which the quark and lepton multiplets are introduced in the spinor SO(5) representation in order to be compatible with SO(11) GHGUT \cite{Funatsu:2019xwr,Funatsu:2019fry,Funatsu:2020znj,Funatsu:2020haj,Funatsu:2021ryn,Funatsu:2021gnh,Funatsu:2021yrh}, with the running of the gauge couplings studied in \cite{Maru:2022mbi}. The former gauge group, SU(6), has been studied in a flat extra dimensional setting in both supersymmetric \cite{Hall:2001zb,Burdman:2002se,Haba:2004qf} and non-supersymmetric \cite{Lim:2007jv} contexts. In general, the constraining nature of the SU(6) symmetry leads to the problematic appearance of light exotic fermions (see below) and early models also suffered from massless down-type quarks and charged leptons. The latter issue was recently addressed by localizing the SM fermions on a 4D brane~\cite{Maru:2019bjr,Maru:2019lit,Maru:2022hex} (and introducing additional mirror fermions).

Recently, some of the authors of this paper put forward an SU(6) GHGUT in a warped extra dimension that manages to reproduce the full SM spectrum from a minimal amount of 5D fields in lowest possible representations while avoiding light exotic fermions \cite{Angelescu:2021nbp,Angelescu:2021qbr}. Here we will scrutinize the phenomenology of this model and other variants thereof and also see how it performs in {\it explaining} the observed flavor hierarchies in nature.  It is already well known that extra dimensional models that feature a warped metric can be very successful in addressing the flavor hierarchies  \cite{Grossman:1999ra,Gherghetta:2000qt,Huber:2000ie,Huber:2003tu}. Small Yukawa couplings are obtained due to the overlap integrals over the extra dimension of fermions, being localized differently, with the Higgs field (a mechanism that predates the Randall-Sundrum (RS) model \cite{Arkani-Hamed:1999ylh}). Similarly, the hierarchies in the CKM matrix find a very natural explanation in these models. 

Additionally, we will confront the models with various flavor constraints. Bounds from Flavor Changing Neutral Currents (FCNCs) are also suppressed by the small overlap functions in the extra dimension, a mechanism that has been called RS-GIM mechanism~\cite{Gherghetta:2000qt,Huber:2000ie,Agashe:2003zs,Agashe:2004cp,Agashe:2004rs} (an analogue to the GIM mechanism \cite{Glashow:1970gm} at work in the SM in suppressing FCNCs). Nevertheless, RS-GIM is not nearly as effective as its SM counterpart and $\Delta=2$ flavor constrains have been studied in \cite{Agashe:2004ay,Agashe:2004cp,Casagrande:2008hr,Csaki:2008zd,Blanke:2008yr,Agashe:2008uz,Blanke:2008zb,Bauer:2009cf,Albrecht:2009xr,Keren-Zur:2012buf}. Constraints on dipole operators in warped space models have also been well studied in the past \cite{Agashe:2006iy,Gedalia:2009ws,Csaki:2010aj,Keren-Zur:2012buf,Blanke:2012tv,Beneke:2012ie,Konig:2014iqa,Beneke:2015lba,Moch:2015oka}. Since we are unifying the EW and strong interactions, including leptons and quarks in a correlated way, the existing works on flavor constraints in warped extra dimensional settings might not be very well suited for the model at hand. Moreover, many of the past flavor studies of gauge-Higgs unification have been done in custodial models based on the SO(5) bulk gauge symmetry which is qualitatively different from our SU(6) model, lacking a custodial symmetry. The non-custodial EW gauge group analogue of SU(6) is a bulk SU(3) ($\supset$ SU(2)\textsubscript{L}) for which flavor constraints have not been studied to the same extent as in the custodial analogue. Finally, our GHGUT contains new bosons, as the scalar leptoquark, that contributes to loop-mediated FCNCs.

The lack of custodial symmetry means constraints from EWPT will be more stringent (although the 5D nature of the Higgs will relax them compared to the usual IR-brane localized Higgs boson scenario, see Section~\ref{sec:EWPT}). However, these constraints do not seem as relevant in the vanilla models, since they are overshadowed by the much stronger flavor constraints. This has resulted in rich model building through the use of various flavor symmetries within extra dimensional models in order to weaken the flavor bounds both in the lepton sector \cite{Csaki:2008qq,Perez:2008ee,Chen:2008qg,Chen:2009gy,Kadosh:2010rm,delAguila:2010vg,Hagedorn:2011pw,Hagedorn:2011un,vonGersdorff:2012tt,Kadosh:2013nra,Ding:2013eca,Frank:2014aca,Chen:2015jta,Frigerio:2018uwx} (see also \cite{Goertz:2021ndf,Goertz:2021xlx} for a recent review) and in the quark sector \cite{Cacciapaglia:2007fw,Fitzpatrick:2007sa,Santiago:2008vq,Csaki:2009bb,Csaki:2009wc,Wojcik:2019kfq}. 

In this work we will not pursue these additional flavor symmetries and be content with the resulting little hierarchy that exists in the Higgs potential from evading flavor bounds. Beyond that, such model building with flavor symmetries would seem fruitless for GHGUTs since SU(6) GHGUT contains extra light colored scalar degrees of freedom alongside the Higgs boson, dictated by the unification of electroweak and strong symmetries, that provide a hard bound on the fundamental IR scale of the model (a bound that is not present in models of electroweak gauge-Higgs unification), see below. As in 4D GUTs, proton stability is an important issue for GHGUTs. In extra dimensional models \cite{Agashe:2004bm,Agashe:2004ci}, including our SU(6) GHGUT model \cite{Angelescu:2021nbp}, the proton is rendered stable by a global baryon number symmetry  which allows the new colored bosons to reside at the TeV scale with the bound being driven by collider searches. The overall resulting tension of the IR-scale being $1-2$ orders of magnitude above the EW scale is however tiny compared to the full GUT hierarchy problem. The approach we take therefore is similar in spirit to \cite{Barnard:2014tla}, accepting a little hierarchy problem but at the same time providing a compelling model of grand unification that evades stringent flavor constraints and explains the observed hierarchies in fermion masses and mixings. Finally, the light new scalars furnish striking experimental signatures to test the model in the future.

We organize the paper in the following sections: Section~\ref{sec:WarpedGaugeFields} provides a concise review on the basis of gauge fields and fermions in a warped extra dimensions (and introduces our conventions), which can safely be skipped by readers familiar with warped models. In Section~\ref{sec:Elements} we provide the various tools needed for a phenomenological analysis of models of gauge-Higgs (grand) unification and in particular SU(6) GHGUTs that may also be skipped by familiar readers.  Section~\ref{sec:Results} introduces the specific SU(6) GHGUT model that is naturally suited to provide a full model of the observed flavor hierarchies in nature without fine-tuned parameters. In Section~\ref{sec:Flavor} we discuss the flavor constraints that we will consider and in particular in Subsection~\ref{sec:Results2} we present the major results of the paper, namely the various flavor bounds on our model. In Section~\ref{sec:EWPT} we analyze EWPT and, finally, in Section~\ref{sec:Potential} we discuss the extended scalar potential of SU(6) GHGUT and the particular features of the scalar spectrum. We conclude in Section~\ref{sec:Conclusion} while two appendices contains the more technical aspects of the presented calculations underlying our results.

%%%%%%%%%%%%%%%%%%%%%%%%%%%%%%%%%%%%%%%%%%%%%
%%%%%%%%%%%%%%%%%%%%%%%%%%%%%%%%%%%%%%%%%%%%%
\section{Gauge fields and fermions in warped space}\label{sec:WarpedGaugeFields}

\noindent In this section, we explore the basics behind gauge-Higgs unification in warped space, see \cite{Csaki:2004ay,Csaki:2005vy,Sundrum:2005jf,Contino:2010rs,Gherghetta:2010cj,Ponton:2012bi,Csaki:2015xpj} for a comprehensive review of extra dimensions. We work in conformal coordinates in which the warped metric is given by
\begin{align}
    \text{d}s^2=\Big(\frac{R}{z}\Big)^2(\eta_{\mu\nu}\text{d}x^\mu \text{d}y^\nu -\text{d}z^2),
\end{align}
where $z\in[R,R^\prime]$, and $R\sim 1/M_{\textrm{pl}}$ ($R^\prime\sim 1/\textrm{TeV})$ is the position of the UV (IR) brane. The idea behind gauge-Higgs unification is to embed the Higgs field in the fifth component of a five dimensional gauge field $A_M$, with $M=0,1,2,3,5$, where $A_5$ is a 4D  scalar. As we will see, only under specific circumstances can this fifth component  actually be identified as a physical scalar field in the spectrum.

\subsection{Gauge fields}

We start with the 5D Yang-Mills Lagrangian in the convention where the group generators are normalized such that $\textrm{Tr}(T^a T^b)= \frac{1}{2}\delta_{ab}$
\begin{align}\label{Gauge}
    S_{\textrm{YM}}= \int_R^{R^\prime}\text{d}^4x\text{d}z \sqrt{G}\Big(-\frac{1}{2}G^{MN}G^{AB}\textrm{Tr}(F_{MA}F_{NB}) \Big) = \int_R^{R^\prime}\text{d}^4x\text{d}z \Big(\frac{R}{z}\Big)\Big(-\frac{1}{4}F^a_{AB}F^{AB,a} \Big),
\end{align}
with $G=(R/z)^{10}$ the determinant of the metric. The quadratic terms of the above Lagrangian contain mixing terms between $A_5$ and $A_\mu$. The gauge-fixing term is chosen such as to cancel these terms, reading
\begin{align}
    S_{\textrm{GF}}=\int\text{d}^4x\text{d}z\Big(\frac{R}{z}\Big)  \Bigg(-\frac{1}{2\xi}\Big(\partial_\mu A^\mu-\xi \frac{z}{R}\partial_5\big(\frac{R}{z}A_5\big)\Big)^2\Bigg).
\end{align}
Varying the quadratic action under $\delta A_\mu$ and $\delta A_5$ we find the equations of motion (EOMs)
\begin{align}
    \Bigg[\eta^{\mu\nu} \partial^2  -(1-\frac{1}{\xi})\partial^\mu\partial^\nu\Bigg]A_\nu-z\partial_5\Big(\frac{1}{z}\partial_5A^\mu\Big) =0,\qquad
    \partial^2 A_5 - \xi \partial_5\Big(z\partial_5(\frac{1}{z} A_5)\Big )= 0.
\end{align}
The boundary conditions (BCs) for the fifth dimension are obtained by setting the contributions at the boundaries to zero, arising from integration by parts
\begin{equation}
    \int \text{d}^4x \Big[\frac{R}{z}\Big(\partial_\mu A_5-\partial_5A_\mu\Big)\delta A^\mu\Big]^{z=R^\prime}_{z=R} = 0, \quad 
    \int \text{d}^4x\Big[\frac{R}{z}\Big(\partial_\mu A^\mu -\xi z \big(\partial_5(\frac{1}{z}A_5)\big)\Big)\delta A_5\Big]^{z=R^\prime}_{z=R} = 0.
\end{equation}
We shall consider two possible BCs consistent with the above conditions, namely 
\begin{equation}\label{5dgaugebosonBC}
    (+): A_5|_{z=R,R^\prime} = 0, \partial_5 A_\mu|_{z=R,R^\prime} = 0, \quad
    (-) : A_\mu|_{z=R,R^\prime} = 0, \partial_5 \Big(\frac{1}{z} A_5\Big)|_{z=R,R^\prime}=0.
\end{equation}
The $(+)$ refers to a Neumann BC for the $A_\mu$ field, while the $(-)$ refers to a Dirichlet BC for the $A_\mu$ field. For phenomenological purposes it is crucial to identify when we can have a (massless) zero mode. It turns out that this arises for the vector field $A_\mu$ when applying $(+,+)$ BCs at the (UV,\,IR) branes, while a zero mode for the scalar field $A_5$ appears when choosing $(-,-)$ BCs. For mixed BCs no zero modes arise.

To see this in detail, we write both 5D fields as an infinite sum of 4D fields with bulk profiles $f_{n,A/5}$, which can be understood as a separation of variables, called Kaluza-Klein (KK) decomposition
\begin{align}
    A_\mu (x,z)=\sum_n f_{n,A}(z) A_{\mu,n}(x),\quad A_5(x,z)=\sum_n f_{n,5}(z) A_{5,n}(x), 
\end{align}
and demand the 4D fields to obey the usual 4D free EOMs 
\begin{align}
    \Bigg[\eta^{\mu\nu} \partial^2  -(1-\frac{1}{\xi})\partial^\mu\partial^\nu\Bigg]A_{\nu,n} +m_n^2A^{\mu}_n = 0,\qquad
    \partial^2 A_{5,n} +m_{5,n}^2 A_{5,n} = 0.
\end{align}
This leads to the differential equations for the bulk profiles
\begin{align}\label{5dgaugeboson}
    -m_n^2f_{n,A} = z\partial_5 \Big(\frac{1}{z}\partial_5 f_{n,A}\Big), \qquad -m_{5,n}^2 f_{n,5}=\xi\partial_5\Big(z\partial_5\big(\frac{1}{z}f_{n,5}\big)\Big).
\end{align}
From the first equation it follows that a vector boson zero mode ($m_0=0$) has a constant bulk profile which implies $(+,+)$ BCs for the mode to exist, while a scalar zero mode ($m_{5,0}=0$), together with the $(-,-)$ BCs, induces $f_{0,5}\sim z $. Interestingly, for higher modes with vector mass $m_n>0$, the relation $f_{n,5} =\frac{1}{m_n} \partial_5 f_{n,A}$ follows with $m_{5,n}=\sqrt{\xi} m_n$. The $\xi$ dependence indicates that $A_5$ is an unphysical Goldstone mode that provides the longitudinal polarization for the massive vector bosons. Canonically normalizing the kinetic terms of the vector fields leads to
\begin{equation}\label{gaugezeronorm}
    \int_R^{R^\prime} \textrm{d}z \frac{R}{z} f_{n,A} f_{m,A} = \delta_{n,m},
\end{equation}
which determines the normalization of the zero mode
\begin{equation}
    f_{0,A}(z)=\sqrt{\frac{1}{R\log(\frac{R^\prime}{R})}}\,,
\end{equation}
while $f_{n,5}$ are automatically normalized by virtue of the EOMs, inducing
\begin{equation}
    \int_R^{R^\prime} \textrm{d}z \frac{R}{z} f_{n,5} f_{m,5} = \delta_{n,m}.
\end{equation}

In summary, the 4D KK theory up to quadratic terms has the form
\begin{equation}
    \sum_n  \int \text{d}^4x \Big( -\frac{1}{4} F_{\mu\nu,n} F^{\mu\nu}_n  -\frac{1}{2\xi}(\partial_\mu A_n^\mu)^2  + \frac{1}{2}m_n^2 A_{\mu,n}^{2} +\frac{1}{2} \partial_\mu A_{5,n} \partial^\mu A_{5,n} - \frac{1}{2} \xi m_n^2 A_{5,n}^{2}\Big),
\end{equation}
revealing the Goldstone nature of $A_{5,n}$ that provides the longitindal polarization of the massive gauge bosons $m_n$. At higher order we also get the usual interactions between the vector bosons and Goldstone modes. The profiles $f_{n,(+,\pm)}(z)$ and the masses $m_{n,(+,\pm)}$ of the $(+,\pm)$ gauge bosons can be found by solving \eqref{5dgaugeboson} subject to the (+) BC \eqref{5dgaugebosonBC} at the UV brane. The profiles are then given in terms of Bessel functions as
\begin{equation}\label{plusplusprofile}
    f_{n,(+,\pm)}(z)=N_{n,(+,\pm)} z \Big(J_1(m_{n,(+,\pm)}z)-\frac{J_0(m_{n,(+,\pm)}R)Y_1(m_{n,(+,\pm)} z)}{Y_0(m_{n,(+,\pm)}R)}\Big),
\end{equation}
with normalization constant $N_{n,(+,\pm)}$, while the tower of KK masses $m_{n,(+,\pm)}$ is found by solving the IR BC. For a $(+,+)$ gauge boson we have a massless and flat profile for the zero mode with the first KK mode at $m_{1,(+,+)}\sim 2.45/R^\prime$, while $(+,-)$ gauge bosons do not have a zero mode but their first KK excitation is rather light $m_{1,(+,-)}\sim 0.25/R^\prime$.

For $(-,\pm)$ gauge bosons the UV BC is different, resulting in a different profile

\begin{equation}\label{minusminusprofile}
    f_{n,(-,\pm)}(z)=N_{n,(-,\pm)} z \Big(J_1(m_{n,(-,\pm)}z)-\frac{J_1(m_{n,(-,\pm)}R)Y_1(m_{n,(-,\pm)} z)}{Y_1(m_{n,(-,\pm)}R)}\Big).
\end{equation}
The spectrum can be found again by applying the IR BC and will result in a tower of massive gauge bosons with the first KK modes having mass $m_{1,(-,+)}\sim 2.40/R^\prime$ and $m_{1,(-,-)}\sim 3.83/R^\prime$, respectively. We will leave out the index $n=1$ for the first KK mode from here on.

Recall from \eqref{5dgaugeboson} that a $(-,-)$ gauge-boson BC implies a physical massless scalar in the spectrum, the zero mode of $A_{5}(x,z)$, simply denoted by $A_{5}(x)$ in the following, with normalized bulk profile $f_5(z)$,
\begin{equation}\label{A5profile}
    A_5(x,z) \supset f_5(z) A_{5}^a(x) T^a = \sqrt{\frac{2}{R}}\frac{z}{R^\prime} A_{5}^a(x) T^a.
\end{equation}
The idea of gauge-Higgs unification is to embed the Higgs field in such a zero mode. Indeed, a potential for $A_5^a(x)$ is forbidden by 5D symmetry. At one-loop, upon symmetry breaking at the boundaries, a finite potential is generated possibly inducing a 4D vev $\langle A_{5}^a(x)\rangle=v^a$.

In the presence of such a vev, solving the gauge and fermion profiles in warped space is complicated. Fortunately, since the vev in gauge-Higgs unification is embedded within the gauge field, one can actually eliminate it from the bulk with a gauge transformation \cite{Hosotani:2005nz} 
\begin{align}
    & A_M\rightarrow \Omega A_M \Omega^\dagger-\frac{i}{g_5}\Omega \partial_M \Omega^\dagger\,, \qquad
    \Omega(z) = \exp\Big(i g_5 \int_R^z dz^\prime  f_5(z^\prime) v^a T^a\Big)\,,
\end{align}
which removes the $A_5$ background field:
\begin{equation}
    A_5(z)\rightarrow \Omega\Big( A_5(z)  - f_5(z) v^a T^a\Big)\Omega^\dagger.
\end{equation}
This transformation is trivial on the UV brane ($\Omega(z=R)=1$), but on the IR it is given, in terms of the Higgs decay constant $f$, by
\begin{equation}
\label{Higgsdecayconstant}
    \Omega(z=R^\prime) = \exp\Big(i \sqrt{2} v^aT^a/f\Big),\qquad  f= \frac{2\sqrt{R}}{g_5 R^\prime}.
\end{equation}

Consistency requires applying this gauge transformation on the fermion and gauge fields
\begin{align}
    A_\mu (z=R^\prime)&\rightarrow \Omega(z=R^\prime) A_\mu (z=R^\prime)\Omega(z=R^\prime)^\dagger \notag \\
    \Psi(z=R^\prime)&\rightarrow\Omega(z=R^\prime)\Psi(z=R^\prime)\,.
\end{align}
Therefore, the effect of an $A_5$ vev will manifest itself through the mixing of generators {\it on the IR brane}. 
We will only need the exact solutions obtained from such a gauge transformation when computing the (extended) scalar potential in Section~\ref{sec:Potential}, but otherwise we will be more intuitive and treat the vev in the bulk as a small expansion parameter with respect to the IR scale $1/R^\prime$.

\subsection{Fermions}\label{sec:WarpedFermions}
Having discussed the principle of gauge-Higgs unification, we summarize the dynamics of free 5D fermions in warped space \cite{Gherghetta:2000qt,Grossman:1999ra,Csaki:2003sh}, starting from the hermitian action (in spaces without boundaries one can integrate by parts to recover its usual form)
\begin{equation}
    S_{\textrm{Fermion}}=\int \textrm{d}^4x\int_R^{R^\prime} \textrm{d}z \sqrt{G} \Big(\frac{i}{2}(\bar{\Psi} e_a^M\gamma^aD_M\Psi- \overline{D_M\Psi}e_a^M\gamma^a\Psi) - m \bar{\Psi}\Psi\Big).
\end{equation}
The Dirac algebra is generalized to 5D, where the inclusion of $\gamma^5=-i \Big(\begin{array}{cc}
\!\!-1 & 0 \\[-2mm]
0 & \!\! 1\!\!\\
\end{array}\Big)$ as the fifth gamma matrix in $\{\gamma^a,\gamma^b\}=2\eta^{ab}$, with $\gamma^\mu=\Big(\begin{array}{cc}
\!\!0 & \!\sigma^\mu\!\! \\[-2mm]
\!\bar{\sigma}^\mu & \!\! 0\!\!\\
\end{array}\Big)$,  makes 5D fermions non-chiral. Moreover,
$e_a^M$ is the f\"{u}nfbein, $e_a^M\eta^{ab}e_b^M=G^{MN}$, which for warped space is given by $e_M^a=(R/z)\delta ^a_M$ from which the spin connection $\omega_M = \gamma_\mu \gamma_5/(4z)\delta_M^\mu$ follows. The covariant derivative is then given in the absence of gauge interactions by $D_\mu \Psi = (\partial_\mu +\gamma_\mu \gamma_5/(4z))\Psi$, and $D_5\Psi = \partial_5 \Psi$. By spinor algebra, the spin connection cancels between the derivative and its hermitian conjugate, such that the resulting action is
\begin{equation}
    S_{\textrm{Fermion}}=\int \textrm{d}^4x\int_R^{R^\prime} \textrm{d}z \Big(\frac{R}{z}\Big)^4\Big(\frac{i}{2}(\bar{\Psi} \gamma^\mu \overleftrightarrow{\partial_\mu} \Psi+\bar{\Psi} \gamma^5 \overleftrightarrow{\partial_5} \Psi) - \frac{m R}{z} \bar{\Psi}\Psi\Big),
\end{equation}
with $\overleftrightarrow{\partial}=\overrightarrow{\partial}-\overleftarrow{\partial}$. Integrating by parts, which leads to boundary terms along the compact direction, and defining the dimensionless parameter $c=mR$ we obtain
\begin{equation}
    S_{\textrm{Fermion}}=\int \textrm{d}^4x\int_R^{R^\prime} \textrm{d}z \Big(\frac{R}{z}\Big)^4\bar{\Psi}\Big(i \gamma^\mu \partial_\mu +i\gamma^5\partial_5 -\frac{2i}{z}\gamma_5 - \frac{c}{z}\Big)\Psi - \frac{i}{2}\int d^4x\Big(\frac{R}{z}\Big)^4 \big[\bar{\Psi}\gamma^5 \Psi\big]^{z=R^\prime}_{z=R}.
\end{equation}

The above form is useful to derive the 5D fermion propagators, but we will be interested in the KK description and decompose the 5D fermions in their chiral components $\Psi=(\chi, \bar{\psi})^T$, leading to the bulk action with boundary terms
\begin{align}
    S_{\textrm{Fermion}}=&\int \textrm{d}^4x\int_R^{R^\prime} \textrm{d}z \Big(\frac{R}{z}\Big)^4 \Big(i\bar{\chi}\bar{\sigma}^\mu\partial_\mu\chi+i\psi\sigma^\mu\partial_\mu\bar{\psi}-\psi\partial_5 \chi +\bar{\chi}\partial_5 \bar{\psi}+\frac{2-c}{z}\psi\chi-\frac{2+c}{z}\bar{\chi}\bar{\psi}\Big)\notag \\
    + &\frac{1}{2}\int \textrm{d}^4x\Big(\frac{R}{z}\Big)^4 \big[\psi\chi-\bar{\chi}\bar{\psi}\big]^{z=R^\prime}_{z=R}.
\end{align}
Varying the action for $\bar{\chi}$ and $\psi$ results in the bulk coupled EOMs
\begin{align}
    i\bar{\sigma}^\mu\partial_\mu\chi+\partial_5 \bar{\psi} -\frac{c+2}{z}\bar{\psi}=0,\qquad i \sigma^\mu\partial_\mu \bar{\psi}-\partial_5\chi +\frac{2-c}{z}\chi = 0\,,
\end{align}
as well as a boundary term 
\begin{equation}
    \frac{1}{2}\int \text{d}^4x \Big[ \Big(\frac{R}{z}\Big)^4(\psi\delta\chi+\delta\psi \chi-\bar{\chi}\delta\bar{\psi}-\delta\bar{\chi}\bar{\psi})\Big]^{z=R^\prime}_{z=R}=0,
\end{equation}
whose vanishing determines the BCs. We will denote with $(-)$ a Dirichlet BC for the left handed (LH) mode and with $(+)$ a Dirichelt BC for the right handed (RH) mode, implying a Neumann-like BC for the other chirality, leading to (for $z=R,R^\prime$)
\begin{align}
    \Psi(-)&:  \chi(x,z)=0 \implies \partial_5 \psi (x,z) = \frac{2+c}{z} \psi(x,z), \notag \\
    \Psi(+)&:  \psi(x,z)=0 \implies \partial_5 \chi (x,z) = \frac{2-c}{z} \chi(x,z).\hspace{-12mm}
\end{align}

The 5D EOMs are solved via a KK decomposition
\begin{equation}\label{KKdecomp}
    \chi(x,z)=\sum_n f_{n,L}(z)\chi_n(x), \quad \bar{\psi}(x,z)=\sum_n f_{n,R}(z)\bar{\psi}_n(x),
\end{equation}
where each 4D spinor satisfies the 4D Dirac equation with mass $m_n$ 
\begin{align}\label{EOM}
    i\bar{\sigma}^\mu\partial_\mu\chi_n -m_n\bar{\psi}_n=0,\qquad
    i\sigma^\mu\partial_\mu\bar{\psi}_n -m_n\chi_n=0.
\end{align}
The EOMs are then reduced to first order differential equations for the bulk profiles, reading 
\begin{align}
\label{eq:fermEOM}
    f_{n,R}^\prime +m_n f_{n,L} -\frac{c+2}{z}f_{n,R}=0, \qquad
   f_{n,L}^\prime -m_n f_{n,R}+\frac{c-2}{z} f_{n,L} =0.
\end{align}
The profiles themselves are orthonormalized, such that the kinetic terms for the 4D fields are canonically normalized and do not mix, resulting in the conditions
\begin{align}
    \int_R^{R^\prime} \text{d}z \Big(\frac{R}{z}\Big)^4f_{n,L}(z)f_{m,L}(z)= \int_R^{R^\prime} \text{d}z \Big(\frac{R}{z}\Big)^4f_{n,R}(z)f_{m,R}(z)=\delta_{n,m}.
\end{align}

For a zero mode, $m_n=0$, the above equations decouple and we find 
\begin{align}\label{zeromodeFermion}
    f_{0,L} = \frac{1}{\sqrt{R^\prime}}\Big(\frac{z}{R}\Big)^2\Big(\frac{z}{R^\prime}\Big)^{-c}f(c), \qquad
    f_{0,R}=\frac{1}{\sqrt{R^\prime}}\Big(\frac{z}{R}\Big)^2\Big(\frac{z}{R^\prime}\Big)^{c}f(-c),
\end{align}
with the \textit{flavor function}
\begin{equation}
    f(c)=\frac{\sqrt{1-2c}}{\sqrt{1-\left(R^\prime/R\right)^{2c-1}}},
\end{equation}
measuring the overlap of the respective fermion-profile with the IR brane.
Such zero modes only appear for specific BCs: a $\Psi(+,+)$ BC will give rise to a massless LH fermion $\chi_0 (x)$, while a $\Psi(-,-)$ BC results in a massless RH fermion~$\psi_0 (x)$. For mixed BCs no zero mode appears. The SM fermionic content will be described by such massless chiral modes before EWSB. We will omit the zero index for the profiles of the zero modes subsequently. For higher modes with $m_n\neq0$ one can solve the EOMs by separating the coupled equations \eqref{eq:fermEOM} into two independent second order equations of Bessel type
\begin{align}\label{fermion}
    f_{n,L}^{\prime\prime} - \frac{4}{z}f_{n,L}^\prime +(m_n^2-\frac{c^2+c-6}{z^2})f_{n,L} = 0, \qquad
    f_{n,R}^{\prime\prime} - \frac{4}{z}f_{n,R}^\prime +(m_n^2-\frac{c^2-c-6}{z^2})f_{n,R} = 0.
\end{align}
The warped sine and cosine functions~\cite{Falkowski:2006vi}
\begin{align}\label{warpedcosinesin}
    S(z,m,c)\!&=\!\frac{\pi}{2} mR \Big(\frac{z}{R}\Big)^{\!1/2+c}\Big(\!J_{1/2+c}(mR)Y_{1/2+c}(mz)\!-\!J_{1/2+c}(mz)Y_{1/2+c}(mR)\Big),\notag \\C(z,m,c)\!&=\!\frac{\pi m R}{2\cos(c\pi)} \Big(\frac{z}{R}\Big)^{\!1/2+c}\Big(\!J_{-1/2+c}(mR)J_{-1/2-c}(mz)\!+\!J_{1/2+c}(mz)J_{1/2-c}(mR)\Big),
\end{align}
provide convenient solutions in terms of Bessel functions such that the BCs $C(R,m_n,c)= 1$, $C'(R,m_n,c)= 0$, $S(R,m_n,c)=0$ and $S'(R,m_n,c)=m_n$ are satisfied and
from which general massive solutions to equation \eqref{fermion} can be constructed as
\begin{align}
    f_{n,L}(z)=& \Big(\frac{R}{z}\Big)^{c-2} (b_n S(z,m_n,c)-a_n C(z,m_n,c)), \notag \\
    f_{n,R}(z) = & \Big(\frac{R}{z}\Big)^{-c-2} (a_n S(z,m_n,-c)+b_n C(z,m_n,-c)).
\end{align}
The solutions for $f_{n,R}$ can be constructed from $f_{n,L}$ by flipping $c\rightarrow -c$ and applying the coupled EOMs at $z=R$. One can easily retrieve the zero mode solutions, as for $m_n=0$ the cosine becomes $1$. We will need the above solutions in particular when describing the KK excitations of the leptons and quarks  in Appendix \ref{AppendixB}, mediating important decays such as $\mu\rightarrow e\gamma$.

\section{Elements of SU(6) gauge-Higgs grand unification}\label{sec:Elements}
\subsection{Gauge bosons}
We outline here the basic structure of SU(6) GHGUTs . The bulk gauge symmetry is SU(6) which is broken to subgroups on the UV and the IR brane via $(-)$ BCs. In \cite{Angelescu:2021nbp} the UV brane symmetry was chosen to be SU(5) and the IR brane symmetry was $G_{\textrm{SM}}$. This choice is reflected in the BCs for the gauge bosons $A_\mu^a$
\begin{equation}\label{eq:bcs}
\begin{split}
A_\mu&= \left( \begin{array}{cc|ccc|c}
 \textcolor{blue}{(++)} & \textcolor{blue}{(++)} & (+-) & (+-) & (+-) & (--)\\
 \textcolor{blue}{(++)} & \textcolor{blue}{(++)} & (+-) & (+-) & (+-) & (--)\\
 \hline
 (+-) & (+-) & \textcolor{blue}{(++)} & \textcolor{blue}{(++)} & \textcolor{blue}{(++)} & (--)\\
 (+-) & (+-) & \textcolor{blue}{(++)} & \textcolor{blue}{(++)} & \textcolor{blue}{(++)} & (--)\\
 (+-) & (+-) & \textcolor{blue}{(++)} & \textcolor{blue}{(++)} & \textcolor{blue}{(++)} & (--)\\
 \hline
 (--) & (--) & (--) & (--) & (--) & (--)\\
\end{array} \right),
\end{split}
\end{equation}
where a broken symmetry corresponds to a Dirichlet BC $(-)$ for the corresponding gauge field.
In this $6\times 6$ matrix we identify the unbroken groups SU(2)\textsubscript{L} and SU(3)\textsubscript{c} by the highlighted $2\times 2 $ and $3\times 3$ submatrices, respectively. The off-diagonal degrees of freedom are of signature $(+,-)$ and correspond to massive vector bosons that are charged both under SU(2)\textsubscript{L} and SU(3)\textsubscript{c}, carrying the quantum numbers of the usual $X,Y$ gauge bosons of 4D SU(5) GUTs, i.e. forming  an SU(2)\textsubscript{L} doublet of SU(3)\textsubscript{c} triplets $(X^{4/3},Y^{1/3}) \sim \mathbf{(3^\ast,2)_{5/6}}$.
As discussed, only $(-,-)$ gauge boson modes will lead to a corresponding massless scalar at tree-level, although at loop-level it will receive a mass. Thus in SU(6) GHGUT, the scalar spectrum will consist of $11$ scalar modes that decompose as $\bf{(1,2)_{1/2}\oplus (3,1)_{-1/3}\oplus (1,1)_0}$ under the SM gauge group. Apart from the Higgs, we therefore find a scalar leptoquark with quantum numbers $\bf{(3,1)_{-1/3}}$  and a scalar singlet.

Importantly, one can consider an alternative structure in the gauge model to the one considered above, which flips the symmetries on the branes. This \textit{flipped} model has an unbroken SU(5) gauge symmetry on the IR brane and the SM gauge symmetry on the UV brane which results in the gauge BCs
\begin{equation}\label{eq:bcsflipped}
\begin{split}
A_\mu&= \left( \begin{array}{cc|ccc|c}
 \textcolor{blue}{(++)} & \textcolor{blue}{(++)} & (-+) & (-+) & (-+) & (--)\\
 \textcolor{blue}{(++)} & \textcolor{blue}{(++)} & (-+) & (-+) & (-+) & (--)\\
 \hline
 (-+) & (-+) & \textcolor{blue}{(++)} & \textcolor{blue}{(++)} & \textcolor{blue}{(++)} & (--)\\
 (-+) & (-+) & \textcolor{blue}{(++)} & \textcolor{blue}{(++)} & \textcolor{blue}{(++)} & (--)\\
 (-+) & (-+) & \textcolor{blue}{(++)} & \textcolor{blue}{(++)} & \textcolor{blue}{(++)} & (--)\\
 \hline
 (--) & (--) & (--) & (--) & (--) & (--)\\
\end{array} \right).
\end{split}
\end{equation}
We will refer to this model as $G_{\textrm{SM}}^{(\textrm{UV})}$ in reference to its SM gauge symmetry on the UV brane, while we refer to the previous/original model \eqref{eq:bcs} as $G_{\textrm{SM}}^{(\textrm{IR})}$. 
The choice between these two models will have important consequences concerning the gauge coupling running which we leave for a future analysis.\footnote{In the  $G_{\textrm{SM}}^{(\textrm{IR})}$ model, the GUT symmetry would be broken only below the IR/compositeness scale $1/R^\prime \sim\,{\rm TeV}$, where unification would need to happen, while in the $G_{\textrm{SM}}^{(\textrm{UV})}$ model it would be broken already at the Planck scale (in the holographic dual interpretation just the $G_{\rm SM}$ subgroup of the full GUT group would actually be gauged \cite{Randall:2001gc}). Also a (more conventional) scenario where the GUT breaking is realized slightly below the Planck scale via a UV-brane localized scalar sector is worth investigating.} Aside from the running, the only phenomenological difference in the gauge sector is that the nature of the $X,Y$ gauge bosons switches from $(+,-)$ to $(-,+)$ which means they become heavier and do not couple to UV localized fermions (to good approximation) in constrast with the $(+,-)$ $X,Y$ gauge bosons from the $G_{\textrm{SM}}^{(\textrm{IR})}$ model that are light and have to a good approximation universal couplings to the three SM fermion generations. Of course, when including the fermions, the flipping of the gauge symmetries can have more important consequences for the flavor of the model, which we will explore in this paper in detail. In particular, the model we will identify as the phenomenologically most viable in Section~\ref{sec:Results} will be of $G_{\textrm{SM}}^{(\textrm{UV})}$ type.
\subsection{Fermions}
The fermion sector of the setup comes in full SU(6) representations, with the minimal embedding consisting of a $\bf{1}, \bf{6}, \bf{15}$ and a $\bf{20}$, which are the four smallest SU(6) representations \cite{Angelescu:2021nbp}. If one does not want to introduce brane matter fields, this is indeed a necessary condition and, as it turns out, a sufficient condition for viable SU(6) GHGUT. The $\bf{20}$ is a necessary bulk fermion in SU(6) GHGUT as it is the smallest representation containing a fermion with the quantum numbers of the up-type quark, ${\bf (3,1)_{2/3}}$, that connects through the $A_5$ Higgs to a fermion with the quantum numbers of the doublet quark, ${\bf (3,2)_{1/6}}$. In other words, the covariant derivative of a $\bf{20}$ contains a Yukawa coupling for the up-quark. The same reasoning can be applied to the Yukawa coupling for the down-quark and the electron for which the $\bf{15}$ is the smallest bulk fermion. Note that although the $\bf{15}$ contains an up-type quark, it does not connect to a doublet and therefore cannot give a Yukawa coupling for the up quark, indicating again the necessity of having a $\bf{20}$. The necessity of a $\bf{6}$ follows from the requirement of a neutrino yukawa coupling. Therefore on general grounds, one needs minimally a $\bf{20},\bf{15}$ and a $\bf{6}$. 

Such a setup, however, would lead to heavy neutrinos (see Section~\ref{sec:A5scalarsec}). This can be minimally solved by introducing an additional $\bf{1}$, as will be demonstrated. 
Furthermore, the construction as such would lead to serious problems since we would have two distinct quark doublets, one in the $\bf{20}$ for the up quark Yukawa and one in the $\bf{15}$ for the down quark Yukawa. Similarly, we would have two lepton doublets, one in the $\bf{15}$ for the electron Yukawa and one in the $\bf{6}$ for the neutrino Yukawa. We solve this issue by introducing brane masses on the UV/IR boundaries that connect the doublets such that one physical massless doublet survives while the other decouples, similar to what happens in conventional models of gauge-Higgs unification.

After this brief overview, we will enter in more detail the general embedding of the SM fermions into these four SU(6) bulk fermions, $\bf{20},\bf{15},\bf{6}$, and $\bf{1}$, which is given by
\begin{align}
\label{Mix-embedding}
     {\bf 20} \rightarrow & {\bf 10} = q^\prime {\bf ( 3,2)}_{1/6}  \oplus {\bf (3^*,1)}_{-2/3} \oplus e^{\prime c} {\bf (1,1)}_1  \notag \\ 
     & {\bf 10^*} = {\bf (3^*,2)}_{-1/6}  \oplus u {\bf (3,1)}_{2/3} \oplus {\bf (1,1)}_{-1}, \notag \\
    {\bf 15} \rightarrow &{\bf 10} = q {\bf (3,2)}_{1/6}  \oplus {\bf (3^*,1)}_{-2/3} \oplus e^c {\bf (1,1)}_1 \notag\\  
    & {\bf 5} = d^\prime {\bf  (3,1)}_{-1/3}\oplus l^{\prime c} {\bf (1,2)}_{1/2}, \notag \\
    {\bf 6} \rightarrow & {\bf 5} = d  {\bf (3,1)}_{-1/3} \oplus l^c  {\bf (1,2)}_{1/2} \notag \\
    & {\bf 1} = \nu^c {\bf (1,1)}_0, \notag \\
        {\bf 1} \rightarrow & {\bf{1}} = \nu^{\prime c} {\bf (1,1)}_{0}\,,
\end{align}
where we display the decomposition of the multiplets under SU(5)$\subset$ SU(6).
Since fermions in 5D come in Dirac representations, we have for each bulk fermion a LH and a RH component. We denote in \eqref{Mix-embedding} the 5D fermions by the usual symbols of the SM fermions they contain. In the following, we will work with the LH/RH components of the 5D fields and therefore add a L/R subscript to indicate their embedding, employing $\Psi^c_L\equiv(\Psi^c)_L=(\Psi_R)^c$\footnote{Therefore the label L/R designates the transformation properties under the Lorentz group. For the SM leptons, embedded as conjugates, this means the singlet and doublet electron are denoted by respectively $e^c_L$ and $l^c_R$.}. 
The primed fermions contain no zero modes, but carry the same SM charges as some of the SM zero modes and will mix with the latter once boundary terms are added, a phenomenon well-known in models of gauge-Higgs unification. As we will explore below, this effect will lead to kinetic mixing and as a consequence the physical mass eigenstates will reside in both the primed and unprimed fermions. The notable exception is the RH up-type quark which has no primed partner and therefore no mixing will be present. We define as {\it exotics} the remaining fermions without any labels in \eqref{Mix-embedding}, which do not mix with the SM mass eigenstates, and are necessary to complete the bulk fermions in SU(6) representations. In SU(6) GHGUT there are in total four different exotic fields, which after  considering brane masses and EWSB combine into three exotic sectors: one electron-like exotic $\tilde{E}$ residing in the $\bf{20}$, one down-type $\tilde{D}$ exotic in the $\bf{20}$, and one up-type $\tilde{U}$ exotic residing in (three different fields within) the $\bf{15}$ and the $\bf{20}$.

Once the gauge sector and fermion embedding is specified we still need to add boundary masses in order to get the correct SM spectrum. Indeed, without brane masses, not all LH fermions connect to their RH  counterparts resulting in the corresponding SM fermions remaining massless. Due to the restrictive SU(5) gauge symmetry there are only three possible masses that can be written down on the corresponding brane, of which only two are of main interest for SU(6) GHGUT. The first connects the  $\bf{10}$ of the $\bf{20}$ to the $\bf{10}$ of the $\bf{15}$, while the second connects the $\bf{5}$ of the $\bf{15}$ to the $\bf{5}$ of the $\bf{6}$. The former provides a link for the quark doublet and for the electron singlet to their respective { \it primed} partners and is therefore denoted by $M_{q/e}$, while the latter connects the RH down quark and lepton doublet to their respective { \it primed} partners and is correspondingly denoted with $M_{d/l}$\footnote{On the other hand, in case the gauge symmetry on the brane is $G_{\textrm{SM}}$, the brane masses connecting the SU(5) $\bf{10}$s and $\bf{5}$s will actually decompose into all possible terms connecting individual SM fermions in a $G_{\rm SM}$ invariant way.}. There is also a connection between the two singlets in order to obtain the correct neutrino mass, denoted with $M_\nu$, but that one is of less phenomenological interest for our present analysis; the same holds for components of the brane masses in the exotics sector.  Symbolically we denote these connections as
\begin{alignat}{4}
      {\bf 20} &\rightarrow  && \bf{10} \quad \oplus \quad && \bf{10^*} \quad     && \notag \\
     \quad     &  \quad      &&  \Updownarrow   M_{q/e}            && \quad               && \notag \\
      {\bf 15} & \rightarrow && \bf{10} \quad \oplus \quad && \bf{5} \quad \oplus && \notag \\
     \quad     & \quad       &&\quad                       && \hspace{-3pt} \Updownarrow M_{d/l}      && \notag \\
     {\bf 6}   & \rightarrow &&\quad                       && \bf{5} \quad \oplus && \bf{1} \notag \\
     \quad     & \quad       &&\quad                       && \quad               && \hspace{-3pt}   \Updownarrow M_{\nu} \notag \\
      {\bf 1}  & \rightarrow &&\quad                       && \quad               && \bf{1} \notag\qquad \,.
\end{alignat}

Apart from these three SU(5) invariant boundary masses there is also the question of where to place the boundary mass: on the UV brane or on the IR brane. The placement is determined by which one is allowed by the fermion BCs -- indeed most of the boundary masses one could write down will simply vanish due to these BCs. Therefore in total there are 4 different versions of the SU(6) GHGUT: one where $M_{q/e}$ and $M_{d/l}$ are UV brane masses, one where both are IR brane masses and two \textit{mixed} models. In addition, we can consider these four models but with the \textit{flipped} gauge symmetries, $G_\textrm{SM}^{(\textrm{UV})}\leftrightarrow G_\textrm{SM}^{(\textrm{IR})}$, leading to a total of 8 different models. Flavorwise these different models become phenomenologically very distinct. In Table \ref{tab:my-table} we summarize the main problems with these different models as will become clear in Section~\ref{sec:Results}. In the end the most natural model is one with IR brane masses and the SM gauge symmetry on the UV which will be the main subject of analysis in this paper. 

\begin{table}[h]
\centering
\begin{tabular}{|l|l|l|}
\hline                                                                                       &$G_\textrm{SM}^{(\textrm{UV})}$ & $G_\textrm{SM}^{(\textrm{IR})}$ \\ \hline
\begin{tabular}[c]{@{}l@{}}IR brane masses: $M_{q/e}, M_{d/l}$\end{tabular}   &                \qquad \cmark        & Light exotics:    \xmark     \\ \hline
\begin{tabular}[c]{@{}l@{}}UV brane masses:  $M_{q/e}, M_{d/l}$\end{tabular} & Large FCNCs:   \xmark        & Large FCNCs:  \xmark         \\ \hline
\begin{tabular}[c]{@{}l@{}}IR brane masses:  $M_{q/e}$\\ UV brane masses: $M_{d/l}$\end{tabular}  & CKM matrix :   \xmark        & CKM matrix :  \xmark         \\ \hline
\begin{tabular}[c]{@{}l@{}}IR brane masses:  $M_{d/l}$\\ UV brane masses: $M_{q/e}$\end{tabular}  & CKM matrix :  \xmark         & CKM matrix : \xmark          \\ \hline
\end{tabular}
\caption{Summary of the different incarnations of SU(6) GHGUT. The mixed models fail to reproduce the CKM matrix while the UV-brane mass models suffer from large FCNCs. The IR-brane model with the SM on the UV brane is the most suitable as the IR variant suffers from light exotic fermions.}
\label{tab:my-table}
\end{table}

\subsection{Boundary conditions}\label{sec:BC}

As mentioned, in order to make the model phenomenologically viable, we need to introduce the boundary mass terms that will connect the bulk fermions. These terms will modify the usual BCs for the bulk fermions. The general derivation of BCs for fermions on an interval was studied in \cite{Csaki:2003sh,Angelescu:2019viv}, out of which we will only need the simple case of a brane mass. 

We consider two 5D fermions $\Psi_1,\Psi_2$
\begin{equation}
    \Psi_1=\begin{pmatrix}
    \chi_1 \\ \bar{\psi}_1
    \end{pmatrix}, \qquad
    \Psi_2=\begin{pmatrix}
    \chi_2 \\ \bar{\psi}_2
    \end{pmatrix}.
\end{equation}
In order to have non-vanishing brane masses betweens these two 5D fermions, they must obey opposite BCs. Chosing a $[+]$ BC for $\Psi_1$ on the IR brane, leading to $\psi_1(z=R^\prime)=0$, and a $[-]$ BC for $\Psi_2$, inducing $\chi_2(z=R^\prime)=0$, a non-vanishing invariant mass term can be written on the IR brane, reading
\begin{equation}
    \int \textrm{d}^4x \Big(\frac{R}{R^\prime}\Big)^4(M\psi_2 \chi_1+\textrm{h.c.})_{z=R^\prime}.
\end{equation}
The effect of such an IR brane mass can be studied by pushing the Lagrangian into the bulk to $z=R^\prime-\epsilon$ \cite{Csaki:2003sh}, such that the bulk EOMs contain a dirac delta  
%%%%%%%%%%%%%%%%%%
\begin{align}
    i\bar{\sigma}^\mu \partial_\mu \chi_1 +\partial_5 \bar{\psi}_1-\frac{(c+2)}{z}\bar{\psi}_1+M^* \bar{\psi}_2 \delta(y-R^\prime+\epsilon)=0, \notag \\
    i\bar{\sigma}^\mu \partial_\mu \bar{\psi}_2 -\partial_5 \chi_2+\frac{(2-c)}{z}\chi_2+M \chi_1 \delta(y-R^\prime+\epsilon)=0.
\end{align}
%%%%%%%%%%%%%%%%%%
Integrating these two equations over the extra dimension results in two jump conditions
\begin{align}
    [\bar{\psi}_1]_{|R^\prime-\epsilon}=-M^* \bar{\psi}_{2 |R^\prime-\epsilon}, \notag \\
    [\chi_2]_{|R^\prime-\epsilon}=M \chi_{1 |R^\prime-\epsilon}.
\end{align}
In order for the defining BCs at $z=R^\prime$ to be respected, at $z=R^{\prime -}\!\!\equiv\! \lim_{\epsilon \to 0} [R^{\prime}\!-\!\epsilon]$ we obtain the BCs
\begin{align}
    \bar{\psi}_{1|R^{\prime -}}=M^* \bar{\psi}_{2|R^{\prime -}}, \notag \\
    \chi_{2|R^{\prime -}} = -M \chi_{1|R^{\prime -}}.
\end{align} 
Assuming similar brane masses and BCs as before, but now on the UV brane,
\begin{equation}
    \int \textrm{d}^4x (M\psi_2 \chi_1+\textrm{h.c.})_{z=R}\,,    
\end{equation}
a similar derivation leads to
\begin{align}
    \bar{\psi}_{1|R^+}=-M^* \bar{\psi}_{2|R^+}, \notag \\
    \chi_{2|R^+}=M \chi_{1|R^+}.
\end{align}

\subsection{Kinetic mixing}\label{kineticmixing}

We now explore the mixing that results from the modified BCs due to the brane masses. Indeed, because of the brane masses connecting the different bulk fermions, the original SM zero modes (unprimed fields in \eqref{Mix-embedding}) and non-zero modes that carry the same SM charges (primed fields) mix, with the physical zero-mode eigenstates residing in a linear combination of both. For example, consider $e_L^c$ (which corresponds to the SU(2)\textsubscript{L} singlet due to the charge conjugation) within the bulk $\bf{15}$, which features a LH zero mode (corresponding to (+,+) BCs) that will mix with the $e_L^{\prime c}$ in the bulk $\bf{20}$. Symbolically, the 4D electron field $e_L^{c,0}(x)$ is embedded within these 5D fields as
\begin{equation}
    e_L^c(x,z)=f_{e_L^c}(z) e_L^{c,0} (x)+..., \qquad e_L^{\prime c}(x,z)=f_{e_L^{\prime c}}(z) e_L^{c,0} (x)+...\,.
\end{equation}
The addition of a brane mass on the IR connecting these two fields changes the BC for the primed field into $e^{\prime c}_L(z=R^\prime)=M_e e^{c}_L(z=R^\prime)$, which allows us to find the two profiles
\begin{align}
f_{e_L^c}(z)=&\frac{1}{\sqrt{R^\prime}}\Big(\frac{z}{R}\Big)^2\Big(\frac{z}{R^\prime}\Big)^{-c_{15}}f(c_{15}),\notag \\
f_{e_L^{\prime c}}(z)=&\frac{1}{\sqrt{R^\prime}}\Big(\frac{z}{R}\Big)^2\Big(\frac{z}{R^\prime}\Big)^{-c_{20}}M_e f(c_{15}),
\end{align}
using equation \eqref{zeromodeFermion}. In Appendix \ref{AppendixA} we list all the profiles for our eventual model with IR-brane localized masses. Note that the generalization to three generations requires these profiles to be vectors in generation space and $M_e$ is consequently a $3\times3$ matrix while
\begin{equation}
    f_{c_{15}}\equiv\textrm{diag}(f(c_{15,1}),f(c_{15,2}),f(c_{15,3}))
\end{equation}
is a diagonal matrix denoting the localization of the three different generations. Inspecting the 4D effective kinetic term of $e_L^{c,0} (x)$ we see how the brane mixing modifies the normalization of the 4D fields, resulting in 
\begin{align}\label{eq:electronNorm}
    \mathcal{L}\supset&i\bar{e}_L^{c,0} \int d z \Big((\frac{1}{\sqrt{R^\prime}}\Big(\frac{z}{R}\Big)^2\Big(\frac{z}{R^\prime}\Big)^{-c_{15}}f_{c_{15}})^2+(\frac{1}{\sqrt{R^\prime}}\Big(\frac{z}{R}\Big)^2\Big(\frac{z}{R^\prime}\Big)^{-c_{20}}M_e f_{c_{15}})^2 \Big) \gamma^\mu \partial_\mu e_L^{c,0} \notag \\
    =& i\bar{e}_L^{c,0}(1+f_{c_{15}}M_e^\dagger f_{c_{20}}^{-2} M_e f_{c_{15}}) \gamma^\mu \partial_\mu e_L^{c,0} \notag \\
    \equiv& i\bar{e}_L^{c,0}K_{e_L^c}\gamma^\mu \partial_\mu e_L^{c,0}\,,
\end{align}
which is not canonically normalized for non-zero brane mass $M_e$.

However, with the kinetic matrix $K_{e_L^c}$ being hermitian, we can redefine the zero modes as $e_L^{c,0}\rightarrow K_{e_L^c}^{-1/2}e_L^{c,0}$ to go from the flavor basis to the kinetic basis. Such transformation on the fermion fields - before diagonalizing the mass matrix - is crucial to obtain the correct phenomenology in GHU models.
For example if one starts with a mass matrix $\mathcal{M}_{e^c}$ after EWSB for the electron in the flavor basis (see Section~\ref{sec:A5scalarsec} for scalar/Higgs couplings), one first has to transform this matrix to the kinetic basis via the redefinition introduced above -- and its counterpart $K_{l_R^c}^{-1/2}$ for the lepton doublet $l_R^c$ mixing with $l_R^{\prime c}$ -- follwed by the usual bi-unitary mass diagonalization of the (conjugate) leptons with rotation matrices $U_{{L/R},e^c}$, to obtain the physical spectrum
\begin{equation}\label{eq:diagonalization}
    \mathcal{M}_{e}^{SM}=U_{L,e^c}^\dagger K_{l_R^c}^{-1/2\dagger} \mathcal{M}_{e^c} K_{e_L^c}^{-1/2} U_{R,e^c}.
\end{equation}
Here, $\mathcal{M}_{e}^{SM}$ is the diagonal matrix containing the masses of the charged leptons. After a similar transformation in the neutrino sector we can find the PMNS matrix as the product $V_{\textrm{PMNS}}\equiv U_{L,\nu}^\dagger U_{L,e} = U_{L,\nu^c}^T U_{L,e^c}^*$ in terms of the rotation matrices of the conjugate lepton mass matrices $\mathcal{M}_{e^c},\mathcal{M}_{\nu^c}$ and similarly for the CKM matrix in the quark sector, $V_{\textrm{CKM}}=U_{L,u}^\dagger U_{L,d}$. 
Below, we will discuss the various resulting gauge and scalar couplings in our model in the flavor basis. The couplings in the mass basis are then obtained by subsequent rotations to the kinetic and mass basis.

\subsection{Gluon and photon couplings}
We begin by exploring the couplings of the fermions to the gluon $G_\mu$ and photon $A_\mu$. The profiles of the zero mode gluon $f_{0,G}(z)$ and photon $f_{0,A}(z)$ are flat along the extra dimension, in consequence of the intact color and electromagnetic symmetry due to the (+,+) nature of these gauge bosons. We will therefore prefer to indicate these profiles by the more general $f_{0,(+,+)}(z)$ to distinguish them from different types of gauge bosons that we will encounter in later sections. This flatness means the couplings are identical to those of the SM. This can be easily confirmed for example by looking at the gluon couplings to the quarks in the flavor basis. Consider the couplings to the LH quarks
\begin{align}\label{eq:GluonLagrangian}
    \mathcal{L}\supset g_5\big( \bar{q}_{L}^\alpha \gamma^\mu q_{L}^\beta +\bar{q}_{L}^{\prime\alpha} \gamma^\mu  q_{L}^{\prime \beta} )T^a_{\alpha\beta}G_\mu^a,
\end{align}
where $T^a_{\alpha\beta}$ are the color generators. Notice that due to kinetic mixing we have to include both the LH doublet from the $\bf{15}$ and from the $\bf{20}$, connected via brane masses. 

The usual normalization of the gauge boson profile \eqref{gaugezeronorm} is such that the kinetic terms are canonically normalized, however we will opt for a different normalization such that the gauge boson interactions with the fermions are simpler, namely 
\begin{equation}
    \int \textrm{d} z \Big(\frac{R}{z}\Big) f_{n,(+,+)}(z)^2 = R \log(R^\prime/R).
\end{equation}
This makes the zero mode profiles particularly simple, $f_{0,(+,+)}(z)=1$. The resulting kinetic terms can in turn be canonically normalized by redefining the gauge fields with a net effect of replacing the 5D coupling $g_5$ with $g_5/(\sqrt{R\log(R^\prime/R)})$ which can be identified with the SM gauge coupling\footnote{At the classical level, this creates the problem in SU(6) GHGUT of having a strong coupling $g_s$ equal to $g$ and a hypercharge coupling $g^\prime=\sqrt{3/5}g$ with the wrong Weinberg angle. Note that this situation is not any different from 4D GUTs: we are relying on the running of the gauge couplings at the quantum level to give the correct measured low-energy couplings. We leave the study of this evolution for future work.}. The coupling of the LH quark, within the $\bf{15}$, to a gluon is then given by
\begin{align}
    S\supset g_s\int \text{d}z\text{d}^4x\Big(\frac{R}{z}\Big)^4(\bar{q}_L\gamma^\mu q_L) G_\mu= &g_s \int \text{d}z \Big(\frac{R}{z}\Big)^4  f_{q_L}^2(z) \int \text{d}^4x (\bar{q}_L^0\gamma^\mu {q_L^0}) G^{0}_\mu+... \notag \\
    =& g_s \int \text{d}^4x  (\bar{q}_L^0\gamma^\mu {q_L^0}) G^{0}_\mu+...,
\end{align}
where we neglect all but the zero modes of the KK tower of the fermions and gauge bosons. Therefore in generation space and in the flavor basis, the couplings will be independent of the localization of the generations and equal to $g_s \mathbf{1}_{3\times 3}$. Since the rotation to the mass basis is unitary, the couplings remain invariant and SM-like. Taking into account that due to the brane masses, the LH quark is also partly living in the $\bf{20}$, the full coupling is  $g_s K_{q_L} \mathbf{1}_{3\times 3} $, with $K_{q_L}$ equal to the kinetic mixing matrix of the LH quarks. Thus after the canonical normalization of the kinetic terms, these couplings will again be $g_s \mathbf{1}_{3\times 3}$, just as in the SM.

In contrast, in the couplings of the fermions to the first KK mode of the gluons and photons (and of the $Z/W$ bosons) the 5D nature of the model is exposed, giving rise to Flavor Changing Neutral Currents (FCNCs) due to the $z$ dependent profile of these modes along the extra dimension. This will make the couplings dependent on the fermion localization or bulk-mass parameter $c$. Therefore in the flavor basis the couplings will still be diagonal but non-universal in generation space. Then, after rotating into the canonically normalized mass basis, the coupling matrix will have off-diagonal entries, inducing FCNCs in the quark sector and the lepton sector. Here we will develop formulae for these couplings. The signature of these KK modes for the photon and gluon is  $(+,+)$, meaning they correspond to unbroken gauge symmetries. The results are also applicable to the KK bosons of $Z/W$ bosons - the effects of EWSB on the proporties of the latter modes are negligible to good approximation. Using the profile $f_{(+,+)}(z)$ from \eqref{plusplusprofile} with the corresponding normalization
%%%%%%%%%%%%%%%%%%%%%%%%%%%
\begin{equation}\label{normKK}
    N_{(+,+)}=\frac{\pi m_{(+,+)}Y_0(m_{(+,+)}R) \sqrt{2\log\frac{R^\prime}{R}}}{2\sqrt{\frac{Y_0^2 (m_{(+,+)}R)}{Y_0^2 (m_{(+,+)}R^\prime)}-1}}\sim\frac{\sqrt{2\log\frac{R^\prime}{R}}}{R^\prime J_1(m_{(+,+)}R^\prime)},
\end{equation}
and the corresponding first KK mass $m_{(+,+)}\sim2.45/R^\prime$, we find the coupling of the (LH) zero mode quark to the first KK gluon to be (ignoring kinetic mixing with the $\bf 20$)
\begin{align}
    g_s\int \text{d}z\text{d}^4x\Big(\frac{R}{z}\Big)^4(\bar{q}_L\gamma^\mu q_L) G_{\mu}= &g_s \Bigg(\int \text{d}z \Big(\frac{R}{z}\Big)^4 f_{(+,+)}(z) f^2_{q_L}(z)\Bigg) \int \text{d}^4x (\bar{q}^0_L\gamma^\mu q^0_L) G^{1}_\mu + ... \,.
\end{align}
%%%%%%%%%%%%%%%%%%%%%%%%%%%

It is convenient to define an effective coupling, $\lambda_{(+,+)}(c)$, capturing the deviation from the strong coupling 
%%%%%%%%%%%%%%%%%%%%%%%%%%%
\begin{align}\label{eq:effectivecoupling}
    \lambda_{(+,+)}(c)\equiv&\Bigg(\int_R^{R^\prime} \text{d}z \Big(\frac{R}{z}\Big)^4 f_{(+,+)}(z)f^2_{q_L}(z) \Bigg) \notag \\
    \approx&N_{(+,+)} R^\prime \Bigg(\int_0^1 dt t^{1-2c}J_1(tm_{(+,+)}R^\prime)f^2(c)-\frac{1}{(m_{(+,+)} R^\prime)\log\Big(\frac{2 }{m_{(+,+)}R}\Big)}\Bigg)\notag \\
    \approx&N_{(+,+)} R^\prime(m_{(+,+)}R^\prime)\Bigg(\frac{0.05}{0.48-0.35c}f^2(c)-\frac{1}{(m_{(+,+)} R^\prime)^2\log\Big(\frac{2 }{m_{(+,+)}R}\Big)}\Bigg),
\end{align}
%%%%%%%%%%%%%%%%%%%%%%%%%%%
where we used approximations for the Bessel functions and the numerical value of the integral in the second and last step, respectively, capturing the behavior in full parameter space. We can understand its form from the KK gluon profile, which consists of an IR--localized peak and a small constant part. The second term in the expression reflects this localization independent contribution, while the first term is amplified for an IR localized fermion which has a large overlap with the peak of the KK gluon profile.

\subsection{Z/W couplings}\label{sec:ZWboson}
A second category of gauge fields are the $Z/W$ bosons. Their first KK profiles behave as the first KK wave functions of the gluon and photon, discussed above, to good approximation - the exactness of which we will discuss below. The $Z/W$ boson zero modes undergo a more drastic change when compared to their gluon and photon counterparts. Since electroweak symmetry is broken, the former deviate from flatness close to the IR brane, in contrast to the photon and gluon zero modes.

The effect of EWSB is contained in the covariant derivative of the Higgs boson
\begin{equation}
    S\supset \int \textrm{d}^4x \textrm{d} z \Big(\frac{R}{z}\Big) D_\mu A_5 D^\mu A_5\,,
\end{equation}
which will make the profiles non-trivial. When the Higgs field gets a potential and vev at the one-loop level (see Section~\ref{sec:Potential}), this term induces a non-diagonal mass matrix for the gauge bosons $A=Z/W$ that get a mass from EWSB
\begin{equation}\label{eq:gaugeBosonMasses}
    \mathcal{L}\supset \frac{(m_A)^2}{2} \sum_{n,m=0} f_{nm} A_\mu^n A^{\mu m},
\end{equation}
where we work in the basis where the original $W^{0,\mu}$ and $B^\mu$ have already been rotated by the Weinberg angle into the $\gamma/Z$ bosons and we define the overlaps $f_{nm}$ as
\begin{align}\label{eq:overlap}
    f_{nm} = \int_{R}^{R^{\prime}}\mathrm{d}z\, \left(\frac{R}{z}\right) \left[f_5(z)\right]^2 f_{n,(+,+)}(z) f_{m,(+,+)}(z) \, ,
\end{align}
with $f_5(z)$ (see Eq. \eqref{A5profile}) the bulk profile of the Higgs and $f_{n,(+,+)}(z)$ that of the n'th KK mode of the gauge boson $A=W/Z$ in the gauge basis. On top of this contribution from EWSB, one has to add the mass from the RS geometry for the KK modes and diagonalize the total mass matrix to obtain the physical gauge bosons. 

Applying this to the $Z$ boson and only including the first KK mode with mass $m_{(+,+)}=2.45/R^\prime$, we find that the mass eigenstates, which we denote by $Z_0^\mu,Z_1^\mu$, are related to the gauge eigenstates, denoted by $Z^{\mu}_{(0)},Z^{\mu}_{(1)}$,  by the linear combinations
\begin{align}\label{eq:linComb}
    Z_0^\mu=&Z^{\mu}_{(0)}-f_{01}\Big(\frac{m_Z}{m_{(+,+)}}\Big)^2 Z^{\mu}_{(1)},\notag \\
    Z_1^\mu=&f_{01}\Big(\frac{m_Z}{m_{(+,+)}}\Big)^2 Z^{\mu}_{(0)} +Z^{\mu}_{(1)},
\end{align}
where $f_{01}=5.47$ is the off-diagonal overlap between the zero and first gauge eigenstates. Higher overlaps $f_{0i}$ quickly decouple ($f_{02}=-1.02$ for instance)\footnote{This is in contrast to IR-brane localized Higgs scenarios where the overlap integrals remain equal (up to a sign) for higher modes~\cite{Agashe:2006iy}.}. This allows us to neglect the mixing coming from the higher modes. The next leading effect in fact does not come from the second KK $Z$ gauge boson, but from the massive $(-,-)$ gauge boson $X^\mu$ with mass $m_{1,(-,-)}=3.83/R^\prime$ that corresponds to the abelian subgroup U(1)\textsubscript{X} $\subset$ SU(6)/SU(5).

The above expression for the 4D mass eigenstates can be reformulated in terms of the corresponding 5D profiles for the mass eigenstates which are convenient to compute the corresponding couplings. Indeed, using the KK decomposition for the $Z$ boson and inverting \eqref{eq:linComb} we find
\begin{align}\label{eq:massbasisgauge}
    Z^\mu&=f_{0,(+,+)} Z_{(0)}^\mu + f_{(+,+)} Z_{(1)}^\mu+...\notag \\&= \underbrace{(f_{0,(+,+)}-f_{01}\Big(\frac{m_Z}{m_{(+,+)}}\Big)^2f_{(+,+)}}_{f_{Z^0}(z)})Z_0^\mu + \underbrace{(f_{01}\Big(\frac{m_Z}{m_{(+,+)}}\Big)^2f_{0,(+,+)}+f_{(+,+)})}_{f_{Z^1}(z)})Z_1^\mu+...,
\end{align}
which reflects the relation between the 4D mass and gauge eigenstates from Eq. \eqref{eq:linComb}.

The above expression merits a few observation. First of all, it is justified to neglect the effect of EWSB on $Z_1^\mu$ as it will be further suppressed by $(m_Z/m_{(+,+)})^2$.
Moreover, the second term above for the bulk profile of $Z_0^\mu$ will give a non-flat contribution to the physical $W/Z$-boson and cannot be omitted: it is of phenomenological importance for FCNCs.
Since we already derived the couplings of a fermion to the first KK mode in the previous section, the effective coupling to a $W/Z$ boson simply follows. If we take as an example the coupling of the $Z$ gauge boson zero mode $Z_\mu^0$ to the electron singlet $e_L^c$ in the $\bf{15}$, the 5D coupling reads
\begin{equation}
    S\supset \frac{g}{c_W}\int \text{d}z\text{d}^4x\Big(\frac{R}{z}\Big)^4(\bar{e}_L^c\gamma^\mu e_L^c) Z_\mu = \frac{g}{c_W} \int \text{d}z \Big(\frac{R}{z}\Big)^4 f_{Z^0}(z) f^2_{e^c_L}(z) \int \text{d}^4x (\bar{e}_L^{c,0}\gamma^\mu e_L^{c,0}) Z^{0}_\mu+...\,,
\end{equation}
The deviations of the couplings of the fermions to the Z from the SM case $g_{\textrm{SM}}=\frac{g}{c_W}$ is then determined by the overlap function 
\begin{align}\label{eq:effectiveZW}
    \lambda_{Z^0}(c)=&\Bigg(\int_R^{R^\prime}\text{d}z \Big(\frac{R}{z}\Big)^4 f_{Z^0}(z)f^2_{e^c_L}(z)  \Bigg) \notag \\
    =&\Bigg(1-f_{01}  \lambda_{(+,+)}(c) \Big(\frac{m_Z}{m_{(+,+)}}\Big)^2 \Bigg).
\end{align}
We see that the coupling receives corrections mostly from physics close to the IR brane. In consequence, UV localized fermions ($c>0.5$ for LH modes) will behave very much SM-like. Since part of the electron singlet is living in the $\bf{20}$ due to brane mixing, we also have to add the contribution from the $Z$ boson coupling to this singlet. The couplings in the mass basis can then be found by successive rotation to the kinetic basis and mass basis.

Let us end this section by commenting on the strength of FCNCs coming from the $Z$ boson versus its first KK excitation. By comparing the localization dependence of the coupling to the first KK $Z$ boson $\lambda_{(+,+)}(c)$ to that of the interaction with the $Z$ boson $\lambda_{Z^0}(c)$, we find the following ratio between FCNCs
\begin{equation}
    \frac{1}{f_{01}(m_Z/m_{(+,+)})^2}\Big(\frac{m_Z}{m_{(+,+)}}\Big)^2 = 1/f_{01},
\end{equation}
taking into account a $(m_Z/m_{(+,+)})^2$ mass suppression. We conclude that in general the first KK $Z$ boson will provide a $1/f_{01}\sim 18\%$ correction to the leading contribution, and we will therefore include the first KK $Z$ boson contribution in flavor observables. 

\subsection{X,Y couplings}
Apart from the SM-like gauge bosons, we also encounter the usual SU(5) $X,Y$ gauge bosons in SU(6) GHGUTs. Their couplings are encoded in the covariant derivative. Restricting ourselves to terms involving only SM fermions we find the following current-couplings in the interaction basis
%%%%%%%%%%%%%%%%%%%%%%
\begin{align}\label{eq:XYcurrents}
    \mathcal{L}\supset\frac{g_5}{\sqrt{2}}\Big(& X_{\mu,\alpha}^{\dagger}(\bar{d}_L^{\alpha} \gamma^\mu e_L^{c}+\bar{d}_R^{\alpha} \gamma^\mu e_R^{c}) -Y_{\mu,\alpha}^{\dagger}(\bar{u}_L^{\alpha}\gamma^\mu e_L^{c}+\bar{d}_R^{\alpha}\gamma^\mu \nu^{c}_R) \notag \\
    +&X_{\mu,\alpha}^{\dagger}(\bar{d}_L^{\prime\alpha} \gamma^\mu e_L^{\prime c}+\bar{d}_R^{\prime\alpha} \gamma^\mu e_R^{\prime c}) -Y_{\mu,\alpha}^{\dagger}(\bar{u}_L^{\prime\alpha}\gamma^\mu e_L^{\prime c}+\bar{d}_R^{\prime\alpha}\gamma^\mu \nu^{\prime c}_R)+ \textrm{h.c.} \Big)\,,
\end{align}
%%%%%%%%%%%%%%%%%%%%%%
where we include both SM-like fermions containing the corresponding zero modes and the primed fermions that will mix with the SM fermions due to brane masses. Note that the RH up quark is not embedded within the same $\bf{10}$ as the electron singlet and quark doublet, instead the latter coming with an exotic quark. This prohibits the dangerous diquark couplings that mediate proton decay.\footnote{More generally the model contains a global baryon number conservation prohibiting proton decay \cite{Angelescu:2021nbp}.} 

To evaluate the couplings above, we have to perform overlap functions between the $X,Y$ gauge bosons and the different zero modes in the interaction basis and afterwards rotate to the mass basis. In our models the $X,Y$ gauge bosons come in two variants of BCs: $(+,-)$ modes and $(-,+)$ modes, corresponding to the SM gauge group residing on the IR brane or on the UV brane, respectively. In the case of $(+,-)$ gauge bosons, the first KK mode has a mass $m_{(+,-)}\sim 0.25/R^\prime$ and we find the corresponding form of the profile in equation \eqref{plusplusprofile}. Due to the mass being small with respect to $1/R^\prime$, one can expand the arguments of the Bessel functions in $m_{(+,-)} z$, leading to
%%%%%%%%%%%%%%%%%%%%%%
\begin{equation}
    f_{(+,-)}(z)\approx1+\frac{(m_{(+,-)}z)^2}{2}\log\Big(\frac{m_{(+,-)}R}{2}\Big).
\end{equation}
%%%%%%%%%%%%%%%%%%%%%%
With this approximation the effective coupling to the LH fermion zero modes, with profile $f_L(z)$, becomes (normalized similar to \eqref{eq:effectiveZW}, with a factor $g/\sqrt 2$ pulled out)
%%%%%%%%%%%%%%%%%%%%%%
\begin{align}
    \lambda_{(+,-)}(c)=&\Bigg(\int_R^{R^\prime} \Big(\frac{R}{z}\Big)^4 f_{(+,-)}(z)f_{L}^2(z) \Bigg) \notag \\
    =&\Bigg(1-\frac{m_{(+,-)}^2R^{\prime 2}}{2(3-2c)}\log\Big(\frac{2}{m_{(+,-)}R}\Big)f^2(c)\Bigg).
\end{align}
%%%%%%%%%%%%%%%%%%%%%%
Consequently, UV localized fermions ($c>0.5$ for LH modes) will couple to the $X,Y$ bosons in a fashion very similar to usual 4D GUTs (although the mass is radically different). 

In the models where the SM gauge symmetry resides on the UV brane, $G_\textrm{SM}^{(\textrm{UV})}$, the $X,Y$ gauge bosons become of $(-,+)$ nature and their couplings to fermions get very different. Already from the UV BC, implying a vanishing profile, it follows that the more a fermion is UV localized, the smaller the coupling will become. The $(-,+)$ gauge-boson wavefunctions are given in equation \eqref{minusminusprofile} resulting in the approximately normalized wave function for the first KK mode with mass $m_{(-,+)}=2.40/R^\prime$
%%%%%%%%%%%%%%%%%%%%%%
\begin{equation}
    f_{(-,+)}(z)\approx \frac{\sqrt{2\log\frac{R^\prime}{R}}}{R^\prime J_1(m_{(-+)} R^\prime)}z J_1(m_{(-+)} z).
\end{equation}
%%%%%%%%%%%%%%%%%%%%%%
The effective coupling to zero mode fermions can then be computed from the following overlap function 
%%%%%%%%%%%%%%%%%%%%%%
\begin{align}
    \lambda_{(-,+)}(c)=&\Bigg(\int_R^{R^\prime}\text{d}z \Big(\frac{R}{z}\Big)^4 f_{(-,+)}(z)f_{L}^2(z) \Bigg) \notag \\
    \approx&\frac{\sqrt{2\log\frac{R^\prime}{R}}}{J_1(m_{(-+)} R^\prime)} f^2(c) \Bigg(\int_0^1 dt t^{1-2c}J_1(m_{(-+)} R^\prime t)\Bigg)\notag \\
    \approx &\frac{\sqrt{2\log\frac{R^\prime}{R} }}{J_1(m_{(-+)} R^\prime)}(m_{(-+)} R^\prime) \frac{0.05}{0.48-0.35c}f^2(c),
\end{align}
%%%%%%%%%%%%%%%%%%%%%%
where we use the same approximation for the overlap function as in \eqref{eq:effectivecoupling}. Note that in contrast to interactions with the KK mode of a $(+,+)$ state, this coupling can be turned to zero for a sufficiently UV localized LH fermion $c>0.5$ while the $(+,+)$ and $(+,-)$ modes are non-zero throughout the whole bulk and their couplings to fermions cannot be suppressed (unless for extreme IR localization in the latter case).

In the $G_{\textrm{SM}}^{\textrm{UV}}$ models, therefore the resulting $(-,+)$ $X,Y$ gauge bosons will only play a very minor role in terms of flavor, as the first two generations of fermions will indeed be UV localized and only couple feebly to the $X,Y$ states (moreover their masses are an order of magnitude larger than those of the $(+,-)$ gauge bosons). In contrast, in the $G_{\textrm{SM}}^{\textrm{IR}}$ models, the $X,Y$ bosons act very much like heavier $W/Z$ gauge boson with nearly flat extra dimensional profile and thus couple to the fermions throughout the whole bulk. 

\subsection{\texorpdfstring{$A_5$}{A\_5} scalars}\label{sec:A5scalarsec}
One of the most striking consequence of SU(6) GHGUTs is the presence of an extended scalar sector consisting of a colored triplet and a singlet in addition to the Higgs. We shall omit the singlet for the time being, as it only mediates SM--exotic couplings in contrast to the triplet that mediates SM-SM couplings  (as does the Higgs).

Being part of a five dimensional gauge field, the scalars are included within the fifth component of the covariant derivative
%%%%%%%%%%%%%%%%%%%%%
\begin{equation}
    \mathcal{L}\supset -i g_5\sum_{F=1,6,15,20}\Big(-\bar{\Psi}_{F,L}(T^i_F A^i_5)\Psi_{F,R}+\bar{\Psi}_{F,R}(T^i_F A^i_5)\Psi_{F,L}\Big).
\end{equation}
%%%%%%%%%%%%%%%%%%%%%
The sum over $F$ covers the different bulk representations of the fermions while the sum over $i$ corresponds to the 35 generators in the theory. As discussed, there are 11 physical scalars in our model corresponding to the Higgs, which provides masses to the SM fields after acquiring a vev, the scalar leptoquark and the singlet field  (see Section~\ref{sec:Potential} for the analysis of the scalar potential). We will be working with complex fields for the Higgs, $H$ $\bf{(1,2)_{1/2}}$ and the leptoquark $S$ $\bf{(3,1)_{-1/3}}$ which are embedded together with their complex conjugates within $(T^i_F A^i_5)$. Working out the above Lagrangian and only taking the scalar interactions that induce SM-SM couplings we find
%%%%%%%%%%%%%%%%%%%%%%%%%%%%%
\begin{align}
    \mathcal{L}\supset&-i\frac{g_5}{\sqrt{2}}(\bar{q}_L^{\alpha,i}d^{\prime}_{R,\alpha} +\epsilon^{ji}\bar{u}_R^\alpha q^\prime_{L,j,\alpha}-\bar{l}_R^{c,i} \nu_L^c+\epsilon^{ji}\bar{e}^c_L l^{c \prime}_{R,j} )H_i \notag \\
    &-i\frac{g_5}{\sqrt{2}}(\bar{u}_{R}^\alpha e_L^{c\prime}+\bar{q}_L^{\alpha,i}  l^{c\prime}_{R,i}+\bar{d}_R^{\alpha} \nu_L^c)S_{\alpha}+ \textrm{h.c.},
\end{align}
%%%%%%%%%%%%%%%%%%%%%%%%%%%%%%
with $\epsilon^{12}=1$ and $\langle H \rangle=\frac{1}{\sqrt{2}}(0,v)^T$. 

Again, some of these couplings include the \textit{primed} fermions since they mix with the zero modes after inclusion of boundary terms and thus contribute to the SM mass eigenstate. To understand these couplings in more detail we consider, as an example, the $\bar{l}_R^{c,i} \nu_L^c H_i $ term. The corresponding wavefunctions of the RH and LH zero modes read
%%%%%%%%%%%%%%%%%%%%%%%%%%%%%%%%
\begin{align}
    f_{\nu_L^c}(z) =& \frac{1}{\sqrt{R^\prime}}\Big(\frac{z}{R}\Big)^2\Big(\frac{z}{R}\Big)^{-c_6}f_{c_6}, \notag \\
    f_{l_R^c}(z) =& \frac{1}{\sqrt{R^\prime}}\Big(\frac{z}{R}\Big)^2\Big(\frac{z}{R}\Big)^{c_6}f_{-c_6},
\end{align}
%%%%%%%%%%%%%%%%%%%%%%%%%%%%%%%%
since both the neutrino doublet and singlet are embedded within the bulk $\bf{6}$. The extra-dimensional profile of a real scalar zero mode (see \eqref{A5profile}) is given by $f_5(z) = \sqrt{2/R} (z/R^\prime)$, which leads to the following overlap after separating the 5D fields into their bulk profile and 4D fields 
%%%%%%%%%%%%%%%%%%%%%%%%%%%%%%%%
\begin{align}\label{eq:Higgsprofileoverlap}
 {\cal L} & \supset -\frac{g_5}{\sqrt{2}}\int \text{d}z \Big(\frac{R}{z}\Big)^4\int \text{d}^4x (-\bar{l}_R^{c,i} \nu_L^c)\langle H_i \rangle \notag \\  & = -\frac{g_5}{\sqrt{2}}\frac{v}{\sqrt{2}} \int \text{d} z \sqrt{\frac{2}{R}}\Big(\frac{z}{R^\prime}\Big)\Big(\frac{1}{R^\prime}f_{c_6}f_{-c_6}\Big) \int \text{d}^4x (\bar{\nu}_R^{c,0} \nu_L^{c,0}) + \dots  \notag \\  &=  -\frac{g_* v}{2\sqrt{2}} f_{c_6}f_{-c_6}\int \text{d}^4x (\bar{\nu}_R^{c,0} \nu_L^{c,0}) + \dots\,,
\end{align}
%%%%%%%%%%%%%%%%%%%%%%%%%%%%%%%%
where $g_*=g_5/\sqrt{R}$ is the dimensionless counterpart of the five-dimensional coupling and we recover a regular 4D mass term for the neutrino. Unsurprisingly, the localization of the bulk field $\bf{6}$ is determining this mass. However the same localization is also determining the localization of the RH down quark and conjugate LH doublet, and fitting the corresponding mass would result in a too heavy neutrino. This is why we need a singlet $\bf{1}$ that will mix with the zero mode neutrino to recover a light mass eigenstate such that after kinetic normalization small neutrino masses are obtained as will be explicitly shown in Section~\ref{sec:Results}. 

The couplings with the triplet can be computed in a similar way, for example the $-\frac{g_5}{\sqrt{2}}(\bar{d}_R^{\alpha} \nu_L^c S_\alpha)$ term leads to the 4D Yukawa interaction 
%%%%%%%%%%%%%%%%%%%%%%%%%%%%%%%%
\begin{align}
 {\cal L} \supset -\frac{g_5}{\sqrt{2}}\int \text{d}z \Big(\frac{R}{z}\Big)^4 \int \text{d}^4x(\bar{d}_R \nu_L^c)S = & -\frac{g_5}{2} \int \text{d} z \sqrt{\frac{2}{R}}\Big(\frac{z}{R^\prime}\Big)\Big(\frac{1}{R^\prime}f_{c_6}f_{-c_6}\Big) \int \text{d}^4x (\bar{d}_R^{0} \nu_L^{c,0})S \notag +... \\= & -\frac{g_*}{2} f_{c_6}f_{-c_6}\int d^4x (\bar{d}_R^{0} \nu_L^{c,0})S + \dots\, .
\end{align}
%%%%%%%%%%%%%%%%%%%%%%%%%%%%%%%%
Notice that we find the same couplings as for the Higgs-neutrino Yukawa coupling derived above. This is a more general observation: the Yukawa couplings of the triplet scalar will be identical to the Higgs-Yukawa couplings before mass diagonalization, as they are embedded in the same SU(5) multiplet. This will lead to a strong flavor-hierarchical pattern in the triplet-Yukawa terms, similar to the Higgs-Yukawa couplings. As a consequence, the scalar triplet will not play a leading role in flavor constraints as its couplings to the first two generations are very suppressed. For loop processes, when including the KK excitations of the fermions, this conclusion will be changed where the leptoquark is the most constraining mediator for the $\mu\rightarrow e \gamma$ decay.

\section{Flavor of SU(6) gauge-Higgs grand unification}\label{sec:Results}

Having discussed the building blocks and couplings of SU(6) GHGUT, we will analyse the most compelling incarnation of the model and test if it can reproduce the flavor hierarchies observed in nature. This will allow us to understand why other configurations do not work. The main objective will thus be to reproduce the masses of the SM fermions, the CKM matrix and the PMNS matrix while having the smallest possible IR scale and avoiding any large hierarchies in the input parameters. We will also include a realistic neutrino sector in our scans. To this end, we choose an arbitrary benchmark of allowed neutrino masses of $7\,\textrm{meV},\ 11\,\textrm{meV}$, and $51$\,\textrm{meV} in the normal hierarchy, however varying them within the allowed range would not change the analysis notably. 

The absolute values of the elements of the CKM matrix contain the following strong hierarchies \cite{ParticleDataGroup:2020ssz}
%%%%%%%%%%%%%%%%%%%%%%%%%
\begin{equation}\label{CKM}
    V_{\textrm{CKM}}\sim\begin{pmatrix}
    |V_{ud}| & |V_{us}| & |V_{ub}| \\
    |V_{cd}| & |V_{cs}| & |V_{cb}| \\
     |V_{td}|& |V_{ts}| & |V_{tb}| \\
    \end{pmatrix} =
    \begin{pmatrix}
    0.9740 & 0.2265 & 0.0036\\
    0.2264 & 0.9732 & 0.0405 \\
    0.00885 & 0.0398 & 0.9992 \\
    \end{pmatrix} \sim 
    \begin{pmatrix}
    1 & \lambda & \lambda^3\\
    \lambda & 1 &\lambda^2 \\
    \lambda^3 & \lambda^2 & 1 \\
    \end{pmatrix}, 
\end{equation}
%%%%%%%%%%%%%%%%%%%%%%%%%
with the Wolfenstein parameter $\lambda\approx 0.23$,
which we will attempt to reproduce in our model. We also make sure that  we obtain approximately the correct amount of CP violation sourced from the CKM matrix, as quantified by the Jarlskog invariant $J=\textrm{Im}(V_{us}V_{cb}V_{ub}^* V_{cs}^*)\sim 3\times 10^{-5}$. The PMNS matrix on the other hand does not contain large hierarchies,
%%%%%%%%%%%%%%%%%%%%%%%%%
\begin{equation}
    V_{\textrm{PMNS}}\sim\begin{pmatrix}
    0.82 & 0.55 & 0.15 \\
    0.37 & 0.58 & 0.71 \\
    0.40 & 0.59 & 0.69 
    \end{pmatrix},
\end{equation}
%%%%%%%%%%%%%%%%%%%%%%%%%
where we take the central values from \cite{Esteban:2020cvm}. Concretely, in the scans over parameter space, we will require that all fermion masses and the entries of the CKM and PMNS matrices do not deviate more than 20\% from their target values. While this exceeds the experimental uncertainties, it facilitates the scan and demanding smaller deviations does not change our results significantly. We also require the brane-mass matrix entries $0.2<|(M_f)_{ij}|<2$ to be order one.

\subsection{IR brane mass model}\label{sec:IRbranemodel}
Let us start with the model that was identified in Table \ref{tab:my-table} as the most promising one where we consider all brane masses to reside on the IR brane. This will allow us to understand the issues with the other embeddings. Concerning the gauge sector, we will study the  model with the unbroken SU(5) symmetry on the IR brane and the SM gauge group on the UV, $G_{\textrm{SM}}^{\textrm{(UV)}}$, since the model with the SM group on the IR brane, $G_{\textrm{SM}}^{\textrm{(IR)}}$, suffers from light exotics as we will see. The BCs, where as usual we specify those for the LH components of the bulk fermion, become unique and simply follow from the requirement to have only IR-brane masses
%%%%%%%%%%%%%%%%%%%%%%%%%
\begin{align}
\label{IR-embedding}
    {\bf 20} \rightarrow & q^\prime{\bf ( 3,2)}_{1/6}^{+,-}  \oplus {\bf (3^*,1)}_{-2/3}^{+,-} \oplus  e^{c \prime} {\bf (1,1)}_1^{+,-}  \notag \\ 
     & {\bf (3^*,2)}_{-1/6}^{+,-}  \oplus u {\bf (3,1)}_{2/3}^{-,-} \oplus{\bf (1,1)}_{-1}^{+,-}, \notag \\
    {\bf 15} \rightarrow  & q {\bf (3,2)}_{1/6}^{+,+} \oplus {\bf (3^*,1)}_{-2/3}^{-,+} \oplus e^c {\bf (1,1)}_1^{+,+} \notag \\  
    & d^\prime {\bf (3,1)}_{-1/3}^{-,+}\oplus  l^{c \prime}{\bf (1,2)}_{1/2}^{-,+}, \notag \\
   {\bf 6} \rightarrow & d  {\bf (3,1)}_{-1/3}^{-,-} \oplus l^c  {\bf (1,2)}_{1/2}^{-,-} \oplus \nu^c {\bf (1,1)}_0^{+,+}, \notag \\
   {\bf 1} \rightarrow &  \nu^{c \prime}{\bf (1,1)}_{0}^{+,-}.
\end{align}
%%%%%%%%%%%%%%%%%%%%%%%%%
The boundary masses consist of three terms 
%%%%%%%%%%%%%%%%%%%%%%%%%
\begin{equation}
\label{UVflipped-IRbrane}
    S_{IR} = -\int \text{d}^4 x \Big(\frac{R^\prime}{R}\Big)^4\big(
    M_{q/e} \psi_{{\bf{20}},10}\chi_{{\bf{15}},10}+M_{d/l}\psi_{{\bf{6}},5}\chi_{{\bf{15}},5}+M_\nu \psi_{{\bf{1}},1}\chi_{{\bf{6}},1}+\text{h.c.}\big).
\end{equation}
%%%%%%%%%%%%%%%%%%%%%%%%%
Note that, given the chosen BCs, these are the only non--vanishing brane masses that one can write down, with the only exception being a UV-brane mass in the up-type exotic sector
%%%%%%%%%%%%%%%%%%%%%%%%%
\begin{align}
\label{UVflipped-UVbrane}
   S_{UV}  =& -\int\text{d}^4 x
    \big(M_{\tilde{u}} \psi_{{\bf{15}},(3^*,1)}\chi_{{\bf{20}},(3^*,1)} + \text{h.c.} \big).
\end{align}
%%%%%%%%%%%%%%%%%%%%%%%%%
This UV-brane mass will not impact the flavor aspects of the model but is important in the Higgs potential (see Section~\ref{sec:Potential}). The full solution for the fermion zero-mode profiles can be found in Appendix \ref{AppendixA}. 

The hierarchies in the CKM matrix can be naturally reproduced in this model -- indeed, looking at the mass matrices in the flavor basis, we find (see Eq.~\eqref{UVflipped-quark-masses})
%%%%%%%%%%%%%%%%%%%%%%%%%
\begin{align}
& {\cal M}_u = \frac{g_* v}{2\sqrt{2}} f_{c_{15}}M_{q/e}^\dagger f_{-c_{20}}, \notag \\
& {\cal M}_d = -\frac{g_* v}{2\sqrt{2}} f_{c_{15}}M_{d/l}^\dagger f_{-c_{6}}.
\end{align}
%%%%%%%%%%%%%%%%%%%%%%%%%
It is well known \cite{Huber:2000ie} that the bi-unitary diagonalization of the above matrices will feature transformation matrices that carry distinct hierarchies, following from the hierarchies in the mass matrices. For instance in the down sector we generically find
\begin{equation}\label{CKMrotL}
U_{L,d}\sim
    \begin{pmatrix}
    1 && \frac{f(c_{15,1})}{f(c_{15,2})} && \frac{f(c_{15,1})}{f(c_{15,3})}\\
    \frac{f(c_{15,1})}{f(c_{15,2})} && 1 &&\frac{f(c_{15,2})}{f(c_{15,3})} \\
    \frac{f(c_{15,1})}{f(c_{15,3})} && \frac{f(c_{15,2})}{f(c_{15,3})} && 1
    \end{pmatrix}, \quad 
   U_{R,d}\sim
    \begin{pmatrix}
    1 && \frac{f(-c_{6,1})}{f(-c_{6,2})} && \frac{f(-c_{6,1})}{f(-c_{6,3})}\\
    \frac{f(-c_{6,1})}{f(-c_{6,2})} && 1 &&\frac{f(-c_{6,2})}{f(-c_{6,3})} \\
    \frac{f(-c_{6,1})}{f(-c_{6,3})} && \frac{f(-c_{6,2})}{f(-c_{6,3})} && 1
    \end{pmatrix},
\end{equation}
with the rotation matrices in the up sector being analogous. Therefore, by taking the localizations of the $\bf{15}$ to fulfill
\begin{equation}\label{CKMcond}
    f(c_{15,1})/f(c_{15,2})\sim\lambda, \quad f(c_{15,2})/f(c_{15,3})\sim\lambda^2, \quad f(c_{15,1})/f(c_{15,3})\sim\lambda^3,
\end{equation}
the LH rotation matrices in the up and down sector will carry the same hierarchies as the CKM matrix \eqref{CKM} (see also \cite{Casagrande:2008hr}). The latter matrix, being the product $V_{\textrm{CKM}}=U_{L,u}^\dagger U_{L,d}$, will then take over these hierarchies. 

This also explains why \textit{mixed} models, where both UV-- and IR--brane masses are present, generally do not succeed in having a hierarchical CKM matrix. For instance if the $\bf{15}$ and $\bf{6}$ are connected through a UV-brane matrix $M_{d/l}$ instead, the mass matrix in the flavor basis will be modified to
\begin{equation}
    \mathcal{M}_d^\prime = -\frac{g_* v}{2\sqrt{2}} f_{c_{15}}\Big(\frac{R}{R^\prime}\Big)^{-c_{15}}M_{d/l}^\dagger\Big(\frac{R}{R^\prime}\Big)^{c_{6}} f_{-c_{6}},
\end{equation}
which will make the LH down rotation matrix of a different structure than the LH up rotation matrix resulting in a CKM matrix with all entries being ${\cal O}(1)$. 

$\textit{UV}$ models, on the other hand, where both $M_{d/l}$ and $M_{q/e}$ are on the UV brane can reproduce the CKM matrix. However such a setup requires the heavier generations of fermions to be localized closer to the UV brane, while the lighter generations are localized more in the IR as the mass generation now happens through the UV brane. Flavor constraints will make these models highly tuned as the lighter generations now couple strongly to KK excitations that are still IR localized, inducing large FCNCs. In this case, we find the IR scale gets pushed to $\sim \mathcal{O}(10^3)$~TeV. Finally, the model with IR--brane masses and the SM gauge group on the IR brane can be discarded because it contains light exotics. In particular, the exotic lepton with charge +1 embedded in the $\bf{20}$ would have $(-,+)$ BCs and thus become very light for the localization of the bulk $\bf{20}$ that is necessary to have light up and charm quarks ($c_{20,1/2}<-0.5$).

Going back to the most viable IR brane mass model, we will assume IR localization for the $\bf{15}$ hosting the third generation LH quark doublet ($c_{15,3}<0.5$) in order to reproduce the large top mass. The multiplets hosting the doublets of the two lighter generations will then naturally be UV localized $c_{15,{1/2}}>0.5$ by virtue of CKM relations \eqref{CKMcond}, which will help to suppress FCNCs. The remaning bulk localizations for the $\bf{20}$ and $\bf{6}$ are uniquely determined by the requirement of reproducing the correct SM quark masses resulting in the constraints
%%%%%%%%%%%%%%%%%%%%%%%%%
\begin{align}\label{numericalvalues2}
    &\frac{m_c}{m_t}\sim\lambda^2 \frac{f(-c_{20,2})}{f(-c_{20,3})}, \quad \frac{m_u}{m_t}\sim\lambda^3 \frac{f(-c_{20,1})}{f(-c_{20,3})}\notag \\&  \frac{m_b}{m_t}\sim \frac{f(-c_{6,3})}{f(-c_{20,3})}, \quad \frac{m_s}{m_t}\sim \lambda^2 \frac{f(-c_{6,2})}{f(-c_{20,3})}, \quad \frac{m_d}{m_t}\sim \lambda^3 \frac{f(-c_{6,1})}{f(-c_{20,3})}\,,
\end{align}
%%%%%%%%%%%%%%%%%%%%%%%%%
implying one free parameter in the flavor sector (that could be taken as $c_{15,3}$) that rescales the overall localization of the left-handed fermions.\footnote{We note that, in our setup, only the multiplets containing the 3rd generation quark doublet and the RH top quark are IR localized, whereas the remaining multiplets are UV localized. Therefore this model keeps most of the attractive features of the MCHM with the lighter fermions being mostly elementary and the heavier ones composite.} Of course, the anarchic nature of the brane matrices $M_{q/e}$ and $M_{d/l}$ causes these relations to be only approximate.

Moving to the lepton sector, we find the mass matrix of the charged (conjugate) leptons in the flavor basis \eqref{eq:UV-flipped-leptonic-masses}
%%%%%%%%%%%%%%%%%%%%%%%%%
\begin{equation}\label{masselectron}
  \mathcal{M}_{e^c} = -\frac{g_* v}{2\sqrt{2}} f_{-c_{6}}M_{d/l}f_{c_{15}}.
\end{equation}
%%%%%%%%%%%%%%%%%%%%%%%%%
We see that there is a degeneracy of electron and down masses, just as in SU(5) 4D GUTs. Since the gauge symmetry on the UV brane is $G_\textrm{SM}$, it is an ideal brane to introduce  SU(5) breaking effects that can correct this mass relation. It is indeed possible to lift the degeneracy of the electron and down quarks by introducing a brane kinetic term: for example, we can add a UV--localized kinetic term for the 5D fermion $l_R^c$ in the $\bf{6}$ with coefficient $\kappa$, reading
%%%%%%%%%%%%%%%%%%%%%%%%%
\begin{equation}
    S_{\textrm{kin},UV}=\int \textrm{d}^4x \, i  \kappa R \, \bar{l}_R^c \bar{\sigma}^\mu \partial_\mu l_R^c |_{z=R}.
\end{equation}
%%%%%%%%%%%%%%%%%%%%%%%%%
The effect of such a brane localized term will be that the UV BCs will be modified to \cite{Csaki:2003sh}
%%%%%%%%%%%%%%%%%%%%%%%%%
\begin{equation}
    \bar{l}^c_{L} |_{z=R}=i\kappa\bar{\sigma}^\mu \partial_\mu l^c_{R}|_{z=R}, \quad \partial_5 l^c_{R}|_{z=R}=-i\sigma^\mu \partial_\mu \bar{l}^c_{L}|_{z=R}  + \frac{2+c_{6}}{R}l^c_L|_{z=R}.
\end{equation}
%%%%%%%%%%%%%%%%%%%%%%%%%

Expanding the 5D fields in their KK decomposition \eqref{KKdecomp} and using the EOMs \eqref{EOM} results in the modified BC
%%%%%%%%%%%%%%%%%%%%%%%%%
\begin{equation}
    f_{l^c_L}(z=R)=\kappa m_n R f_{l^c_R}(z=R).
\end{equation}
%%%%%%%%%%%%%%%%%%%%%%%%%
In the limit of vanishing kinetic term $\kappa=0$ we recover the usual condition for a $(-,-)$ 5D fermion, while for a general non-zero $\kappa$, the zero-mode solution is unaffected by the UV kinetic term with the exception of the absolute normalization of the kinetic term. Indeed, the inclusion of a UV kinetic term will change the kinetic mixings \eqref{kineticterms} as follows
%%%%%%%%%%%%%%%%%%%%%%%%%
\begin{align}
    K_{l^c_R} = 1 + f_{-c_{6}} M_{d/l} f_{-c_{15}}^{-2} M_{d/l}^\dagger f_{-c_{6}}\rightarrow & 1 +\kappa R f_{l^c_R}^2(z=R)+  f_{-c_{6}} M_{d/} f_{-c_{15}}^{-2} M_{d/l}^\dagger f_{-c_{6}}\notag \\ \approx & 1+(-2c_{-6}-1)\kappa +  f_{-c_{6}} M_{d/l} f_{-c_{15}}^{-2} M_{d/l}^\dagger f_{-c_{6}},
\end{align}
%%%%%%%%%%%%%%%%%%%%%%%%%
where the approximation holds for UV-localized fermions with $c<-0.5$\footnote{For IR localized fermions, unsurprisingly, the effect of a UV kinetic term is negligible.}. Therefore one can break the down-electron degeneracy and approximately obtain the correct charged lepton masses for the UV kinetic term\footnote{We chose here a diagonal matrix for simplicity but stress that, due to the mild hierarchies between charged-lepton and down-type quark masses, a rather anarchic general matrix could be employed, too. In fact, employing generic brane-kinetic terms for all fermions would not spoil the generation of flavor hierarchies in the setup and in that sense all parameters can be natural while {\it predicting} rather degenerate down-quark and lepton masses, at the same time accommodating very different up-type quark masses.}
%%%%%%%%%%%%%%%%%%%%%%%%%
\begin{equation}
    \kappa=\begin{pmatrix}
    \frac{(m_d/m_e)^2 -1}{-2c_{6,1}-1} & 0& 0 \\
    0 & \frac{(m_s/m_\mu)^2-1}{-2c_{6,2}-1} & 0 \\
    0 & 0  & \frac{(m_b/m_\tau)^2-1}{-2c_{6,3}-1} 
    \end{pmatrix}.
\end{equation}
%%%%%%%%%%%%%%%%%%%%%%%%%

Concerning the neutrino sector, we find the following mass matrix in the flavor basis (see \eqref{eq:UV-flipped-leptonic-masses})
%%%%%%%%%%%%%%%%%%%%%%%%%
\begin{equation}
\mathcal{M}_{\nu^c} = \frac{g_* v}{2\sqrt{2}} f_{-c_{6}}f_{c_{6}}.
\end{equation}
%%%%%%%%%%%%%%%%%%%%%%%%%
Unsurprisingly, the masses of the neutrinos are purely determined by the localizations of the bulk $\bf{6}$, which already determines the masses of the down-type quarks and charged leptons. Therefore the singlet $\bf{1}$ is crucial as it allows the RH neutrino to mostly reside in the singlet resulting in a small neutrino mass. Indeed, incorporating the kinetic mixing, the neutrino mass matrix in the canonically normalized basis reads
%%%%%%%%%%%%%%%%%%%%%%%%%
\begin{equation}
    \mathcal{M}_{\nu^c}^{\textrm{can}}=\frac{g_* v}{2\sqrt{2}} K_{l^c_R}^{-1/2 \dagger} f_{-c_6} f_{c_6} K_{\nu^c_L}^{-1/2}.
\end{equation}
%%%%%%%%%%%%%%%%%%%%%%%%%
Neglecting mixing implies the neutrino masses
%%%%%%%%%%%%%%%%%%%%%%%%%
\begin{equation}\label{massneutrino}
    m_{\nu,i} \simeq \frac{g_* v}{2\sqrt{2}} f(-c_{6,i}) f(c_{1,i})/M_{\nu,ii}
\end{equation}
%%%%%%%%%%%%%%%%%%%%%%%%%
for UV localized singlets ($c_1>0.5$). Small neutrino masses can then be achieved by taking the singlet $\bf{1}$ very UV localized. The ratio of masses is therefore determined by the $c_{1,i}$ parameters,
%%%%%%%%%%%%%%%%%%%%%%%%%
\begin{equation}
  \frac{m_{\nu_i}}{m_{\nu_j}} \simeq \frac{f(-c_{6,i})f(c_{1,i})}{f(-c_{6,j})f(c_{1,j})}.
\end{equation}
%%%%%%%%%%%%%%%%%%%%%%%%%

Since the flavor function is exponentially sensitive to the localization parameter, only slight differences in the localizations of the singlets $\bf{1}$ are necessary in order to reproduce realistic neutrino masses. Concerning the PMNS matrix, it  equals the product of the two LH lepton rotation matrices $V_{PMNS}=U_{L,\nu}^\dagger U_{L,e}=U_{L,\nu^c}^T U_{L,e^c}^*$ (see under Eq. \eqref{eq:diagonalization}). It follows (similarly from Eq. \eqref{CKMrotL}) that these matrices are determined by the $\bf{6}$ localizations
%%%%%%%%%%%%%%%%%%%%%%%%%
\begin{equation}
    (U_{L,\nu})_{ij} \sim (U_{L,e})_{ij}\sim \frac{f(-c_{6,i})}{f(-c_{6,j})}, \quad i\leq j.
\end{equation}
%%%%%%%%%%%%%%%%%%%%%%%%%
This is ideal to explain $V_{\textrm{PMNS}}$ since the localizations of the $\bf{6}$ do not contain any strong hierarchies. This reflects the fact that the RH down localizations could be chosen almost degenerate, as was noted e.g. in \cite{Santiago:2008vq} where the localizations were in fact chosen to be the same to protect the RH down sector from FCNCs. Indeed there are no strong hierarchies remaining: the down-type quark and lepton mass hierarchies are almost fully accounted for by the localizations of the $\bf{15}$ and the ratios between the localizations of the $\bf{6}$ can therefore finally be extracted from \eqref{numericalvalues2} as
%%%%%%%%%%%%%%%%%%%%%%%%%
\begin{equation}
    \frac{f(-c_{6,2})}{f(-c_{6,1})}\sim 4, \quad \frac{f(-c_{6,3})}{f(-c_{6,2})}\sim 2.
\end{equation}
%%%%%%%%%%%%%%%%%%%%%%%%%
As mentioned, these small hierarchies mean the model is well suited to explain the unhierarchical PMNS matrix and we take the view that its specific features are the result of the anarchic brane masses on the IR. Moreover, this feature is not spoiled for different benchmark neutrino masses. In fact, even for scenarios where the lightest neutrino is much lighter than the other two, requiring large hierarchies in the $f(c_{1,i})$'s, the PMNS matrix will not be disturbed since these hierarchies will mostly affect the RH neutrino rotation matrices and therefore will not enter into $V_{\textrm{PMNS}}$. Therefore, we generically obtain an anarchic structure for the PMNS matrix.

Having accounted for the observed flavor hierarchies, we perform a scan over parameter space and select 5000 points that reproduce these hierarchies (or lack thereof for the PMNS matrix) within $20\%$. In the next section we discuss and test the flavor constraints on the model.

\section{Flavor constraints}\label{sec:Flavor}

Having identified a successful model to explain the flavor hierarchies, we can now go over to the flavor constraints. As in general RS models, the flavor structure of GHGUTs is particularly rich. We will discuss two broad categories of observables, flavor violation in the quark sector due to meson mixing and flavor violation in the lepton sector in charged lepton observables. Previously, these observables have been proven to provide the most stringent constraints on extra dimensional models, which is why we focus on them in this work. In the literature, flavor  effects in both the quark and lepton sectors have been well studied \cite{Agashe:2004ay,Agashe:2004cp,Casagrande:2008hr,Csaki:2008zd,Blanke:2008yr,Agashe:2008uz,Blanke:2008zb,Bauer:2009cf,Albrecht:2009xr,Keren-Zur:2012buf,Agashe:2006iy,Gedalia:2009ws,Csaki:2010aj,Blanke:2012tv,Beneke:2012ie,Konig:2014iqa,Beneke:2015lba,Moch:2015oka}, but independently from another and often in the context of an IR-brane localized Higgs. In GHGUT, one can no longer study these effects separately since quarks and leptons are unified in the same multiplets and therefore their localization is intertwined.

\subsection{Meson mixing}
Due to the presence of KK gauge bosons with non-universal interactions, tree-level FCNCs can arise. These effects are naturally suppressed by the mass of the corresponding bosons $m_{(+,+)}\sim 2.45/R^{\prime}$. However, with $\mathcal{O}(1)$ couplings this suppression would not be nearly enough to evade the current bounds on FCNC couplings from meson mixing. Indeed $\mathcal{O}(1)$ couplings would imply lower bounds on the mediating gauge bosons at the level of $\sim 10^4$~TeV. However, due to the RS-GIM mechanism, which ensures that off-diagonal couplings are suppressed by the same small overlap functions that generate small yukawa couplings, the actual bounds are substantially smaller.

FCNCs give notable effects in $\Delta F = 2$ flavor observables such as $B_s-\bar{B}_s$, $B_d-\bar{B}_d$, $D-\bar{D}$ and $K-\bar{K}$ mixing, where the latter is generally the most constraining. Due to the large SU(3) coupling, the dominant contribution comes from the tree-level exchange of the lightest KK gluon, although in principle the KK photon and the electroweak gauge bosons and their KK modes also contribute. The relevant Lagrangian, featuring off-diagonal couplings to the first KK gluon in the mass basis, reads
%%%%%%%%%%%%%%%%%%%%%%
\begin{equation}
    \mathcal{L}=(g_{L,q}^{ij}\bar{q}_L^{\alpha,i}\gamma_\mu q_{\beta,L}^{j}+g_{R,q}^{ij}\bar{q}_R^{\alpha,i}\gamma_\mu q_{\beta,R}^{j})(T^a)_{\alpha}^{\beta}G^{a,1}_\mu,
\end{equation}
%%%%%%%%%%%%%%%%%%%%%%
where $q=u,d$ and $i,j$ are generation indices and below we omit a zero mode superscript for the SM fermions. The full expressions for these couplings in the mass basis are given by 
\begin{align}
    g_{L,u}=&g_s U_{L,u}^\dagger K_{q_L}^{-1/2\dagger}\Big(\lambda_{(+,+),c_{15}}+f_{c_{15}} M_{q/e}^\dagger f^{-1}_{c_{20}} \,\lambda_{(+,+),c_{20}}\,f^{-1}_{c_{20}}M_{q/e}f_{c_{15}}\Big)K_{q_L}^{-1/2}U_{L,u} \notag  \\
    g_{R,u}=&g_s U_{R,u}^\dagger \lambda_{(+,+)c_{20}} U_{R,u}\notag \\
    g_{L,d}=&g_s U_{L,d}^\dagger K_{q_L}^{-1/2\dagger}\Big(\lambda_{(+,+),c_{15}}+f_{c_{15}} M_{q/e}^\dagger f^{-1}_{c_{20}} \,\lambda_{(+,+),c_{20}}\,f^{-1}_{c_{20}}M_{q/e}f_{c_{15}}\Big)K_{q_L}^{-1/2}U_{L,d} \notag  \\
    g_{R,d}=&g_s U_{R,d}^\dagger K_{d_R}^{-1/2\dagger}\Big(\lambda_{(+,+),-c_6}+f_{c_{-6}} M_{d/l} f^{-1}_{-c_{15}} \,\lambda_{(+,+),-c_{15}}\,f^{-1}_{-c_{15}}M_{d/l}^\dagger f_{-c_{6}}\Big)K_{d_R}^{-1/2}U_{R,d}\,,
\end{align}
with $\lambda_{(+,+),c}$ the coupling to the first KK mode, given in Eq. \eqref{eq:effectivecoupling}.

After integrating out the KK gluon with mass $m_{(+,+)}=2.45/R^\prime$ we obtain the effective 4D Hamiltonian (focusing on the down sector)
%%%%%%%%%%%%%%%%%%%%%%%%%%%
\begin{align}
     \mathcal{H}= & C_1^{ij}(m_{(+,+)}) (\bar{d}_L^{\alpha,i}\gamma_\mu d_{L,\alpha}^j)(\bar{d}_L^{\beta,i}\gamma^\mu d_{L,\beta}^j)+\tilde{C}_1^{ij}(m_{(+,+)}) (\bar{d}_R^{\alpha,i}\gamma_\mu d_{R,\alpha}^j)(\bar{d}_{R}^{\beta,i}\gamma^\mu d_{R,\beta}^j)\notag \\
     +&C_4^{ij}(m_{(+,+)}) (\bar{d}_L^{\alpha,i} d_{R,\alpha}^j)(\bar{d}_R^{\beta,i} d_{L,\beta}^j)+C_5^{ij}(m_{(+,+)}) (\bar{d}_L^{\alpha,i} d_{R,\beta}^j)(\bar{d}_R^{\beta,i} d_{L,\alpha}^j),
\end{align}
with 
\begin{align}
    &C_1^{ij}(m_{(+,+)})=\frac{g_{L,d}^{ij} g_{L,d}^{ij}}{6m_{(+,+)}^2} , \quad \tilde{C}_1^{ij}(m_{(+,+)})=\frac{g_{R,d}^{ij} g_{R,d}^{ij}}{6m_{(+,+)}^2} , \notag \\  &C_4^{ij}(m_{(+,+)})= -\frac{g_{L,d}^{ij}g_{R,d}^{ij}}{m_{(+,+)}^2}, \quad C_5^{ij}(m_{(+,+)}) =\frac{g_{L,d}^{ij}g_{R,d}^{ij}}{3m_{(+,+)}^2}.
\end{align}
%%%%%%%%%%%%%%%%%%%%%%%%%%%
The coefficients of these operators are experimentally well constrained \cite{UTfit:2007eik}, especially in the Kaon sector, for which $i,j=1,2$. The imaginary component of $C_4^{21}(\Lambda)$ experiences the most stringent bound, requiring a new physics scale of $\sim 10^4$ TeV for order one couplings. The limits quoted in \cite{UTfit:2007eik} are calculated at the new physics scale (identified with the bound on $\Lambda$ assuming order one coupling) and considering only one effective operator present at a time. Since these coefficients depend considerably on the renormalization scale, they should be translated to our new physics scale $\sim m_{(+,+)}$. Considering this, we will use the bounds quoted in \cite{Csaki:2008zd}, that are evaluated at $3$ TeV. Moreover the $C_4^{21}(\Lambda)$ coefficient only receives contributions from the KK gluon exchange and not from photon or other electroweak  gauge bosons. The approach of only taking into account gluon exchange is therefore especially well motivated.

\subsection{Tree-level lepton flavor violation}
We will now consider observables in the lepton sector that arise at tree-level, namely $\mu^+\rightarrow e^+ e^- e^+$ decay and $\mu - e$ conversion. The relevant Lagrangian contains the couplings of the leptons to the $Z$ boson (and its first KK mode) in the mass basis, reading

\begin{equation}\label{eq:effZlagrangian}
    \mathcal{L}=\frac{g}{c_W}\Big(g_L^{ij}\bar{e}_L^i\gamma_\mu e_L^{j}+g_R^{ij}\bar{e}_R^{i}\gamma_\mu e_R^{j}\Big)Z^{0,\mu}+\frac{g}{c_W}\Big(g_L^{\prime ij}\bar{e}_L^i\gamma_\mu e_L^{j}+g_R^{\prime ij}\bar{e}_R^{i}\gamma_\mu e_R^{j}\Big)Z^{1,\mu}.
\end{equation}
Although the leptons are embedded as conjugate fields in our model, we will convert these couplings to those of unconjugated fields for phenomenology calculations.
The coefficients are given by 
\begin{align}
    g_L=& (-1/2+s_W^2)\Big(U_{L,e^c}^\dagger K_{l^c_R}^{\dagger, -1/2} \Big(\lambda_{Z^0,-c_6}+f_{-c_6}M_{d/l}f^{-1}_{c_{15}}\,\lambda_{Z^0,-c_{15}}\,f^{-1}_{c_{15}}M_{d/l}^\dagger f_{-c_6}\Big)K_{l^c_R}^{-1/2} U_{L,e^c}\Big)^* \notag \\
    g_R=& (s_W^2)\Big(U_{R,e^c}^\dagger K_{e^c_L}^{\dagger,-1/2} \Big(\lambda_{Z^0,-c_6}+f_{-c_6}M_{d/l}f^{-1}_{c_{15}}\,\lambda_{Z^0,-c_{15}}\,f^{-1}_{c_{15}}M_{d/l}^\dagger f_{-c_6}\Big)K_{e^c_L}^{-1/2} U_{R,e^c}\Big)^*,
\end{align}
where we used the couplings derived in equation \eqref{eq:effectiveZW} (and added a complex conjugate to obtain the couplings to the unconjugated leptons), while the equivalent couplings to the first KK $Z$ boson can be obtained by the substitution $\lambda_{Z^0,c}\rightarrow\lambda_{(+,+),c}$. We neglect the fermion mixing with KK modes as its leading effect is suppressed not only by $(vR^\prime)^2$, but by an additional factor of $\log(R^\prime/R)$\cite{Casagrande:2008hr}. 

\subsubsection{\texorpdfstring{$\mu^+\rightarrow e^+ e^- e^+$}{mu3e} decay}
Flavor violation in  $\mu\rightarrow 3e$ can be  parametrized with the effective Lagrangian~\cite{Kuno:1999jp,Chang:2005ag}
%%%%%%%%%%%%%%%%%%%%%%%%%%%%
\begin{align}\label{basiclagrangian}
    \mathcal{L}=&-\frac{4G_F}{\sqrt{2}}\Big( g_1(\bar{e}_R \mu_L)(\bar{e}_R e_L) +g_2(\bar{e}_L \mu_R)(\bar{e}_L e_R) + g_3(\bar{e}_R\gamma^\mu \mu_R)(\bar{e}_R\gamma_\mu e_R)\notag \\ 
    & +g_4(\bar{e}_L\gamma^\mu \mu_L)(\bar{e}_L\gamma_\mu e_L) +g_5(\bar{e}_R\gamma^\mu  \mu_R)(\bar{e}_L\gamma_\mu e_L) + g_6(\bar{e}_L\gamma^\mu \mu_L)(\bar{e}_R\gamma_\mu e_R) \notag \\ 
    & + m_\mu A_R \bar{e}_R\sigma^{\mu\nu}\mu_L F_{\mu\nu} + m_\mu A_L \bar{e}_L\sigma^{\mu\nu}\mu_R F_{\mu\nu} + \textrm{h.c.}\Big).
\end{align}
%%%%%%%%%%%%%%%%%%%%%%%%%%%%
In the above Lagrangian we recognize the loop-level dipole operators proportional to $A_L$ and $A_R$, which will be subdominant with respect to the tree-level contact interactions. However on their own these dipole operators give the striking $\mu\rightarrow e\gamma$ signature which we will discuss in the next section. Neglecting small KK-photon contributions, the remaining operators are generated at tree level by $Z$ boson (zero-- and KK--mode) exchange, therefore the scalar contact interactions are absent ($g_1=g_2=0$) in the model at hand, resulting in the branching ratio \cite{Kuno:1999jp,Chang:2005ag} 
%%%%%%%%%%%%%%%%%%%%%%%%%%%%
\begin{equation}
    \textrm{Br}(\mu\rightarrow 3e)= 2(g_3^2+g_4^2)+g_5^2+g_6^2\,.
\end{equation}
%%%%%%%%%%%%%%%%%%%%%%%%%%%%
Dating from 1988, the current experimental bound still reads \cite{BELLGARDT19881}
\begin{equation}
    \textrm{Br}(\mu\rightarrow 3e)<10^{-12}
\end{equation}
 and is expected to be improved by the upcoming Mu3e experiment to  \cite{Perrevoort:2017uex}
 \begin{equation}
     \textrm{Br}(\mu\rightarrow 3e)<10^{-16}\,.
 \end{equation}
 
Integrating out the $Z$ boson in Eq.~\eqref{eq:effZlagrangian} we find the effective Lagrangian
%%%%%%%%%%%%%%%%%%%%%%%%%%%%
\begin{align}
    \mathcal{L}= -\frac{4 G_F}{\sqrt{2}}\Big(&g_L^{ij} g_L^{kl} (\bar{e}^i_L\gamma_\mu e^j_L)(\bar{e}^k_L\gamma^\mu e^l_L) +g_R^{ij} g_R^{kl} (\bar{e}^i_R\gamma_\mu e^j_R)(\bar{e}^k_R\gamma^\mu e^l_R)  \notag \\
     & 2 g_L^{ij} g_R^{kl} (\bar{e}^i_L\gamma_\mu e^j_L)(\bar{e}^k_R\gamma^\mu e^l_R)\Big),
\end{align}
%%%%%%%%%%%%%%%%%%%%%%%%%%%%
with the first KK mode leading to a similar result. Including both contributions and matching to Eq.~\eqref{basiclagrangian}, we obtain the couplings contributing to the $\mu\rightarrow 3e$ process 
%%%%%%%%%%%%%%%%%%%%%%%%%%%%
\begin{align}
    g_3=2\Big(g_R^{12} g_R^{11}+g_R^{\prime12} g_R^{\prime11}\big(\frac{m_Z}{m_{(+,+)}}\big)^2\Big), \notag \\
    g_4=2\Big(g_L^{12} g_L^{11}+g_L^{\prime12} g_L^{\prime11}\big(\frac{m_Z}{m_{(+,+)}}\big)^2\Big), \notag \\
    g_5=2\Big(g_R^{12} g_L^{11}+g_R^{\prime12} g_L^{\prime11}\big(\frac{m_Z}{m_{(+,+)}}\big)^2\Big), \notag \\
    g_6=2\Big(g_L^{12} g_R^{11}+g_L^{\prime12} g_R^{\prime11}\big(\frac{m_Z}{m_{(+,+)}}\big)^2\Big).
\end{align}

\subsubsection{\texorpdfstring{$\mu-e$}{mue} conversion}

For $\mu-e$ conversion in nuclei we consider quark-lepton effective operators in the effective Lagrangian \cite{Kuno:1999jp,Chang:2005ag}

\begin{align}\label{eq:muonconversioneff}
    \mathcal{L}=&-\frac{2G_F}{\sqrt{2}}\Big(\bar{e}(s-p\gamma^5)\mu\sum_{q}\bar{q}(s_q-p_q\gamma^5)q+\bar{e}\gamma^\alpha(v-a\gamma^5)\mu\sum_{q}\bar{q}\gamma_\alpha(v_q-a_q\gamma^5)q+ \textrm{h.c.}\Big).
\end{align}
Effects in the light-quark sector will be very small and therefore $v_q=T_3-2Q\sin^2\theta_W$ and $a_q=T_3$, as in the SM. We also omit possible tensor couplings as they lead to  non-coherent transitions and are therefore suppressed by approximately the number of nucleons. For coherent muon conversion, only the scalar and vector couplings are relevant. These can then be converted from the quark level to the nucleon level, in order to obtain the conversion rate~\cite{Kuno:1999jp,Chang:2005ag}
%%%%%%%%%%%%%%%%%%%%%%%%%%%%%%
\begin{align}\label{muoneconversion}
    \textrm{Br}(\mu\rightarrow e)_N=&\frac{G_F^2 F_p^2 m_\mu^5\alpha^3 Z^4_{\textrm{eff}}}{2\pi^2Z\Gamma_{\textrm{capt}}}\times \notag \\ &\Big(|4eA_L Z+ (s-p)S_N +(v-a) Q_N|^2+|4eA_RZ+(s+p)S_N +(v+a)Q_N|^2 \Big),
\end{align}
%%%%%%%%%%%%%%%%%%%%%%%%%%%%%%
where $\Gamma_{\textrm{capt}}$ denotes the muon capture rate and $Q_N, S_N$ are defined as
%%%%%%%%%%%%%%%%%%%%%%%%%%%%%%
\begin{align}
    S_N=s_u(2Z+N)+s_d(2N+Z), \notag \\
    Q_N=v_u(2Z+N)+v_d(2N+Z). 
\end{align}
%%%%%%%%%%%%%%%%%%%%%%%%%%%%%%
The parameters for $^{48}_{22}\textrm{Ti}/^{27}_{13}\textrm{Al}/^{197}_{79}\textrm{Au}$ nuclei are $F_p\sim 0.55/0.66/0.16$, $Z_{\textrm{eff}}\sim 17.61/11.62/33.5$, and $\Gamma_\textrm{capt}\sim (2.6/0.71/13.07)\times 10^6 \textrm{ sec}^{-1}$ \cite{Chiang:1993xz,Kitano:2002mt,Angelescu:2020uug}. The two strongest experimental constraint are obtained for $^{48}_{22} \textrm{Ti}$ and $^{197}_{79}\textrm{Au}$ and read \cite{Klapdor-Kleingrothaus:1999xyb,SINDRUMII:2006dvw}
%%%%%%%%%%%%%%%%%%%%%%%%%%%%%%
\begin{align}
    \textrm{Br}&(\mu\rightarrow e)_{\textrm{Ti}}<6.1\times 10^{-13},\notag\\
    \textrm{Br}&(\mu\rightarrow e)_{\textrm{Au}}<9.1\times 10^{-13}.
\end{align}
The conversion rate for Titanium is slightly smaller and as a result we will only work with the latter. The upcoming experiments COMET and Mu2e \cite{Diociaiuti:2020yvo} will probe $\mu - e$ conversion in Aluminium with an expected sensitivity at 90\% confidence level (CL) of
%%%%%%%%%%%%%%%%%%%%%%%%%%%%%%
\begin{equation}
    \textrm{Br}(\mu\rightarrow e)_{\textrm{Al}}<8\times 10^{-17}.
\end{equation}
%%%%%%%%%%%%%%%%%%%%%%%%%%%%%%
We will also include this upcoming experiment in our analysis. 

The relevant couplings have three different origins, cominng from the $Z$ boson and its KK modes, the $X,Y$ gauge bosons, and the scalar triplet. We can safely neglect the scalar triplet as its couplings are too small to be competitive with the other two sources. The contributions of the $X,Y$ gauge bosons will be heavily suppressed in the models where the gauge symmetry on the UV brane is $G_{\textrm{SM}}$, both due to the heavier mass of the $X,Y$ gauge bosons and by their small couplings to the first generation fermions, that are UV localized. However for the models with the SM gauge symmetry on the IR brane, these contributions can be important and need to be accounted for. The contributions from the $Z$ boson are obtained from Eq. \eqref{eq:effZlagrangian}, including the respective quark couplings, leading to \begin{align}
    \mathcal{L}=-\frac{4G_F}{\sqrt{2}} 2 \Big(&g_L^{ij}g_{L,q}^{kl} (\bar{e}_L^i\gamma^\mu e_L^j)(\bar{q}_L^k\gamma_\mu q_L^l)+g_L^{ij}g_{R,q}^{kl} (\bar{e}_L^i\gamma^\mu e_L^j)(\bar{q}_R^k\gamma_\mu q_R^l) \notag \\
    +& g_R^{ij}g_{L,q}^{kl} (\bar{e}_R^i\gamma^\mu e_R^j)(\bar{q}_L^k\gamma_\mu q_L^l)+g_R^{ij}g_{R,q}^{kl} (\bar{e}_R^i\gamma^\mu e_R^j)(\bar{q}_R^k\gamma_\mu q_R^l)\Big),
\end{align} (and similar for the KK $Z$ boson).
Comparing to Eq. \eqref{eq:muonconversioneff} we can read off the values for $v$ and $a$
%%%%%%%%%%%%%%%%%%%%%%%%%%%%%%
\begin{align}
    v=&\Big(g_L^{12}+g_L^{\prime12}\Big(\frac{m_Z}{m_{(+,+)}}\Big)^2\Big)+\Big(g_R^{12}+g_R^{\prime12}\Big(\frac{m_Z}{m_{(+,+)}}\Big)^2\Big), \notag \\
    a=&\Big(g_L^{12}+g_L^{\prime12}\Big(\frac{m_Z}{m_{(+,+)}}\Big)^2\Big)-\Big(g_R^{12}+g_R^{\prime12}\Big(\frac{m_Z}{m_{(+,+)}}\Big)^2\Big).
\end{align}
%%%%%%%%%%%%%%%%%%%%%%%%%%%%%%

\subsection{Loop-level lepton flavor violation}
As a complementary probe to the tree level flavour-violating processes discussed so far, we will now consider the loop-level process $\mu \rightarrow e \gamma$, which is mediated by penguin diagrams. In 5D, these are finite at one-loop order and have been studied in a fully 5D framework in \cite{Csaki:2010aj}. We will prefer to work in a KK picture as it makes the calculation more transparant, as was done in \cite{Agashe:2006iy}, where only the dominating Higgs loop was included. Below, we will review the calculation and include other processes, such as the $Z$-loop and $W$-loop contribution and most relevant for our SU(6) GHGUT, the scalar leptoquark. Since these processes require the careful treatment of KK fermions, we include the details of the calculation in Appendix \ref{AppendixB} and Appendix \ref{AppendixC}. 

The amplitude for the process $\mu(p) \rightarrow e(p^\prime) \gamma(q) $ reads  ${\cal A}= e\, \epsilon_\mu^*(q)M^\mu$ \cite{Lavoura:2003xp}. Gauge invariance dictates that it  must remain invariant under $\epsilon_\mu\rightarrow \epsilon_\mu+q_\mu$, leading to the general form 
%%%%%%%%%%%%%%%%%%%%%%%%%%%%%%
\begin{equation}\label{amplitude}
    M^\mu= \bar{u}_{p^\prime}(C_L \Sigma_L^\mu+C_R \Sigma_R^\mu)u_p/m_\mu,
\end{equation}
%%%%%%%%%%%%%%%%%%%%%%%%%%%%%%
where
%%%%%%%%%%%%%%%%%%%%%%%%%%%%%%
\begin{align}
    \Sigma_L^\mu=& (p^\mu+p^{\prime,\mu})P_L-\gamma^\mu(m_eP_L+m_\mu P_R),\notag \\
    \Sigma_R^\mu=& (p^\mu+p^{\prime,\mu})P_R-\gamma^\mu(m_eP_R+m_\mu P_L),
\end{align}
%%%%%%%%%%%%%%%%%%%%%%%%%%%%%%
with $C_{L/R}$ the model dependent coefficients that are calculated in Appendix \ref{AppendixC} for SU(6) GHGUT.
Orthogonal directions to $(p^\mu+p^{\prime,\mu})$ in \eqref{amplitude} are parameterized by $q^\mu$, however for on-shell processes these disappear due to $q^2=0$ and $\epsilon_\mu^* q^\mu=0$. One can then rewrite this amplitude using the Gordon decomposition
%%%%%%%%%%%%%%%%%%%%%%%%%%%%%%
\begin{equation}\label{dipole}
    M^\mu= \bar{u}_{p^\prime}i\frac{\sigma^{\mu\nu}q_\nu}{m_\mu} (C_L P_L+C_R P_R)u_p.
\end{equation}
%%%%%%%%%%%%%%%%%%%%%%%%%%%%%%
Finally, the decay width for the flavour violating process is given by 
%%%%%%%%%%%%%%%%%%%%%%%%%%%%%%
\begin{equation}
    \Gamma(\mu \to e\gamma) =\frac{(m_\mu^2-m_e^2)^3(|C_L|^2+|C_R|^2)}{16\pi m_\mu^5}
\end{equation}
%%%%%%%%%%%%%%%%%%%%%%%%%%%%%%
This is to be compared to the dominating $\mu\rightarrow e \nu \bar{\nu}$ decay, with $\Gamma(\mu \rightarrow e \nu \bar{\nu})=m_\mu^5 G_F^2/192\pi^3$. The branching ratio for the flavor violating process is then
%%%%%%%%%%%%%%%%%%%%%%%%%%%%%%
\begin{equation}\label{eq:BRs}
    \textrm{Br}(\mu \rightarrow e \gamma)= \frac{12\pi^2 (C_L^2+C_R^2)}{(G_F m_\mu^2)^2}.
\end{equation}
%%%%%%%%%%%%%%%%%%%%%%%%%%%%%%

The most stringent constraint on this branching ratio stems from the MEG experiment \cite{MEG:2016leq}, reading
%%%%%%%%%%%%%%%%%%%%%%%%%%%%%%
\begin{figure}[!t]
  \begin{subfigure}[b]{0.5\textwidth}
    \includegraphics[width=\textwidth]{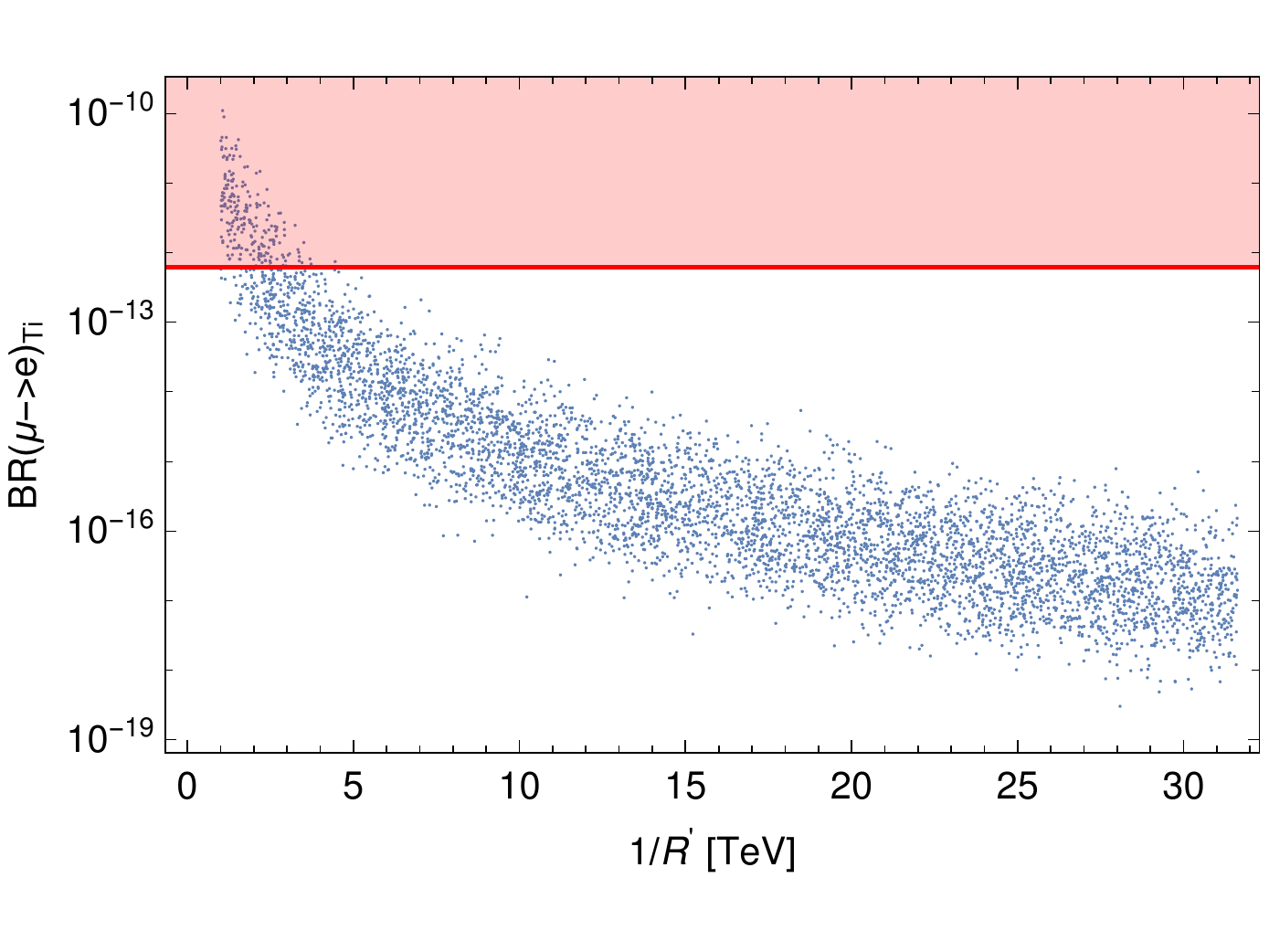}
  \end{subfigure}
  \hfill
  \begin{subfigure}[b]{0.5\textwidth}
    \includegraphics[width=\textwidth]{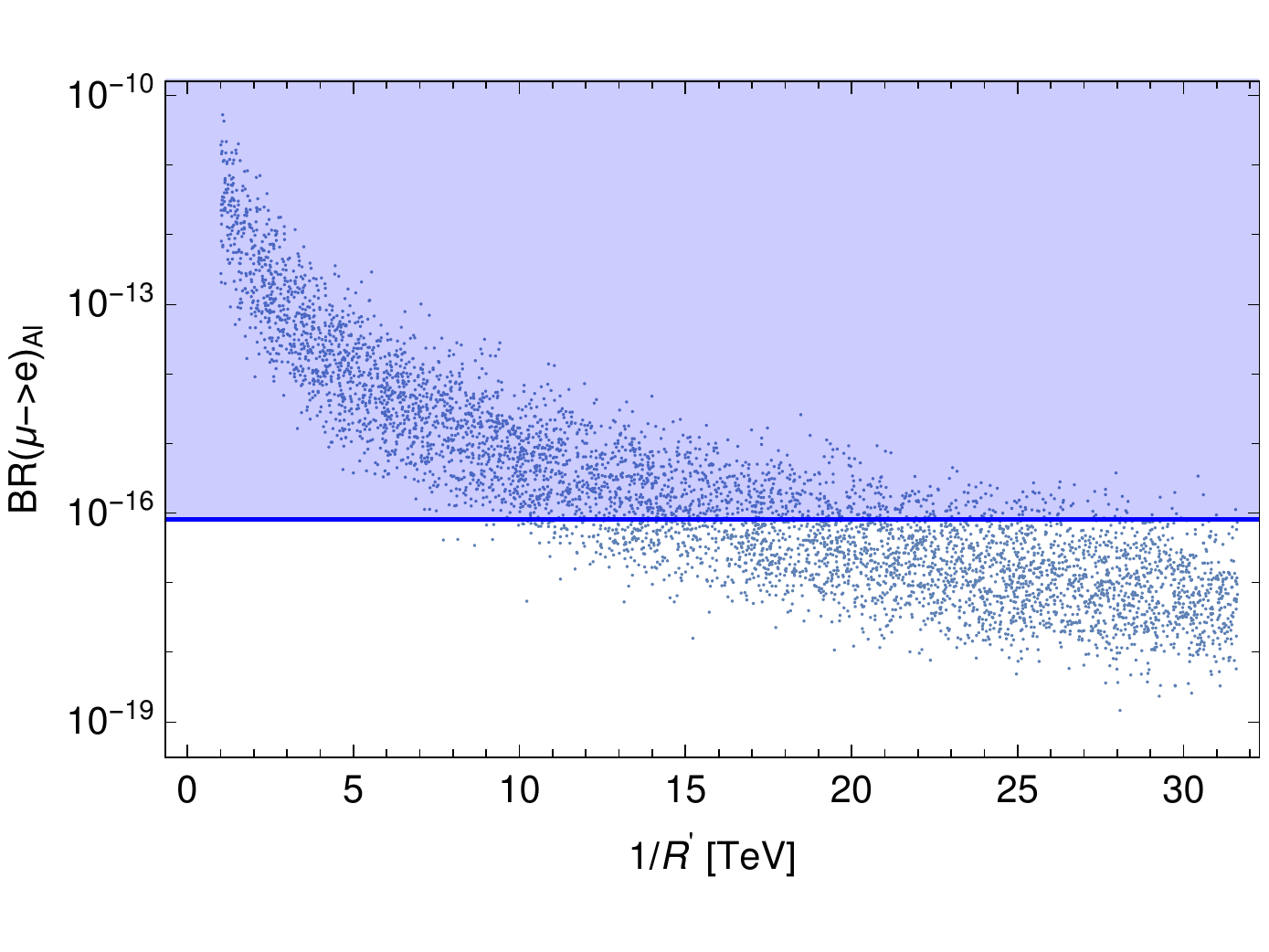}
  \end{subfigure}
  \caption{Current bound on $\mu - e$ conversion in Titanium (left) and future bound on $\mu - e$ conversion in Aluminium (right), displayed together with the predictions of our scan as a function of $1/R^\prime$. The points in the red (blue) region are (expected to be) excluded.}
  \label{fig:lepton1}
\end{figure}
%%%%%%%%%%%%%%%%%%%%%%%%%%%%%%
\begin{equation}
    \textrm{Br}(\mu \rightarrow e \gamma) = 4.2\times 10^{-13},
\end{equation}
%%%%%%%%%%%%%%%%%%%%%%%%%%%%%%
at 90\% CL. In the upcoming years, an update from MEG II \cite{MEGII:2018kmf} is expected, with a projected sensitivity of
%%%%%%%%%%%%%%%%%%%%%%%%%%%%%%
\begin{equation}
    \textrm{Br}(\mu \rightarrow e \gamma) = 6\times 10^{-14}.
\end{equation}
%%%%%%%%%%%%%%%%%%%%%%%%%%%%%%
As mentioned, the flavor violating decay can be induced from four different loop diagrams, namely $W$, $Z$, Higgs, and leptoquark loops. We will treat all four effects and compare them for SU(6) GHGUT (see Appendix \ref{AppendixB} and Appendix \ref{AppendixC} for the details).

\subsection{Results}\label{sec:Results2}

In Fig.~\ref{fig:lepton1}, we illustrate the constraints from $\mu-e$ conversion. The current bound from the Titanium atom is still rather weak for GHGUT models and can be basically neglected for $1/R^\prime>3$ TeV. However upcoming experiments in Aluminium can completely exclude the model for $1/R^\prime<10$ TeV and probe the parameter space into the $25$ TeV range. Our results agree with previous studies of $\mu-e$ conversion in extra dimensional models \cite{Agashe:2006iy}, see also \cite{Goertz:2021ndf,Goertz:2021xlx,Agashe:2013kxa}. 

\begin{figure}[!t]
  \begin{subfigure}[b]{0.5\textwidth}
    \includegraphics[width=\textwidth]{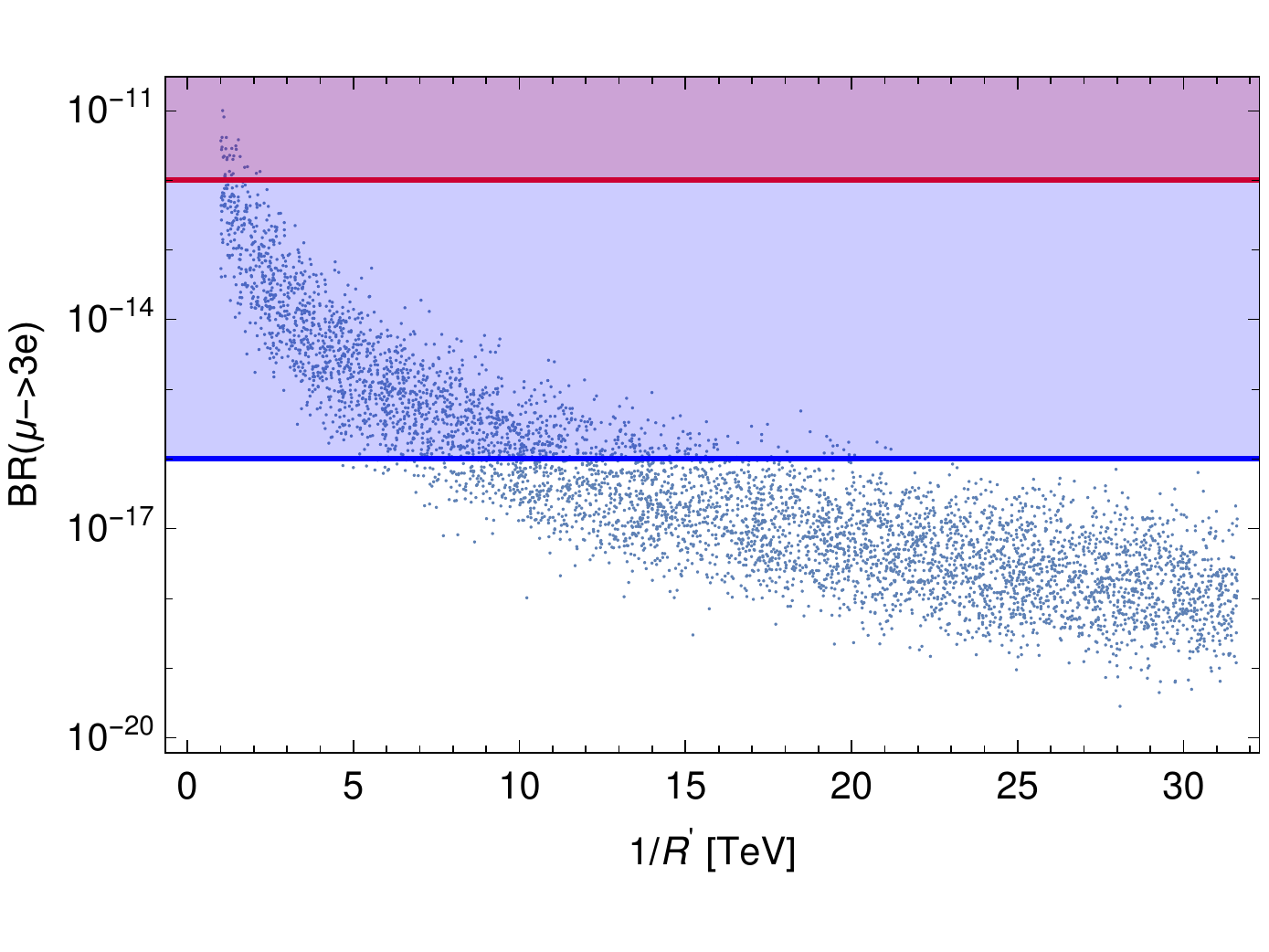}
  \end{subfigure}
  \hfill
  \begin{subfigure}[b]{0.5\textwidth}
    \includegraphics[width=\textwidth]{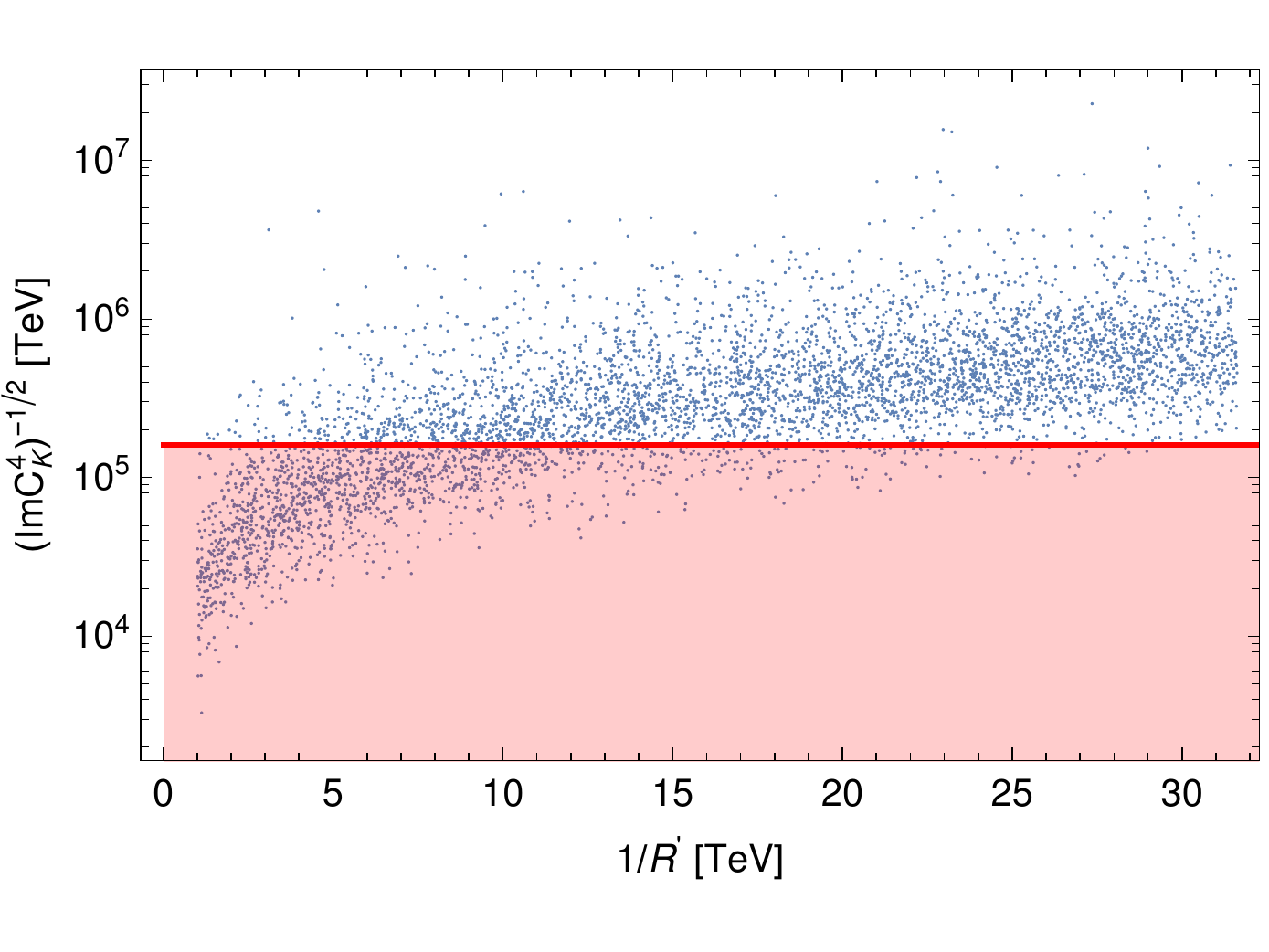}
  \end{subfigure}
  \caption{Current (red) and future (blue) constraints on the $\mu \rightarrow 3e$ decay (left) and $(\textrm{Im} C^4_K)^{-1/2}$ (right), displayed together with the predictions of our scan as a function of $1/R^\prime$.}
  \label{fig:lepton2}
\end{figure}

Fig.~\ref{fig:lepton2} shows the current and upcoming bound on the branching ratio for the decay $\mu \rightarrow 3e$ and the constraints from $K-\bar{K}$ meson mixing in the quark sector. The current bound on $\mu \rightarrow 3e$ is again weak but the upcoming constraint should probe up to the $15$ TeV region. The constraint from $K-\bar{K}$ meson mixing is rather weak, too. Excluding even an IR scale as low as $1/R^\prime \lesssim 3$ TeV is difficult. Bounds from $\mu - e$ conversion therefore furnish the more reliable tree level constraint on the model. Our results agree relatively well with previous studies of $K-\bar{K}$ meson mixing \cite{Csaki:2008zd,Bauer:2009cf}, with differences owing to a different fermion sector.

\begin{figure}[!t]
  \begin{subfigure}[b]{0.5\textwidth}
    \includegraphics[width=\textwidth]{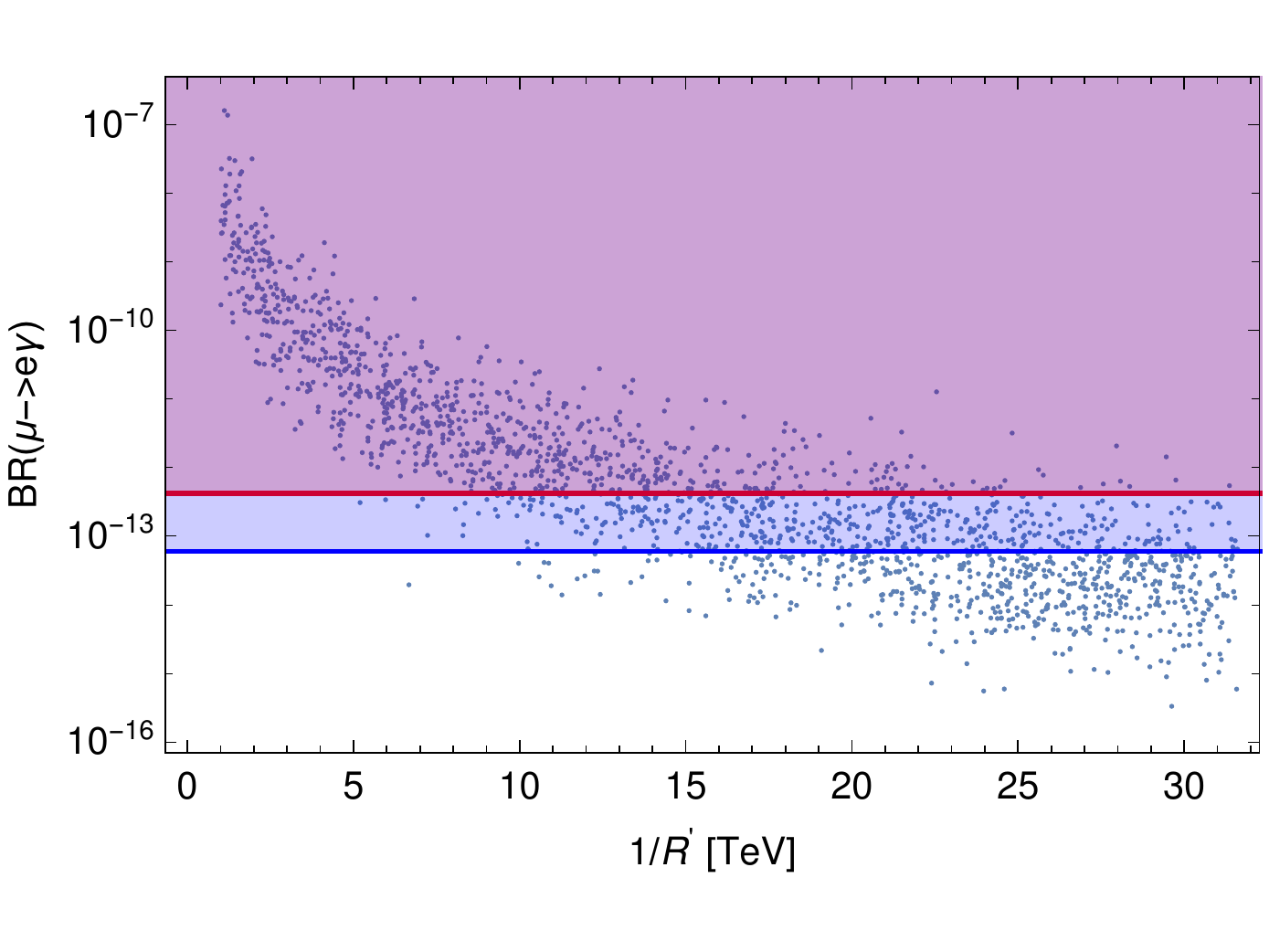}
  \end{subfigure}
  \hfill
  \begin{subfigure}[b]{0.5\textwidth}
    \includegraphics[width=\textwidth]{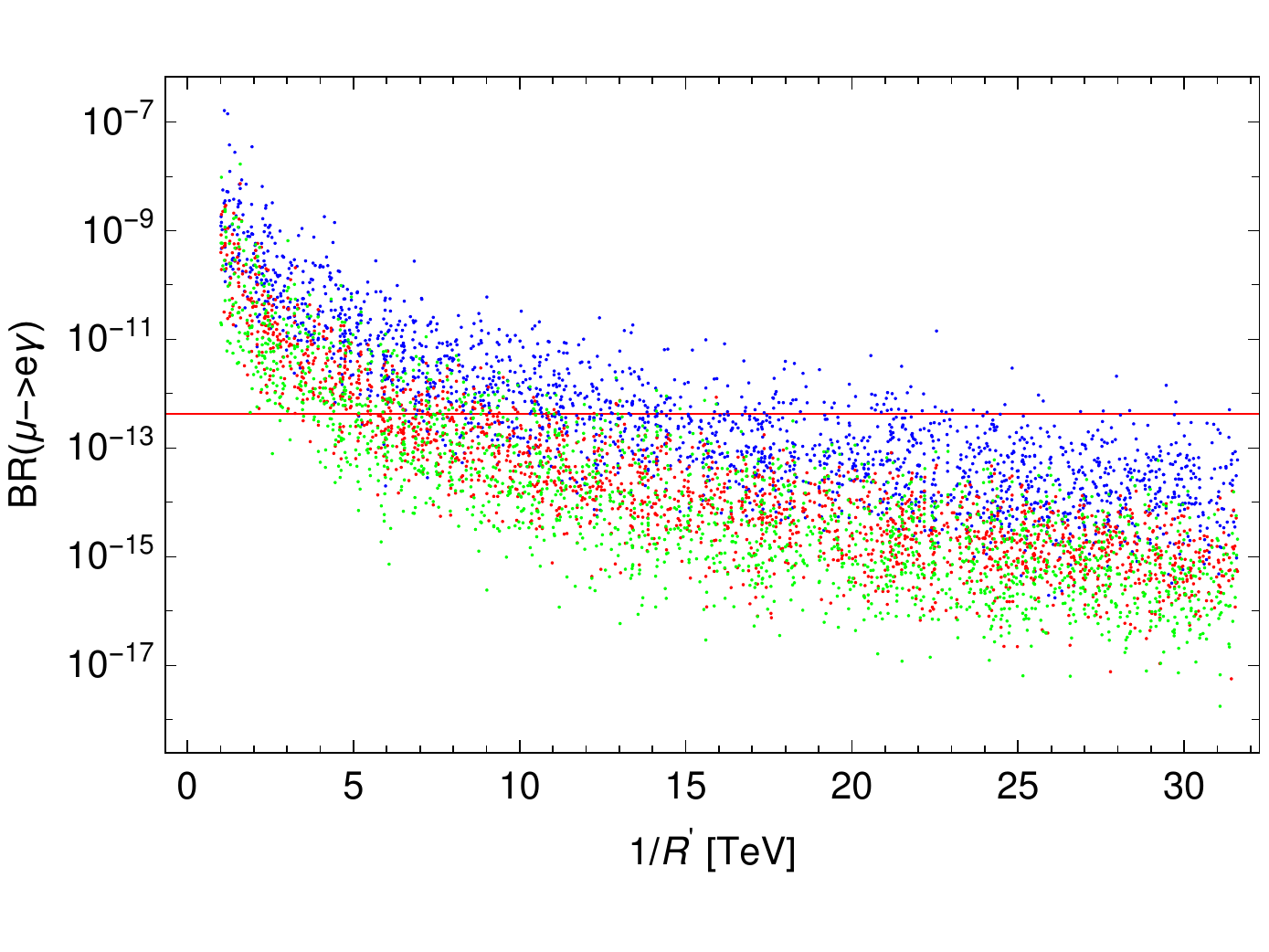}
  \end{subfigure}
  \caption{Current (red) and future (blue) constraints on the $\mu \rightarrow e\gamma $ decay, displayed together with the predictions of our scan (left), and relative size of the leptoquark (blue), Higgs (red), and $Z$ boson (green) loop contributions (right), as a function of $1/R^\prime$. The $W$ boson contribution is negligible and not shown.}
  \label{fig:lepton3}
\end{figure}

Finally, on the left panel of Fig.~\ref{fig:lepton3}, the current and upcoming bounds on the branching ratio of $\mu \rightarrow e \gamma$ are displayed. This process already reliably excludes IR scales $1/R^\prime$ lower than $7$~TeV, which is in agreement with \cite{Agashe:2006iy}. It provides therefore currently the most stringent constraint on the IR scale of the model from flavor. The future improvement by an order of magnitude on the branching ratio means the parameter space for $1/R^\prime < 10$ TeV will be completely excluded. There is a rather broad range of branching ratios for a specific IR scale. This reason behind this variation is that the branching ratio is largely determined by the lightest KK mode mediating the decay and for any specific IR scale $1/R^\prime$, there can be moderate variations in the mass of this mode. On the right panel of Fig.~\ref{fig:lepton3}, the breakdown of the $\textrm{Br}(\mu \rightarrow e \gamma)$ in terms of the different loop contributions is shown (under the assumption that only one mediator is present). The leptoquark contribution is the most relevant one, with the Higgs boson and $Z$ boson contribtuions being a bit smaller. The $W$ boson contribution can be safely neglected over the whole parameter space.

\section{Electroweak precision}\label{sec:EWPT}

Having obtained bounds from the flavor sector in the last section, we investigate in this section effects of SU(6) GHGUT on electroweak precision parameters, namely the oblique parameters $S,T,U$. It will turn out that the constraints on $1/R^{\prime}$ from the flavor considerations above are more stringent than the ones obtained from EWPT in this section, however the latter are less model dependent. These oblique corrections have already been studied in several models of warped extra dimensions, for example in the context of brane fermions \cite{Csaki2002}, IR brane localized Higgs scenarios with and without custodial symmetry \cite{Carena2003,Delgado2007,Casagrande:2008hr}, bulk Higgs scenarios \cite{Agashe2003b}, and also GHU models with custodial symmetry \cite{Agashe:2004rs}, but none of these analyses apply directly to our model. In the following, we show an explicit calculation for the oblique parameters in our setup, but the obtained results apply also to other GHU scenarios without custodial symmetry.

\subsection{Deriving the effective Lagrangian}

To connect to the electroweak parameters we match our model onto an 4D effective theory. The most general 4D Lagrangian for the electroweak gauge bosons is given by~\cite{Csaki2002}
%%%%%%%%%%%%%%%%%%%%%%%%%
\begin{align}
    \mathcal{L} = &- \frac{1}{4} \left(1-\Pi_{\gamma\gamma}^{\prime}(0)\right) F_{\mu\nu}F^{\mu\nu} - \frac{1}{2} \left(1-\Pi_{WW}^{\prime}(0)\right) W_{\mu\nu}W^{\mu\nu} \notag \\
    &- \frac{1}{4} \left(1-\Pi_{ZZ}^{\prime}(0)\right) Z_{\mu\nu}Z^{\mu\nu} + \frac{s_Wc_W}{2} \Pi_{\gamma Z}^{\prime}(0) F_{\mu\nu}Z^{\mu\nu} \notag \\
    &+ \left(m_W^2 + \Pi_{WW}(0) \right) W_{\mu}^{+} W^{-,\mu} + \frac{1}{2} \left( m_Z^2 + \Pi_{ZZ}(0) \right) Z_{\mu} Z^{\mu} \, .
\end{align}
%%%%%%%%%%%%%%%%%%%%%%%%%
Here $m_{W,Z}$ are the SM masses at tree level and $s_W=\sin(\theta_W)$ and $c_W=\cos(\theta_W)$ are the sine and cosine of the weak mixing angle $\theta_W$, respectively. The vacuum polarization amplitudes $\Pi(0)$ and their derivatives $\Pi^{\prime}(0) = \frac{\partial}{\partial q^2}\Pi\rvert_{q^2=0}$ incorporate the effects of BSM physics, with the fermion vertices normalized to their (tree level) SM values. We consider the case of oblique corrections in which all vertex corrections are equal and can therefore be absorbed in the common gauge boson polarizations $\Pi$. This is in general not the case in RS scenarios, but the RS-GIM mechanism ensures that the differences are small~\cite{Casagrande:2008hr}. An exception to this could be IR localized fermions, and their effects, for example in the $Zb\Bar{b}$ coupling, have to be considered independently. It turns out these constraints are in general not competitive with flavor constraints. Thus we focus only on the common oblique part and neglect the differences between fermions.

In RS models the coefficients $\Pi$ can be computed to first order in a tree level calculation by integrating over the extra dimension. All effects result from the fact that EWSB in GHU mixes the gauge bosons with their KK states and, in the case of the $Z$ boson, also mixes the $Z$ boson with the $X^\mu$ associated with U(1)\textsubscript{X} (see Section~\ref{sec:ZWboson}). In this subsection we will explicitly derive the corrections entering EWPT from the $Z$ boson, with the calculation for the $W$ boson proceeding similarly. Then in the next subsection we compare the corrections coming from the $Z$ and $W$ bosons with the current bounds on the oblique parameters $S,T$ and $U$.

Because the wavefunction of the photon zero mode is flat, it gives no contribution, i.e. $\Pi_{\gamma\gamma}^{\prime}(0)=\Pi_{\gamma\gamma}(0)=0$, and in this model there is no kinetic mixing between the $Z$ boson and the photon at tree level, $ \Pi_{\gamma Z}^{\prime}(0)$=0.

Considering only the first two modes of the $Z$ boson as well as the first mode of $X^\mu$, the action, including EWSB from \eqref{eq:gaugeBosonMasses}, takes the form
%%%%%%%%%%%%%%%%%%%%%%%%%
\begin{align}
    S &\supset \int \textrm{d}^4x \bigg[ -\frac{1}{4} Z_{\mu\nu}^{(0)}Z^{(0),\mu\nu} -\frac{1}{4} Z_{\mu\nu}^{(1)}Z^{(1),\mu\nu} -\frac{1}{4} X_{\mu\nu}^{(1)}X^{(1),\mu\nu} \notag \\
    &\qquad + \frac{1}{2} \begin{pmatrix} Z_{\mu}^{(0)} & Z_{\mu}^{(1)} & X_{\mu}^{(1)} \end{pmatrix} \begin{pmatrix}
    m_Z^2 & f_{01} m_Z^2 & -f_{01}^{X} \frac{g_Z g_{X} v^2}{12}\\
    f_{01} m_Z^2 & m_{(+,+)}^2 & -f_{11}^{X} \frac{g_Z g_{X} v^2}{12} \\
    -f_{01}^{X} \frac{g_Z g_{X} v^2}{12} & -f_{11}^{X} \frac{g_Z g_{X} v^2}{12} & m_{(-,-)}^2
    \end{pmatrix} \begin{pmatrix} Z^{(0),\mu} \\ Z^{(1),\mu} \\ X^{(1),\mu} \end{pmatrix} \bigg] \, ,
\end{align}
%%%%%%%%%%%%%%%%%%%%%%%%%
where $m_Z = g_Z v/2$, $g_Z = \sqrt{g^2+g^{\prime 2}}$ and $f_{01}=5.47$ is the overlap between the zero mode and first KK excitation of the $Z$ boson with the Higgs wavefunction, as defined in Section~\ref{sec:ZWboson}. There are also corresponding overlap integrals $f_{01}^{X}=4.94$ and $f_{11}^{X}=29.47$ between the wavefunction of the first mode of $X^{\mu}$ and the zero and first mode of the $Z$ boson respectively. % We included only the gauge eigenstates $Z^{(0)},Z^{(1)}$ and $X^{(1)}$, since the physical $Z$ boson, which results from diagonalizing the mass matrix, is to first order in $(R^\prime v)^2$ given as a linear combination of the lowest modes. 
The overlaps between the zero mode of the $Z$ boson and higher modes are much smaller so that we can neglect them at this order in $(R^\prime v)^2$. Furthermore, $g_X$ denotes the coupling of U(1)\textsubscript{X}. Diagonalizing the mass matrix, one finds the action for the mass eigenstate $Z_0$, reading
%%%%%%%%%%%%%%%%%%%%%%%%%
\begin{align}
    S &\supset \int \textrm{d}^4 x \bigg[ -\frac{1}{4} Z_{0,\mu\nu}Z_0^{\mu\nu} + \frac{1}{2} m_Z^2\left(1- f_{01}^2\frac{m_Z^2}{m_{(+,+)}^2}- \left(f_{01}^X\right)^2\frac{g_X^2 v^2}{36 m_{(-,-)}^2}\right) Z_{0,\mu} Z_0^{\mu} \bigg]\, .
\end{align}
%%%%%%%%%%%%%%%%%%%%%%%%%
Additionally, one has to correctly normalize the fermion interaction terms to their SM values. 

To achieve this, we use Eq. \eqref{eq:effectiveZW} to rescale the $Z$ boson field by
%%%%%%%%%%%%%%%%%%%%%%%%%
\begin{equation}
    Z_{0,\mu} \to \lambda_{Z^0}^{-1} Z_{0,\mu} = \left(1 - \lambda_{(+,+)} f_{01} \frac{m_Z^2}{m_{(+,+)}^2} \right)^{-1} Z_{0,\mu} \, .
\end{equation}
%%%%%%%%%%%%%%%%%%%%%%%%%
As explained above we focus here on the lighter generations of UV localized fermions, for which $\lambda_{(+,+)} \equiv \lambda_{(+,+)}(c>0.5)$ is independent of the fermion localization, representing the common oblique corrections. Note that we neglect the effect of $X^{(1)}$ on the fermion couplings since the corresponding overlap $\lambda_{(-,-)}$ between UV localized fermions and a $(-,-)$ field is very much suppressed. In total this results in the action 
%%%%%%%%%%%%%%%%%%%%%%%%%
\begin{align}
    S &\supset \int \textrm{d}^4x \bigg[ -\frac{1}{4} \left(1 -\left(- 2 \lambda_{(+,+)} f_{01} \frac{m_Z^2}{m_{(+,+)}^2}\right)\right) Z_{0,\mu\nu}Z_0^{\mu\nu} \notag \\
    &\qquad + \frac{1}{2} m_Z^2 \left(1 + \left(-f_{01}^2 + 2 \lambda_{(+,+)} f_{01} \right) \frac{m_Z^2}{m_{(+,+)}^2} - \left(f_{01}^X\right)^2\frac{g_X^2 v^2}{36 m_{(-,-)}^2} \right) Z_{0,\mu} Z_0^{\mu} \bigg] \, ,
\end{align}
%%%%%%%%%%%%%%%%%%%%%%%%%
from which we can read off 
%%%%%%%%%%%%%%%%%%%%%%%%%
\begin{align}
    \Pi_{ZZ}^{\prime}(0) &= - 2 \lambda_{(+,+)} f_{01} \frac{m_Z^2}{m_{(+,+)}^2} \, ,\\
    \Pi_{ZZ}(0) &= m_Z^2 \left[ \left(-f_{01}^2 + 2 \lambda_{(+,+)} f_{01} \right) \frac{m_Z^2}{m_{(+,+)}^2} - \left(f_{01}^X\right)^2\frac{g_X^2 v^2}{36 m_{(-,-)}^2} \right] \, .
\end{align}
%%%%%%%%%%%%%%%%%%%%%%%%%
Doing a similar calculation for the $W$ boson results in 
%%%%%%%%%%%%%%%%%%%%%%%%%
\begin{align}
    \Pi_{WW}^{\prime}(0) &= - 2 \lambda_{(+,+)} f_{01} \frac{m_W^2}{m_{(+,+)}^2} \, , \\
    \Pi_{WW}(0) &= m_W^2 \left(-f_{01}^2 + 2 \lambda_{(+,+)} f_{01} \right) \frac{m_W^2}{m_{(+,+)}^2} \, .
\end{align}

\subsection{\texorpdfstring{$S,T,U$}{STU} parameters}

Rescaling the vacuum polarizations by $\Pi_{WW} = g_W^2 \Pi_{11}, \Pi_{ZZ} = g_Z^2 \Pi_{33}$, etc., we can use the standard definitions of the $S,T,U$ parameters \cite{Peskin1992}
%%%%%%%%%%%%%%%%%%%%%%%%%
\begin{align}
    S &= 16 \pi (\Pi_{33}^{\prime}(0)-\Pi_{3Q}^{\prime}(0)) \, ,\\
    T &= \frac{4\pi}{s_W^2c_W^2m_Z^2} (\Pi_{11}(0)-\Pi_{33}(0)) \, ,\\
    U &= 16 \pi (\Pi_{11}^{\prime}(0)-\Pi_{33}^{\prime}(0))
\end{align}
%%%%%%%%%%%%%%%%%%%%%%%%%
to obtain at leading order in $(vR^{\prime})^2$
%%%%%%%%%%%%%%%%%%%%%%%%%
\begin{align}
    S &= \frac{4 \pi v^2}{m_{(+,+)}^2} \left(-2 \lambda_{(+,+)} f_{01}\right) \, , \label{eq:SPara} \\
    T &= \frac{4 \pi v^2}{4 c_W^2 m_{(+,+)}^2} \left(f_{01}^2 - 2 \lambda_{(+,+)} f_{01} \right) + \frac{4\pi v^2}{36 s_W^2 c_W^2 m_{(-,-)}^2} \left(f_{01}^X\right)^2 \frac{g_{X}^2}{g_{Z}^2} \, , \label{eq:TPara} \\
    U &= 0 \, .
\end{align}
%%%%%%%%%%%%%%%%%%%%%%%%%
Note that $S,T$ are both positive since $\lambda_{(+,+)}f_{01}<0$, which is the case for all other RS scenarios (except when the fermions are on the brane \cite{Csaki2002}). The result depends on the coupling $g_X$, which has to be calculated by running down the unified coupling from the unifying scale. A reasonable expectation is that $g_X \sim g_Z, g_W$, from which it follows that the contribution of $X^{(1)}$ in \eqref{eq:TPara} is about $10 \%$ that of $Z^{(1)}$. As a conservative bound we drop this term in the comparison with the experimental constraints and leave the exact analysis of the running for a future work.

From \eqref{eq:effectivecoupling} it follows to leading order, for the UV localized fermions we consider here, $\lambda_{(+,+)} \sim - \frac{1}{\sqrt{L}}$, with $L=\log(R^{\prime}/R)$ the logarithm of the warp factor. In fact it is convenient to use the same formula to estimate the scaling of $f_{01}$, by realizing that the integral can be recovered by replacing the Higgs wavefunction in \eqref{eq:overlap} with a LH zero mode fermion wavefunction with $c=-1/2$. This leads to $f_{01} \sim \sqrt{L}$ and one can show that similarly $f_{01}^{X} \sim \sqrt{L}$, but with a smaller numerical coefficient. Together, these relations imply $S \sim L^0$ and $T \sim L^1$ to leading order, as in the brane Higgs scenarios without custodial symmetry \cite{Casagrande:2008hr}. In fact one could try to use~\eqref{eq:overlap} for a brane localized Higgs by just replacing the Higgs wavefunction by a delta function on the IR brane. However, in brane Higgs scenarios one can no longer neglect the contribution of higher KK modes, as their overlap with the Higgs stays the same~\cite{Agashe:2006iy}. Taking this into account, we find that the constraints in GHU are significantly weaker. More quantitatively, in our scenario, the $T$ ($S$) parameter gets reduced by a factor approximately $0.4$ ($0.8$) with respect to non--custodial brane Higgs scenarios. It is also worth mentioning that the smallness of the $T$ parameter is partly due to the fact, that the weak mixing angle of the KK modes is considered the same as that of the zero modes. This might change if the running of the couplings is taken into account, which we will explore further in a future work.

\begin{figure}[t]
  \begin{subfigure}[b]{0.5\textwidth}
    \includegraphics[width=\textwidth]{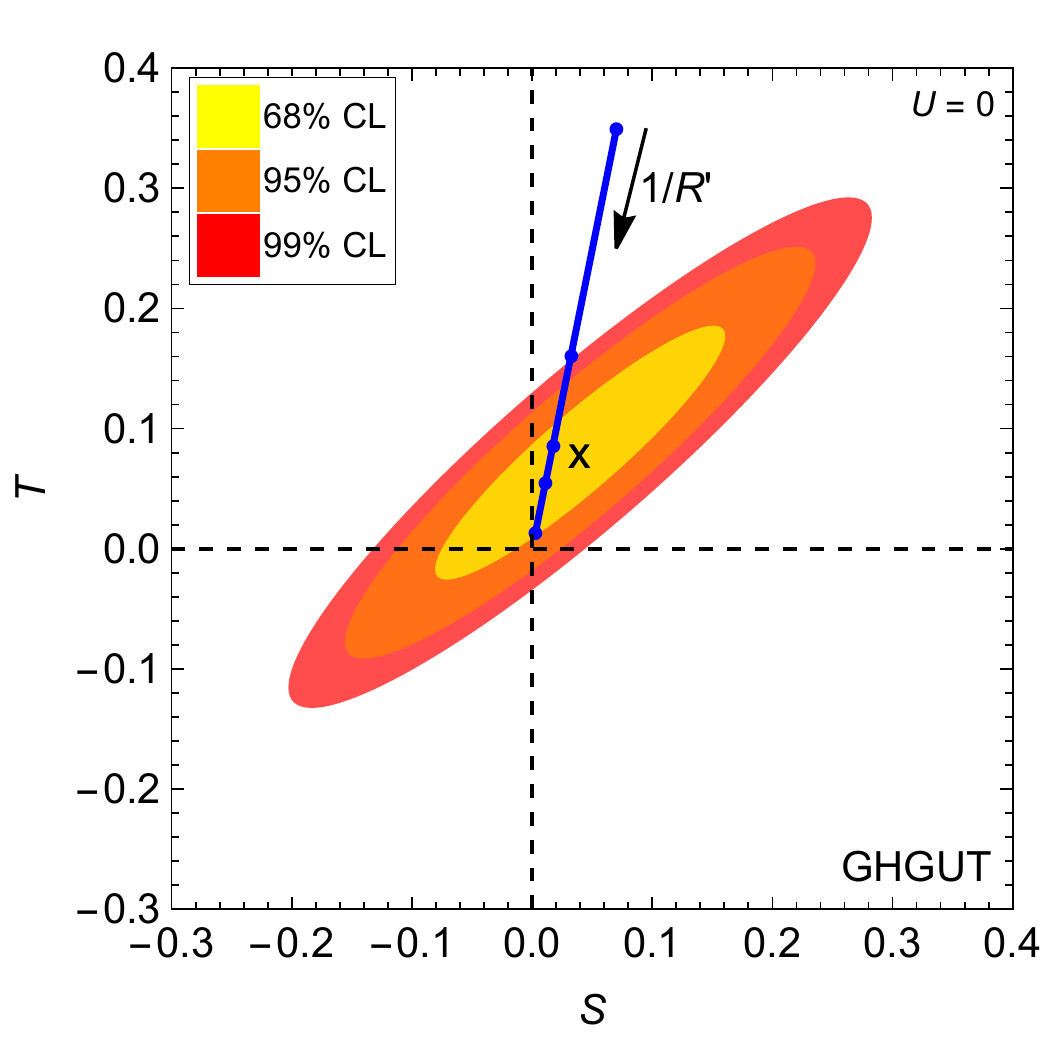}
  \end{subfigure}
  \hfill
  \begin{subfigure}[b]{0.5\textwidth}
    \includegraphics[width=\textwidth]{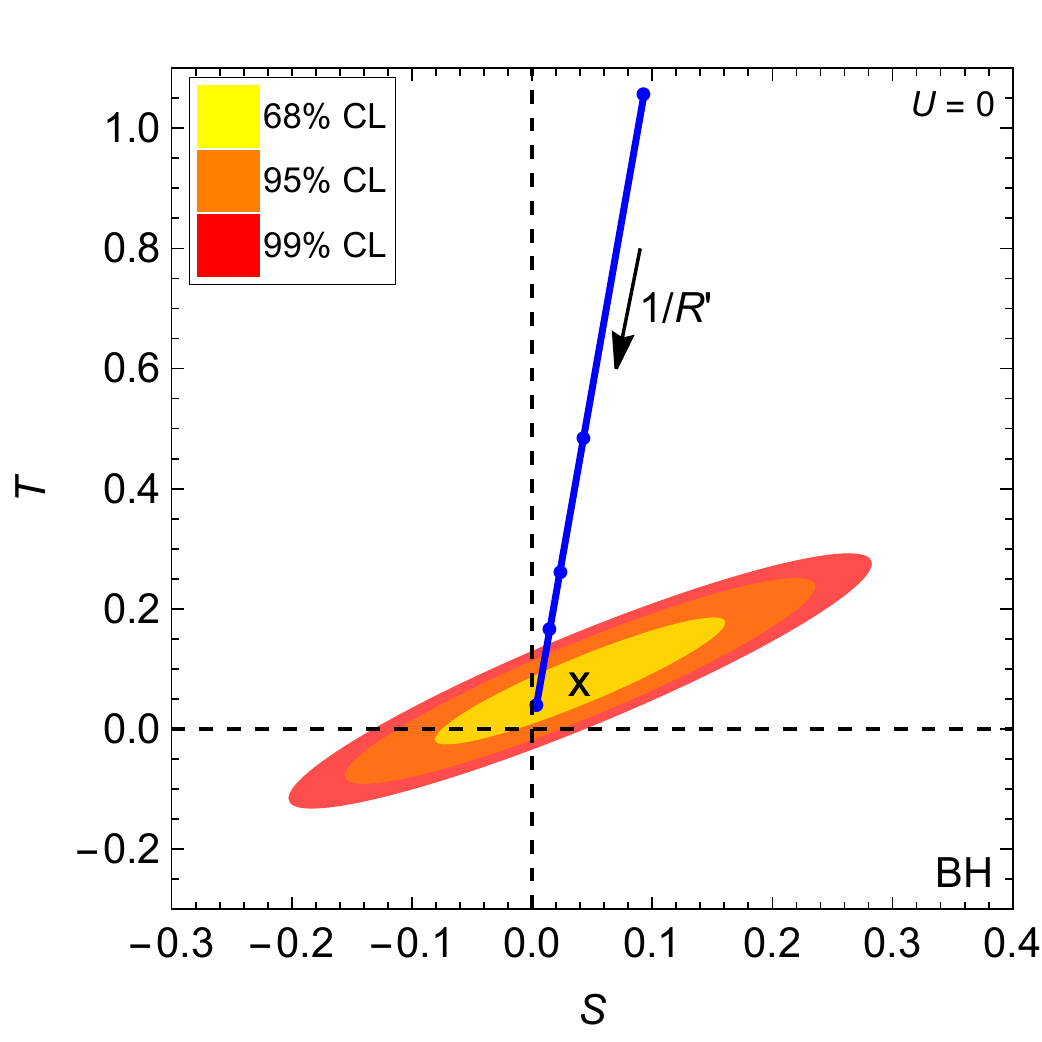}
  \end{subfigure}
  \caption{Regions allowed at $68\%$, $95\%$, and $99\%$ CL in the $S-T$ plane with $U=0$ ($m_{t,\mathrm{ref.}} = 172.5 \, \textrm{GeV}$, $m_{H,\mathrm{ref.}} = 125 \, \textrm{GeV}$) \cite{Haller2018}. The blue line represents the contributions in our GHGUT (left) and in a Brane--Higgs (BH) scenario \cite{Casagrande:2008hr} (right) for $1/R^{\prime} \in \left[2,10\right] \textrm{TeV}$ and $1/R = 10^{18} \, \textrm{GeV}$. Note that both models are without custodial protection. $1/R^{\prime}$ increases in the direction of the arrow and the blue dots represent the values $1/R^{\prime}=2,3,4,5,10\, \textrm{TeV}$. See text for details.}
  \label{fig:EWPTPlot}
\end{figure}

The experimental bounds on the $S$ and $T$ parameters and their correlation matrix are given by \cite{Haller2018}
%%%%%%%%%%%%%%%%%%%%%%%%%
\begin{equation}
    \begin{split}
        S = 0.04 \pm 0.08, \\
        T = 0.08 \pm 0.07,
    \end{split} \qquad \qquad \rho = \begin{pmatrix}
    1.00 & 0.92 \\ 0.92 & 1.00
    \end{pmatrix}\, .
\end{equation}
%%%%%%%%%%%%%%%%%%%%%%%%%
Using these bounds, we give in Fig.~\ref{fig:EWPTPlot} the regions allowed at $68\%$, $95\%$, and $99\%$ CL in the $S-T$ plane. The blue line represents the RS corrections from \eqref{eq:SPara} and \eqref{eq:TPara} for different values of $1/R^{\prime}$. %, where, as explained above, we ignore the contribution of $X^{(1)}$. 
In the global fit the parameter $U$ is set to zero and we used $v=246 \, \textrm{GeV}$ and $s_W^2 = 0.23122$.

% \begin{figure}[ht]
% 	\includegraphics[scale=1]{EWPTImages/EWPTPlot_new.pdf}
% 	\centering
% 	%\vspace{-0.5cm}
% 	\caption{Regions allowed at $68\%$, $95\%$, and $99\%$ confidence level (CL) in the $S-T$ plane with $U=0$ ($m_{t,\mathrm{ref.}} = 172.5 \, \textrm{GeV}$, $m_{H,\mathrm{ref.}} = 125 \, \textrm{GeV}$) \cite{Haller2018}. The blue line represents the contributions in GHU (without custodial protection) for $1/R^{\prime} \in \left[1,30\right] \textrm{TeV}$ and $R = 10^{-18} \, \textrm{GeV}^{-1}$. $1/R^{\prime}$ increases in the direction of the arrow. See text for details.}
%   \label{fig:EWPTPlot}
% \end{figure}

The RS contributions from \eqref{eq:SPara} and \eqref{eq:TPara} pass the CL thresholds at 
%%%%%%%%%%%%%%%%%%%%%%%%%
\begin{align}
    \frac{1}{R^{\prime}} &> 3.0\, \mathrm{TeV} \quad (99\% \, \mathrm{CL})\, , \\
    \frac{1}{R^{\prime}} &> 3.2\, \mathrm{TeV} \quad (95\% \, \mathrm{CL}) \, , \\
    \frac{1}{R^{\prime}} &> 3.7\, \mathrm{TeV} \quad (68\% \, \mathrm{CL}) \, .
\end{align}
%%%%%%%%%%%%%%%%%%%%%%%%%
Note that for moderately large values of $1/R^{\prime}$ the fit compared to the SM is improved.

\section{Scalar potential}\label{sec:Potential}
Having identified an SU(6) model that can be successful in accounting for all the flavor hierarchies, we now put its scalar sector to the test. The one-loop scalar potential can be computed from the gauge boson and fermion spectrum using the Coleman-Weinberg formula  \cite{Falkowski:2006vi}, from which we get (anticipating that our final parameter space will feature a vanishing singlet vev) 
%%%%%%%%%%%%%%%%%%%%%%%%%
\begin{equation}
\label{CW}
    V(h)=\sum_{r} V_r(h)=\sum_{r} \frac{N_r}{(4\pi)^2}\int_0^\infty dp \, p^3\log(\rho_r(-p^2,h)),
\end{equation}
%%%%%%%%%%%%%%%%%%%%%%%%%
where $N_r=-4N_c$ for quarks, $N_r=3$ for gauge bosons, and $\rho_r$ denotes the corresponding spectral function, whose roots at $-p^2=m_{n;r}^2,\, n\in \mathbb{N}$ encodes the physical spectrum. To find these spectral functions one has to solve for the exact gauge boson and fermion masses, considering a general vev for each of the possible massless scalars. As discussed in Section~\ref{sec:WarpedGaugeFields}, this can be more easily done in the \textit{holographic} gauge \cite{Hosotani:2005nz}, where the vev of the scalars is gauged away from the bulk to the IR brane. The largest contribution typically comes from the top quark and $W/Z$ gauge bosons, as indicated by their large coupling to the Higgs. Furthermore, the up-type exotic sector, whose spectrum depends on the Higgs vev, can also contribute to the potential, as found in \cite{Angelescu:2021nbp}. This can easily be understood from considering the equivalent 4D dual formulation of our model in which the exotic sector can be understood as a vector-like fermion in the elementary sector. This new elementary fermion has the quantum numbers of an up-type quark and its connections to the strong sector are identical to those of the up-type quarks from SU(5) invariance. It follows that also for this exotic sector only the third generation couples significantly to the strong sector and is capable of generating a Higgs potential.

For a quantitative discussion, it is useful to have a consistent expansion of the potential. We thus write each of the spectral functions as
\begin{equation}
    \rho_r(-p^2,h)=1+f_{r;1}(-p^2) \sin^2(h/\sqrt{2}f)+f_{r;2}(-p^2) \sin^4(h/\sqrt{2}f),
\end{equation}
with $f=2\sqrt{R}/g_5R^\prime$ the Higgs decay constant defined in \eqref{Higgsdecayconstant}. Denoting $\sin^2(h/\sqrt{2}f)=x$, each contribution to the Higgs potential becomes
%%%%%%%%%%%%%%%%%%%%%%%%%
\begin{align}
    V_r(x)=\frac{N_r}{(4\pi)^2}\int_0^\infty \text{d}p p^3\log(1+f_{r;1}(-p^2)x+f_{r;2}(-p^2)x^2).
\end{align}
%%%%%%%%%%%%%%%%%%%%%%%%%
A consistent expansion up to $x^2$ can then be found as \cite{Csaki:2008zd}
%%%%%%%%%%%%%%%%%%%%%%%%%
\begin{align}
    V_r(x) =& \alpha_r x + \beta_r x^2 + \gamma_r x^2 \log\Big(\frac{2 c_r x}{\Lambda_r^2}\Big),\notag \\
    \alpha_r = & \frac{N_r}{(4\pi)^2}\int_0^\infty \text{d}p p^3 f_{r;1}(-p^2), \notag\\
    \beta_r = & \frac{N_r}{(4\pi)^2}\Bigg(\int_0^\infty \text{d}p p^3 \Big(f_{r;2}(-p^2)-\frac{1}{2}f_{r;1}(-p^2)^2 +\frac{c_r^2}{2\Lambda_r^4 \sinh(p^2/\Lambda_r^2)^2}\Big) - \frac{3}{8}c_r^2\Bigg),\notag \\
    \gamma_r = & \frac{N_r}{(4\pi)^2}\frac{1}{4}c_r^2,
\end{align}
%%%%%%%%%%%%%%%%%%%%%%%%%
where $\Lambda_r$ is an IR regulator coming from the fact that cutting off the expansion at order $\sin^4(v/\sqrt{2}f)$ introduces an IR divergence for zero modes and $c_r=f_{r;1}(-p^2)p^2$ for $p\rightarrow 0$.
\begin{figure}[!tbp]
  \begin{subfigure}[b]{0.5\textwidth}
    \includegraphics[width=\textwidth]{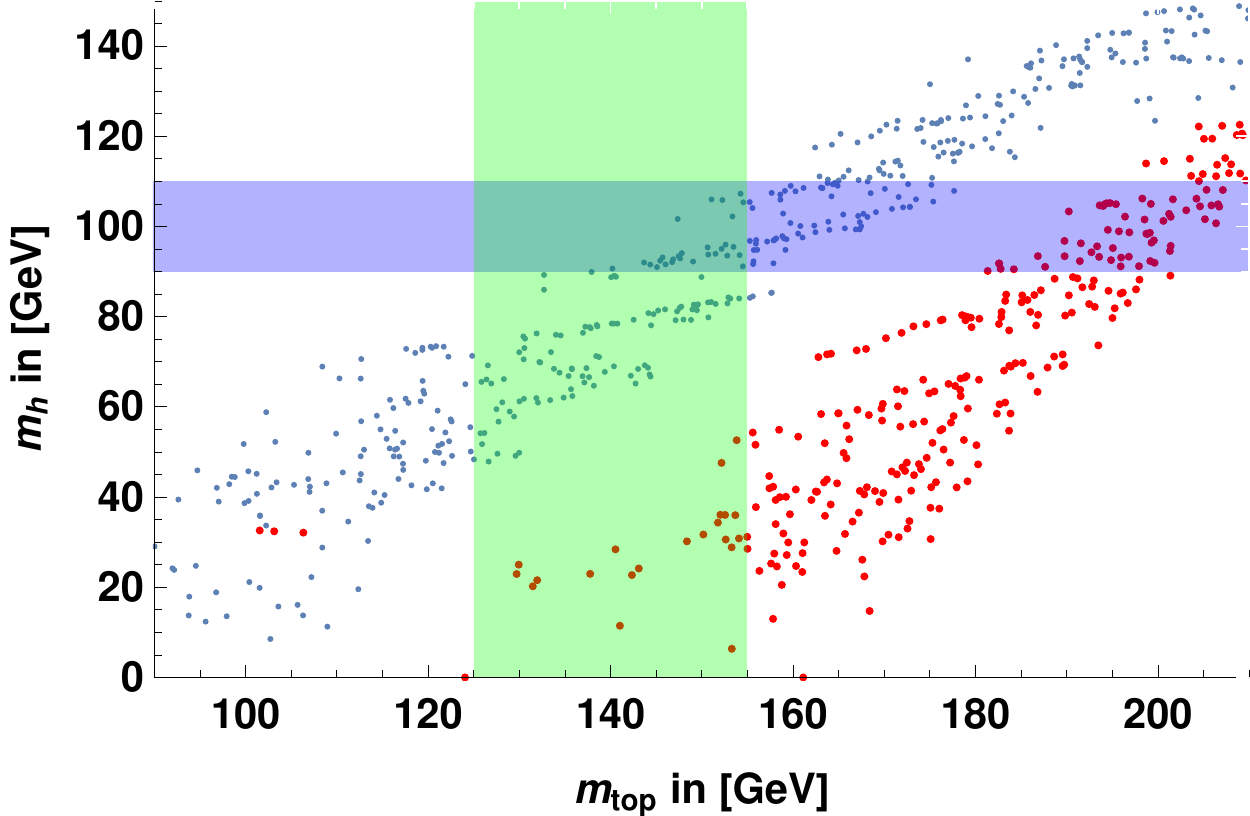}
  \end{subfigure}
  \hfill
  \begin{subfigure}[b]{0.5\textwidth}
    \includegraphics[width=\textwidth]{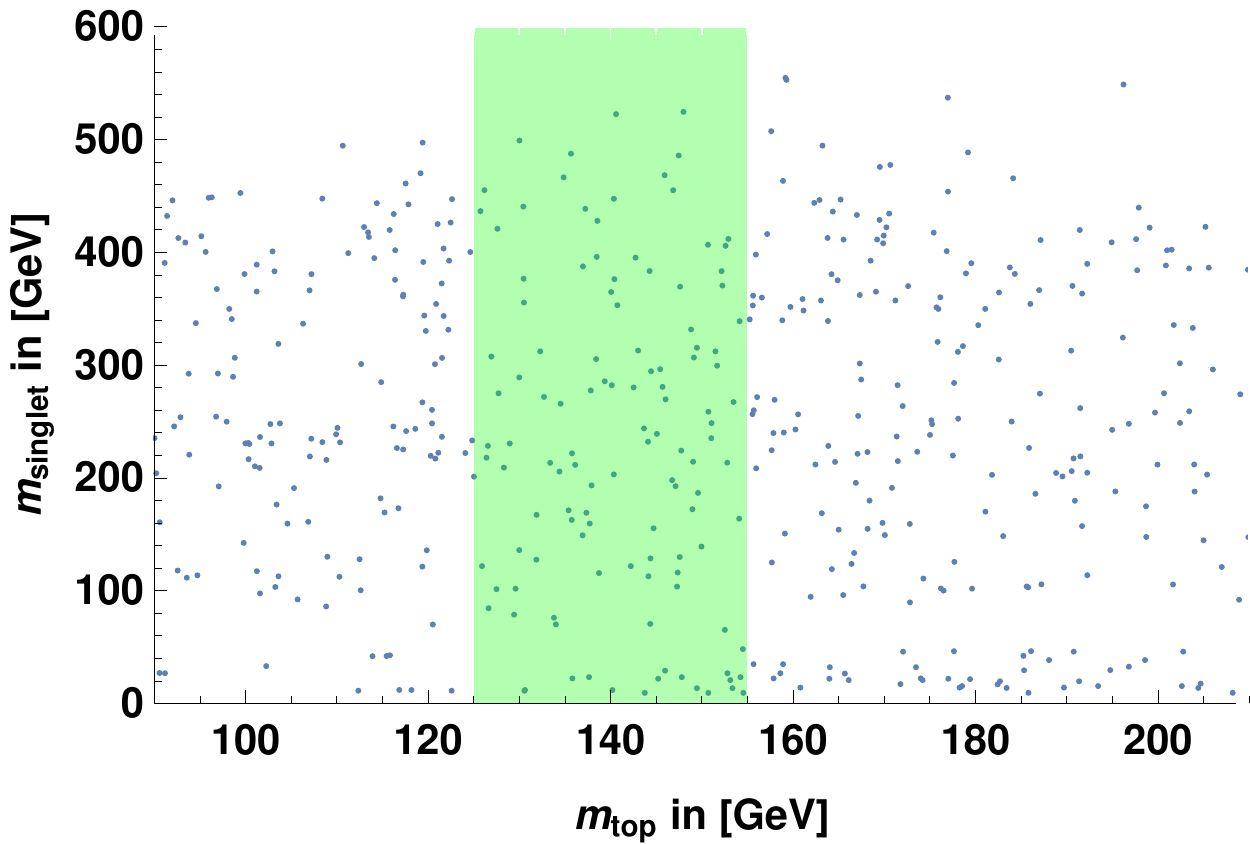}
  \end{subfigure}
  \caption{Left panel: The Higgs mass as a function of the top quark mass for $1/R^\prime=10\,$TeV. Points in red feature $M_{\tilde{u}}>0$ and predict the wrong Higgs mass while the points in blue belong to $M_{\tilde{u}}<0$ and are compatible with the correct Higgs mass. Right panel: The mass of the singlet scalar versus the top quark mass for the points $M_{\tilde{u}}<0$. Points with $M_{\tilde{u}}>0$ generate a vev for the singlet.}\label{Fig:mh}
\end{figure}

In this expansion the Higgs potential becomes particularly simple to analyse, reading\footnote{In contrast to most other gauge-Higgs literature, we find a factor of $2^{1/2}$ inside the trigonometric functions which is compensated by a factor of $2$ and $4$ in $\alpha_r$ and $\beta_r$ respectively. This is due to our unconventional normalization for the generators $\textrm{Tr}(T^a T^b)=\delta^{ab}/2$.}
%%%%%%%%%%%%%%%%%%%%%%%%%
\begin{equation}
    V (h) = \alpha_{\textrm{tot}} \sin^2\Big(\frac{h}{\sqrt{2} f}\Big) + \beta_{\textrm{tot}} \sin^4\Big(\frac{h}{\sqrt{2} f}\Big)+\sin^4\Big(\frac{h}{\sqrt{2} f}\Big) \sum_{r} \gamma_{\textrm{r}}\log \Big(\frac{2c_r\sin^2\Big(\frac{h}{\sqrt{2} f}\Big)}{\Lambda_r^2}\Big),
\end{equation}
%%%%%%%%%%%%%%%%%%%%%%%%%
with $\alpha_{\textrm{tot}}=\sum_r \alpha_r$ and $\beta_{\textrm{tot}}=\sum_r \beta_r$.
Requiring  spontaneous symmetry breaking and a stable potential leads to the conditions $\alpha_{\textrm{tot}}<0$ and $\beta_{\textrm{tot}}>0$, from which the Higgs mass and vev can be computed as
%%%%%%%%%%%%%%%%%%%%%%%%%
\begin{align}\label{Higgsmass}
    \sin(\frac{v}{\sqrt{2}f})^2&=\frac{-\alpha_{\textrm{tot}}}{2\beta_{\textrm{tot}}+\gamma_{\textrm{tot}}}, \notag \\
    m_h&=\frac{(\beta_{\textrm{tot}}+3/2\gamma_{\textrm{tot}})^{1/2}}{f}\sin(\frac{\sqrt{2}v}{f}),
\end{align}
%%%%%%%%%%%%%%%%%%%%%%%%%
with $\gamma_{\textrm{tot}}=\sum_r \gamma_{\textrm{r}}$
The first condition embodies the fine tuning problem in the Higgs potential. For an untuned potential, $|\alpha_{\textrm{tot}}|\sim |\beta_{\textrm{tot}}|\sim f^4 $, there will in general be no large seperation of scale between $v$ and $f$. Tuning consists in cancelling large contributions to $\alpha_{\textrm{tot}}$ from different sectors (mostly the top and the exotics) with each other to generate $\alpha_{\textrm{tot}}\ll \beta_{\textrm{tot}}$ and such that the vev $v$ is stabilised at $v\ll f$. Similarly to the MCHM$_5$, the contributions of the top quark to $\beta$ appear at higher order in the mixing with the strong sector compared to $\alpha$, which can lead to the problem of double tuning~\cite{Panico:2012uw}. This shows the crucial importance of the exotic sector that turns out to contribute to $\beta$ at leading order and can therefore generate a large quartic. As a result, there is no double tuning in our model.

To analyse the EWSB in the model quantitatively, we work with the exact one-loop potential from Eq \eqref{CW}. We restrict ourselves to the third generation fermions plus the $W$ and $Z$ bosons, whose contribution to the scalar potential is dominant. The main parameters playing a role in EWSB are $c_{15,3},c_{20,3},M_{q/e} $ and $M_{\tilde{u}}$, since the $c_{6,3}, c_{1,3}$ and $M_{d/l}, M_\nu$ determine mostly the bottom, tau and neutrino sectors which are only of minor importance. Interestingly, the UV brane mass $M_{\tilde{u}}$ becomes of crucial importance for the Higgs potential as this parameter influences the dependence of the exotic sector on the Higgs background. We note that for each data point employed in the flavor scans we fix $c_{20,3}$ to fine tune the potential to obtain the correct SM vev, satisfying the EWSB condition \eqref{Higgsmass}. Since $c_{20,3}$ also enters the top mass, we expect a strong correlation between the Higgs and top sectors.

The results are presented in Fig.~\ref{Fig:mh}, left panel, where only a fraction of our full scan was used due to the computational costs. Interestingly, there are two branches, one branch with positive exotic brane mass $M_{\tilde{u}}>0$ which predicts a wrong Higgs mass, while the negative brane-mass branch {\it predicts} a Higgs mass in agreement with observation. This points to the crucial importance of the exotic up quark that couples to the Higgs and thus can contribute to its mass. 

\subsection{Color-broken Universe}

After having studied the Higgs potential, we will now analyse the two other directions in field space, along the leptoquark vev $c$ and the singlet vev $s$, which requires a generalization of equation \eqref{CW} to
%%%%%%%%%%%%%%%%%%%%%%%%%
\begin{equation}
    V_r(h,c,s)=\frac{N_r}{(4\pi)^2}\int_0^\infty dp \, p^3\log(\rho_r(-p^2,h,c,s)).
\end{equation}
%%%%%%%%%%%%%%%%%%%%%%%%%
Since we do not need to fine-tune along the color-broken directions in field space, here a (more robust and analytical, but approximate) calculation exploiting the low-energy spectrum of the model will be sufficient. We will find that the colored scalar does not acquire a vev and gets a mass of around $m_{\textrm{LQ}}=0.2/R^\prime$. We can understand this result from a simple analysis of the main contributions to the potential along non-zero $c$. The color-broken Universe consists of 5 massive gluons resulting from the  SU(3)\textsubscript{c} $\times$ U(1)\textsubscript{Y}  $\rightarrow$ SU(2) $\times$ U(1) pattern of symmetry breaking. One gauge boson $Z_c$ is the equivalent of the EW $Z$ boson and the four remaining gauge bosons $W_c$ correspond to the EW $W$ bosons with the straightforward mass relations
%%%%%%%%%%%%%%%%%%%%%%%%%
\begin{align}
    m_{Z_c}=& \frac{g_s c}{2 \cos\theta_{W,c}}, \notag \\
    m_{W_c}=& \frac{g_s c}{2},
\end{align}
%%%%%%%%%%%%%%%%%%%%%%%%%
with $\cos^2\theta_{W,c}=9g_s^2/(12g_s^2+4g^{\prime,2})$ the cosine of the {\it color} Weinberg angle squared. These particles will stabilize the potential along the non-zero color direction in field space. In the color-broken Universe, the most massive fermion now consists of the right handed top which mixes with with the electron singlet forming a state $T_c$. This fermion will tend to destabilize the potential. The yukawa coupling of that particle is (due to SU(5) symmetry) equal to the usual top yukawa $y_t$ resulting in the following mass (for $v=0$)
%%%%%%%%%%%%%%%%%%%%%%%%%
\begin{equation}
    m_{T_c}= \frac{y_t c}{\sqrt{2} }.
\end{equation}
%%%%%%%%%%%%%%%%%%%%%%%%%

Using the simplified relations from \cite{Falkowski:2006vi}, which in broad lines means neglecting the contribution of the primed fields, one can easily estimate the resulting potential that these particles will induce
%%%%%%%%%%%%%%%%%%%%%%%%%
\begin{align}
    V(c)=&4\times\frac{3}{16\pi^2 R^{\prime 4}}\int_0^\infty dp p^3 \log \Big(1+(g_s f R^{\prime})^2 \frac{\sin^2(c/2f)}{\sinh^2(p)} \Big) \notag \\
    + &\frac{3}{16\pi^2 R^{\prime 4}}\int_0^\infty dp p^3 \log \Big(1+(g_s f R^{\prime}/\cos\theta_{W,c})^2 \frac{\sin^2(c/2f)}{\sinh^2(p)} \Big) \notag \\
    -&\frac{4}{16\pi^2 R^{\prime 4}}\int_0^\infty dp p^3 \log\Big(1+(y_t f R^{\prime})^2 \frac{\sin^2(c/\sqrt{2}f)}{\sinh^2(p)} \Big).
\end{align}
%%%%%%%%%%%%%%%%%%%%%%%%%
The question of color being spontaneously broken depends on the relative strength of these terms, in particular the strength of $g_s$ in comparison to $y_t$. Neglecting the contribution of the $Z_c$ boson we find the simple expression for the mass term of the scalar triplet
%%%%%%%%%%%%%%%%%%%%%%%%%
\begin{equation}
    m_{S}^2= \frac{\partial^2 V(c)}{\partial^2 c}|_{c=0}= \frac{3\zeta(3)(3g_s^2-2y_t^2)}{16\pi^2R^{\prime2}} \approx (0.2 /R^\prime)^2,
\end{equation}
%%%%%%%%%%%%%%%%%%%%%%%%%
where we neglect the subleading contribution from EWSB. For simplicity, we also neglect the possible contributions of the exotics, that is to say from fermion towers without mixing with zero modes. As we have seen for the Higgs potential these exotics are crucial to correctly fine-tune the Higgs potential and avoid the double-tuning problem. It turns out that for $M_{\tilde{u}}<0$ the contribution of such exotics can become sizable, although always stabilizing and thereby increasing the mass of the colored triplet. Therefore their inclusion does not impact the overall stability and shape of the potential. In consequence, the conservation of color in this model results from the largeness of the strong coupling in comparison to the top yukawa. The lightness of this scalar is a consequence of the Hosotani mechanism which is a one loop effect. The scalar leptoquark can be looked for at the LHC, with the current bounds being $m_{S}>1.4$ TeV \cite{ATLAS:2021oiz}, as it will decay almost exclusively into $t\tau$. This is since the Yukawa couplings of the scalar are highly hierarchical along generation space, just as the Higgs Yukawa couplings.

Similarly, one can investigate the potential along the singlet vev $s$, although here we constrain ourselves to a numerical analysis. In the non-zero singlet vev $s\neq0$ region, no LH and RH modes get connected and thus all chiral modes remain massless. This means that the contributions to the potential come from KK modes that couple to the singlet vev, which does not allow for an as straightforward analysis as for the colored triplet. Instead we have to rely on numerical methods. We find the dominant contribution coming from the KK towers of the up quark and exotic up quark. The results of our analysis are presented in the right panel of Fig. \ref{Fig:mh}. A rather broad range of masses for the singlet is observed, roughly bounded by
\begin{equation}
    m_{\textrm{singlet}}\lesssim 0.05/R^\prime.
\end{equation}
When restricting ourselves to the branch consistent with a correct Higgs mass, i.e. $M_{\tilde{u}}<0$, we find that no vev is generated for the singlet. A vev for the singlet is induced for $M_{\tilde{u}}>0$ as this can make the up-type exotic contribution destabilizing.

\section{Conclusions}\label{sec:Conclusion}
We have presented a detailed study of SU(6) GHGUT and its phenomenology. We have identified a fermion embedding and a pattern of symmetry breaking that is particularly well-suited to reproduce the observed flavor hierarchies in nature: hierarchical fermion masses, the CKM matrix and the PMNS matrix all find a natural explanation. Furthermore, we computed the leading flavor constraints on the model from meson mixing and charged lepton flavor violation  (from tree-level and loop-level observables). We find $\mu\rightarrow e \gamma$ to be currently the most constraining process pushing the IR scale above $7$ TeV, implying the first KK excitation of the gauge bosons to be at $17$ TeV, which is inaccessible at the LHC but very much in reach of future colliders\cite{Helsens:2019bfw}. The future bounds from $\mu-e$ conversion could safely probe IR scales of the model up to $12$ TeV. In spite of the non-custodial nature of the setup, the constraints from EWPT are not quite as competitive as the flavor bounds, but are less model dependent since the latter could be mitigated, e.g. via imposing flavor symmetries.

The strong bounds on the model mean we are left with a little hierarchy problem in the scalar potential for the Higgs vev. However, assuming a correctly fine-tuned Higgs vev, the resulting Higgs mass, which is a non-trivial prediction of the scalar potential, is found consistent with experimental values. The philosophy, which was also taken in \cite{Barnard:2014tla}, consists in allowing for a slightly tuned Higgs potential but succeeding in evading stringent flavor constraints without ad hoc global symmetries in the flavor sector.
This also aligns rather well with the other constraints, from direct searches and precision tests, which also point more towards scales of $1/R^\prime \gtrsim 5$\,TeV. In this context, it would be interesting to study the quality of unification in dependence on the IR scale, too.

Although the KK gauge bosons and fermions are out of reach of current colliders and a potential signature in $\mu\rightarrow e \gamma$ or other flavor observables will only be an indirect probe of the model, the smoking gun signal of the minimal SU(6)/SU(5) GUT will be the presence of the scalar singlet and leptoquark whose masses are directly proportional to the IR scale of the model $1/R^\prime$, but a considerable factor smaller. Although the leptoquark has ideal quantum numbers to explain some of the recent flavor anomalies, its Yukawas are closely aligned with the Higgs Yukawas, resulting in particular in very small off-diagonal couplings. Interestingly, the singlet could play a crucial role in a model of EW baryogenesis. A detailed analysis of the collider signatures and gauge coupling running is left for future work.  

\subsection*{Acknowledgments}
We are grateful to Simone Blasi for useful discussions.

\appendix
\section{Diagonalization of fermion kinetic and mass matrices}
\subsection{Zero Mode Approximation}\label{AppendixA}
In this appendix we provide the solutions of the IR brane model (see Section \ref{sec:IRbranemodel}) for the wave functions and mass matrices of the fermions (see Section \ref{sec:WarpedFermions} and Section \ref{sec:BC} for the solutions of warped fermions in the presence of brane masses) in the Zero Mode Approximation (ZMA) in which we do not take into account fermion-mass mixing with the KK modes. 

In the quark sector, we find the following profiles
%%%%%%%%%%%%%%%%%%%%%%%%%
\begin{align}
f_{u_R}(z)&=\frac{1}{\sqrt{R^\prime}}\Big(\frac{z}{R}\Big)^2\Big(\frac{z}{R^\prime}\Big)^{c_{20}}f_{-c_{20}},\notag \\
f_{d_R}(z) & =\frac{1}{\sqrt{R^\prime}}\Big(\frac{z}{R}\Big)^2\Big(\frac{z}{R^\prime}\Big)^{c_6}f_{-c_6},\notag \\
f_{d_R^\prime}(z) & =-\frac{1}{\sqrt{R^\prime}}\Big(\frac{z}{R}\Big)^2\Big(\frac{z}{R^\prime}\Big)^{c_{15}}M_{d/l}^\dagger f_{-c_6},
\end{align}
%%%%%%%%%%%%%%%%%%%%%%%%%
where the latter profile enters due to the IR brane mass term $M_{d/l}$ that implies the IR BC $f_{d_R^\prime}(R^\prime) = -M_{d/l}^\dagger f_{d_R}(R^\prime)$, and
%%%%%%%%%%%%%%%%%%%%%%%%%
\begin{align}
f_{q_L}(z)& =\frac{1}{\sqrt{R^\prime}}\Big(\frac{z}{R}\Big)^2\Big(\frac{z}{R^\prime}\Big)^{-c_{15}}f_{c_{15}},\notag \\
f_{q_L^\prime}(z)&=\frac{1}{\sqrt{R^\prime}}\Big(\frac{z}{R}\Big)^2\Big(\frac{z}{R^\prime}\Big)^{-c_{20}}M_{q/e} f_{c_{15}},
\end{align}
%%%%%%%%%%%%%%%%%%%%%%%%%
with the latter profile coming from the IR brane mass term $M_{q/e}$ that implies the IR BC $f_{q_L^\prime}(R^\prime) = M_{q/e} f_{q_L}(R^\prime)$.

From the primed fermions, in which a small admixture of the zero mode can live, we deduce the kinetic mixings in the quark sector
%%%%%%%%%%%%%%%%%%%%%%%%%
\begin{align}
    K_{u_R} &= 1, \notag \\
    K_{d_R} &= 1 + f_{-c_6} M_{d/l} f_{-c_{15}}^{-2}M_{d/l}^\dagger f_{-c_6}, \notag \\
    K_{q_L} &= 1 + f_{c_{15}} M_{q/e}^\dagger f_{c_{20}}^{-2} M_{q/e} f_{c_{15}}.
\end{align}
%%%%%%%%%%%%%%%%%%%%%%%%%
The overlap of the LH and RH modes with the Higgs boson, see Eq.~\eqref{eq:Higgsprofileoverlap}, results in the mass matrices in the flavor basis
%%%%%%%%%%%%%%%%%%%%%%%%%
\begin{align}
\label{UVflipped-quark-masses}
& \mathcal{M}_u = \frac{g_* v}{2\sqrt{2}} f_{c_{15}}M_{q/e}^\dagger f_{-c_{20}}, \notag \\
& \mathcal{M}_d = -\frac{g_* v}{2\sqrt{2}} f_{c_{15}}M_{d/l}^\dagger f_{-c_{6}}.
\end{align}
%%%%%%%%%%%%%%%%%%%%%%%%%

We proceed similarly in the lepton sector, where we find the following profiles for the SM zero modes and their primed partners
%%%%%%%%%%%%%%%%%%%%%%%%%
\begin{align}
f_{l_R^c}(z) & =\frac{1}{\sqrt{R^\prime}}\Big(\frac{z}{R}\Big)^2\Big(\frac{z}{R^\prime}\Big)^{c_6}f_{-c_6},\notag \\
f_{l_R^{\prime c}}(z) & =-\frac{1}{\sqrt{R^\prime}}\Big(\frac{z}{R}\Big)^2\Big(\frac{z}{R^\prime}\Big)^{c_{15}}M_{d/l}^\dagger f_{-c_6},\notag \\
f_{e_L^c}(z)& =\frac{1}{\sqrt{R^\prime}}\Big(\frac{z}{R}\Big)^2\Big(\frac{z}{R^\prime}\Big)^{-c_{15}}f_{c_{15}},\notag \\
f_{e_L^{ \prime c}}(z)&=\frac{1}{\sqrt{R^\prime}}\Big(\frac{z}{R}\Big)^2\Big(\frac{z}{R^\prime}\Big)^{-c_{20}}M_{q/e} f_{c_{15}},\notag \\
f_{\nu_L^c}(z)& =\frac{1}{\sqrt{R^\prime}}\Big(\frac{z}{R}\Big)^2\Big(\frac{z}{R^\prime}\Big)^{-c_{6}}f_{c_{6}},\notag \\
f_{\nu_L^{\prime c}}(z)&=\frac{1}{\sqrt{R^\prime}}\Big(\frac{z}{R}\Big)^2\Big(\frac{z}{R^\prime}\Big)^{-c_{1}}M_\nu f_{c_{6}},
\end{align}
%%%%%%%%%%%%%%%%%%%%%%%%%
where again the IR boundary masses $M_{d/l},M_{q/e}$ and $M_{\nu}$ dictate the profiles of the various primed fields. The lepton kinetic mixing matrices are given by
%%%%%%%%%%%%%%%%%%%%%%%%%
\begin{align}\label{kineticterms}
    K_{l_R^c} &= 1 + f_{-c_6} M_{d/l} f_{-c_{15}}^{-2}M_{d/l}^\dagger f_{-c_6}, \notag \\
    K_{e_L^c} &= 1 + f_{c_{15}} M_{q/e}^\dagger f_{c_{20}}^{-2} M_{q/e} f_{c_{15}},\notag \\
    K_{\nu_L^c} &= 1 + f_{c_6}M_\nu^\dagger f_{c_1}^{-2} M_\nu  f_{c_6},
\end{align}
%%%%%%%%%%%%%%%%%%%%%%%%%
and the mass matrices in the flavor basis are
\begin{align}\label{eq:UV-flipped-leptonic-masses}
& \mathcal{M}_{e^c} = -\frac{g_* v}{2\sqrt{2}} f_{-c_{6}}M_{d/l} f_{c_{15}}, \notag \\
& \mathcal{M}_{\nu^c} = \frac{g_* v}{2\sqrt{2}} f_{-c_{6}}f_{c_{6}}.
\end{align}

\subsection{Beyond the Zero Mode Approximation}\label{AppendixB}

While the previous appendix dealt with the treatment of the zero modes, for some processes, such as $\mu\rightarrow e \gamma$, we need to take into account the first KK modes and their mass mixings. In this appendix we discuss first the correct kinetic normalization of the first two KK modes in the lepton sector and then how to go from the canonically normalized basis to the mass basis and compute all the couplings that are needed for the $\mu\rightarrow e \gamma$ decay.

\subsubsection{Electron singlet}\label{electronsingletAppendix}

We will first examine the case of the $e^c$ and $e^{c\prime}$ states, that are respectively $(+,+)$ and $(+,-)$ modes connected on the IR boundary by the brane mass $M_{q/e}$. From Section~\ref{sec:WarpedFermions} we know the general solution of fermions profiles in warped space, in the flavor basis, is a linear combination of warped sines and warped cosines, in particular see equation \eqref{warpedcosinesin} for their explicit forms. The UV BCs allow us to restrict the solutions to a compact form, reading
%%%%%%%%%%%%%%%%%%%%%%%%%
\begin{align}
    e^c(x,z)=&\sum_n \begin{pmatrix}
-\Big(\frac{R}{z}\Big)^{c-2}\mathbf{C}(z,m_n,c)  \vec{a}_{n} e_{L,n}^c(x)\\
\Big(\frac{R}{z}\Big)^{-c-2}\mathbf{S}(z,m_n,-c) \vec{a}_{n} e_{R,n}^c(x)
\end{pmatrix}, \notag \\
    e^{c \prime}(x,z)=&\sum_n \begin{pmatrix}
-\Big(\frac{R}{z}\Big)^{c^\prime-2}\mathbf{C}(z,m_n,c^\prime)  \vec{a}_{n}^\prime e_{L,n}^{c}(x)\\
\Big(\frac{R}{z}\Big)^{-c^\prime-2}\mathbf{S}(z,m_n,-c^\prime) \vec{a}_{n}^\prime e_{R,n}^{c}(x)
\end{pmatrix}.
\end{align}
%%%%%%%%%%%%%%%%%%%%%%%%%
Here, $\mathbf{C}(z,m_n,c)$ is a diagonal matrix with $c=\textrm{diag}(c_1,c_2,c_3)$, taking into account the full generation space, and the $a_n,a_n^\prime$ are also coefficients in generation space. Moreover, $e_{L,n}^c$ and $e_{R,n}^c$ are the KK-decomposed 4D modes. We expect three charge conjugated LH zero modes $e_{L,1}^c,e_{L,2}^c$ and $e_{L,3}^c$, corresponding to the three SM singlet electrons with no corresponding RH zero modes while the rest of the KK decomposition consists of vector-like massive fermions. 

The IR brane masses will be the source of mixing between generations. Indeed the IR BCs induce the following relations between $\vec{a}_{n}$ and $\vec{a}^\prime_{n}$
%%%%%%%%%%%%%%%%%%%%%%%%%
\begin{align}\label{KKmasses}
    \Big(\frac{R}{R^\prime}\Big)^{c^\prime}\mathbf{C}(R^\prime,m_n,c^\prime)  \vec{a}_{n}^\prime  &= -M \Big(\frac{R}{R^\prime}\Big)^{c}\mathbf{C}(R^\prime,m_n,c)  \vec{a}_{n} \notag \\
    \Big(\frac{R}{R^\prime}\Big)^{-c}\mathbf{S}(R^\prime,m_n,-c) \vec{a}_{n}  &=  M^\dagger \Big(\frac{R}{R^\prime}\Big)^{-c^\prime}\mathbf{S}(R^\prime,m_n,-c^\prime) \vec{a}_{n}^\prime.
\end{align}
%%%%%%%%%%%%%%%%%%%%%%%%%
Requiring the above set of equations to have a non trivial solution $\vec{a}_{n}\neq0\neq \vec{a}^\prime_{n}$ leads to the KK mass spectrum $m_n$, with the first three solutions $m_1=m_2=m_3=0$ corresponding to zero modes. We can now solve for $\vec{a}^\prime_{n}$ as a function of $\vec{a}_{n}$ 
\begin{equation}
    \vec{a}_n^\prime=-C_{c^\prime,R^\prime,n}^{-1} M C_{c,R^\prime,n} \vec{a}_n \equiv K_{e,n} \vec{a}_n,
\end{equation}
where we switch notation and define $C_{c,z,n}\equiv\Big(\frac{R}{z}\Big)^c \mathbf{C}(z,m_n,c)$ and introduce the matrix~$K_{e,n}$.

Having found $\vec{a}_n^{(\prime)}$ and $m_n$ we still need to transform to a basis in which the kinetic terms are canonically normalized. Indeed, for the LH sector we find the kinetic terms
%%%%%%%%%%%%%%%%%%%%%%%%%
\begin{align}
    {\cal L} \supset \bar{e}^c_{L,m}(x) \gamma^\mu\partial_\mu   \Big[\vec{a}^\dagger_m\int \textrm{d}z  (C_{c,z,m}^\dagger C_{c,z,n} +K_{e,m}^\dagger C_{c^\prime,z,m}^\dagger C_{c^\prime,z,n} K_{e,n}) \vec{a}_n \Big] e^c_{L,n}(x),
\end{align}
%%%%%%%%%%%%%%%%%%%%%%%%%
where the expression in brackets is a hermitian matrix over all KK modes with indices $m,n$ and one can therefore define its inverse square root, namely ${(V_L^e)}_{nm}$. Transforming the LH fields by this matrix, $e_{L,n}^c\rightarrow {(V_L^e)}_{nm}e_{L,m}^c$, the kinetic terms are canonically normalized. One has to perform a similar set of transformations in the RH sector with the corresponding matrix ${(V_R^e)}_{nm}$. 

Analogous transformations have to be done on the mass terms 
\begin{align}
    {\cal L} \supset \bar{e}^c_{L,m}(x) \underbrace{\vec{a}^\dagger_m\Big[\int \textrm{d}z  (C_{c,z,m}^\dagger C_{c,z,n} +K_{e,m}^\dagger C_{c^\prime,z,m}^\dagger C_{c^\prime,z,n} K_{e,n}) m_n \vec{a}_n\Big]}_{(M_{E_{KK}})_{mn}}  e^c_{R,n}(x),
\end{align}
where we define the expression in brackets as the matrix $M_{E_{KK}}$.

In summary, we define the LH and RH fermion vectors that combine the different KK modes
%%%%%%%%%%%%%%%%%%%%%%%%%
\begin{align}
    \Psi^c_{e,L}=&(e^c_{L,1},e^c_{L,2},e^c_{L,3},e^c_{L,4},e^c_{L,5},e^c_{L,6},..)\notag \\
    \Psi^c_{e,R}=&(e^c_{R,4},e^c_{R,5},e^c_{R,6},..),
\end{align}
%%%%%%%%%%%%%%%%%%%%%%%%%
where the ellipses stand for neglected higher modes. We only include the lightest set of KK modes which consists of 3 LH and 3 RH fields as can be seen above. These will eventually mix with the doublet fields resulting in a total of 6 vectorlike KK modes. The discussed transformations that canonically normalize the kinetic terms then act as
%%%%%%%%%%%%%%%%%%%%%%%%%
\begin{align}
    \Psi^c_{e,L}\rightarrow V_L^e \Psi^c_{e,L}\notag \\
    \Psi^c_{e,R}\rightarrow V_R^e \Psi^c_{e,R}\,,
\end{align}
%%%%%%%%%%%%%%%%%%%%%%%%%
and transform the KK masses as follows
%%%%%%%%%%%%%%%%%%%%%%%%%
\begin{equation}\label{EKKmass}
    {\cal L} \supset \bar{\Psi}^c_{e,L}M_{E_{\textrm{KK}}}\Psi^c_{e,R}+\text{h.c}\rightarrow \bar{\Psi}^c_{e,L}V_L^{e \dagger} M_{E_{\textrm{KK}}} V_R^e \Psi^c_{e,R}+\text{h.c.}\,.
\end{equation}
\subsubsection{Lepton doublet}

We now perform a similar analysis for the charge conjugated doublets $l^c$ and $l^{c\prime}$ with BCs $(-,-)$ and $(-,+)$, connected on the IR boundary by $M_{d/e}$. The UV BCs now lead to
%%%%%%%%%%%%%%%%%%%%%%%%%
\begin{align}
    l^c(x,z) = \sum_n \begin{pmatrix}
    \Big(\frac{R}{z}\Big)^{c-2} \mathbf{S}(z,m_n,c) \vec{b}_{n} l^c_{L,n}(x)\\
    \Big(\frac{R}{z}\Big)^{-c-2} \mathbf{C}(z,m_n,-c) \vec{b}_{n} l^c_{R,n}(x)
    \end{pmatrix},\notag \\
    l^{c\prime}(x,z) = \sum_n \begin{pmatrix}
    \Big(\frac{R}{z}\Big)^{c^\prime-2} \mathbf{S}(z,m_n,c^\prime) \vec{b}_{n}^\prime l^{c}_{L,n}(x)\\
    \Big(\frac{R}{z}\Big)^{-c^\prime-2} \mathbf{C}(z,m_n,-c^\prime) \vec{b}_{n}^\prime l^c_{R,n}(x)
    \end{pmatrix}\,,
\end{align}
%%%%%%%%%%%%%%%%%%%%%%%%%
while the IR BCs induce the conditions
%%%%%%%%%%%%%%%%%%%%%%%%%
\begin{align}
    \Big(\frac{R}{R^\prime}\Big)^{c} \mathbf{S}(R^\prime,m_n,c)\vec{b}_{n} &= -M_{d/e}\Big(\frac{R}{R^\prime}\Big)^{c^\prime}\mathbf{S}(R^\prime,m_n,c^\prime) \vec{b}_{n}^\prime \notag \\
    \Big(\frac{R}{R^\prime}\Big)^{-c^\prime} \mathbf{C}(R^\prime,m_n,-c^\prime)\vec{b}_{n}^\dagger &= M_{d/e}^\dagger\Big(\frac{R}{R^\prime}\Big)^{-c}\mathbf{C}(R^\prime,m_n,-c) \vec{b}_{n}.
\end{align}
%%%%%%%%%%%%%%%%%%%%%%%%%
Again, requiring non-trivial solutions for $\vec{b}_{n}$ and $\vec{b}_{n}^\prime$ will determine the KK masses $m_n$ (before EWSB). Solving for $\vec{b}_{n}^\prime$ leads to
%%%%%%%%%%%%%%%%%%%%%%%%%
\begin{equation}
    \vec{b}^\prime_n = C^{-1}_{-c^\prime,R^\prime,n} M^\dagger_{d/e} C_{-c,R^\prime,n}\vec{b}_n\equiv K_{l,n}\vec{b}_n.
\end{equation}
%%%%%%%%%%%%%%%%%%%%%%%%%
Similar to the case for the electron singlet, the kinetic terms are non-canonical and can be brought to a canonical basis via unitary rotations $V_{R,L}^l$ that act on the fermion fields 
%%%%%%%%%%%%%%%%%%%%%%%%%
\begin{align}
    \Psi^c_{l,R}=&(l^c_{R,1},l^c_{R,2},l^c_{R,3},l^c_{R,4},l^c_{R,5},l^c_{R,6},..)\notag \\
    \Psi^c_{l,L}=&(l^c_{L,4},l^c_{L,5},l^c_{L,6},..)
\end{align}
%%%%%%%%%%%%%%%%%%%%%%%%%
and on the KK mass matrix ${M_{L_{\textrm{KK}}}}$.

\subsubsection{Neutrino singlet}

The final case to look at is the singlet conjugate neutrinos (which contain the RH neutrino) in the $\bf{6}$ and in the $\bf{1}$ of signature $(+,+)$ and $(+,-)$, connected on the IR brane through a brane mass $M_\nu$. The UV BCs lead to the decompositions
%%%%%%%%%%%%%%%%%%%%%%%%%
\begin{align}
    \nu^c(x,z)=&\sum_n \begin{pmatrix}
-\Big(\frac{R}{z}\Big)^{c-2}\mathbf{C}(z,m_n,c)  \vec{c}_{n} \nu_{L,n}^c(x)\\
\Big(\frac{R}{z}\Big)^{-c-2}\mathbf{S}(z,m_n,-c) \vec{c}_{n} \nu_{R,n}^c(x)
\end{pmatrix},\notag \\
    \nu^{c \prime}(x,z)=&\sum_n \begin{pmatrix}
-\Big(\frac{R}{z}\Big)^{c^\prime-2}\mathbf{C}(z,m_n,c^\prime)  \vec{c}_{n}^\prime \nu_{L,n}^{c}(x)\\
\Big(\frac{R}{z}\Big)^{-c^\prime-2}\mathbf{S}(z,m_n,-c^\prime) \vec{c}_{n}^\prime \nu_{R,n}^{c}(x)
\end{pmatrix}\,,
\end{align}
%%%%%%%%%%%%%%%%%%%%%%%%%
and as usual the IR BCs determine the KK masses, reading
%%%%%%%%%%%%%%%%%%%%%%%%%
\begin{align}
    \Big(\frac{R}{R^\prime}\Big)^{c^\prime}\mathbf{C}(R^\prime,m_n,c^\prime)  \vec{c}_{n}^\prime  &= -M_\nu \Big(\frac{R}{R^\prime}\Big)^{c}\mathbf{C}(R^\prime,m_n,c)  \vec{c}_{n}, \notag \\
    \Big(\frac{R}{R^\prime}\Big)^{-c}\mathbf{S}(R^\prime,m_n,-c) \vec{c}_{n}  &=  M_\nu^\dagger \Big(\frac{R}{R^\prime}\Big)^{-c^\prime}\mathbf{S}(R^\prime,m_n,-c^\prime) \vec{c}_{n}^\prime.
\end{align}
%%%%%%%%%%%%%%%%%%%%%%%%%
Having obtained the masses, we can now solve the system by eliminating $\vec{c}^\prime$ via
%%%%%%%%%%%%%%%%%%%%%%%%%
\begin{equation}
    \vec{c}_n^\prime=-C_{c^\prime,R^\prime,n}^{-1} M_\nu C_{c,R^\prime,n} \vec{c}_n \equiv K_{\nu,n} \vec{c}_n.
\end{equation}
%%%%%%%%%%%%%%%%%%%%%%%%%
Again, the kinetic terms are not yet canonically normalized, which will be achieved by the combinations of transformation matrices $V_L^\nu$ and $V_R^\nu$, acting on the LH and RH vectors
%%%%%%%%%%%%%%%%%%%%%%%%%
\begin{align}
    \Psi^c_{\nu,L}=&(\nu^c_{L,1},\nu^c_{L,2},\nu^c_{L,3},\nu^c_{L,4},\nu^c_{L,5},\nu^c_{L,6},..)\notag \\
    \Psi^c_{\nu,R}=&(\nu^c_{R,4},\nu^c_{R,5},\nu^c_{R,6},..)
\end{align}
%%%%%%%%%%%%%%%%%%%%%%%%%
and and on the KK mass matrix ${M_{\nu_{\textrm{KK}}}}$.
\subsubsection{Mass diagonalization}\label{sec:AppendixCMass}
Having studied the transformation from the flavor basis to the kinetic basis, induced by brane mixing, the final step is going to the mass basis after EWSB. This will mix the charge conjugate electron singlet with the upper SU(2) component of the charge conjugate lepton doublet while the neutrino singlet will mix with the lower SU(2) component. Therefore in the mass basis we will work with the field vectors 
%%%%%%%%%%%%%%%%%%%%%%%%%
\begin{align}
    \Psi_{e,L}^{c,m}&= (\Psi^c_{e,L}, \Psi^c_{l_e,L})= (e^c_{L,1},e^c_{L,2},...,e^c_{L,6},l^c_{e,L,4},l^c_{e,L,5},l^c_{e,L,6},)\notag \\
    \Psi_{e,R}^{c,m}&= (\Psi^c_{l_e,R},\Psi^c_{e,R})= (l^c_{e,R,1},l^c_{e,R,2},..,l^c_{e,R,6},e^c_{R,4},e^c_{R,5},e^c_{R,6})\notag \\
    \Psi_{\nu,R}^{c,m}&= (\Psi^c_{l_\nu,L},\Psi^c_{\nu,R})=(l^c_{\nu,R,1},l^c_{\nu,R,2},...,l^c_{\nu,R,6},\nu^c_{R,4},\nu^c_{R,5},\nu^c_{R,6})\notag \\
    \Psi_{\nu,L}^{c,m}&= (\Psi^c_{\nu,L},\Psi^c_{l_\nu,L})= (\nu^c_{L,1},\nu^c_{L,2},...,\nu^c_{L,6},l^c_{\nu,L,4},l^c_{\nu,L,5},l^c_{\nu,L,6}).
\end{align}
%%%%%%%%%%%%%%%%%%%%%%%%%
Starting from these four, properly normalized, fields we include the effects of EWSB and go to the mass basis. Not only will the zero modes get their SM masses, but the KK masses will also be shifted. Furthermore, this rotation will induce off-diagonal Higgs and gauge interactions between the SM zero modes and the KK modes, that in turn will induce FCNC processes, such as $\mu \rightarrow e \gamma$. 

We start with the 5D Yukawa couplings in the electron sector that originate from the bulk~$\bf{15}$, reading
%%%%%%%%%%%%%%%%%%%%%%%%%
\begin{align}
    \mathcal{L}\supset & -\frac{g_5}{\sqrt{2}}\Big( \bar{e}^c_L H l^{c\prime}_R + \bar{l}^{c\prime}_L H e^c_R + \textrm{h.c.} \Big),
\end{align}
%%%%%%%%%%%%%%%%%%%%%%%%%
with $H(x,z)=f_5(z)(v+h(x) )/\sqrt{2}$ the $A_5$ Higgs. Upon substitution of the $5D$ fields with their KK decomposition, these terms will lead to a tower of interactions including mass terms. We restrict our mass diagonalization to the first KK level. One can see that the first term contains the correct chirality Higgs (H) couplings between the zero modes $e^c_{L,1,2,3}$ and $l^c_{e,R,1,2,3}$ which will give rise to the usual charged lepton Yukawa while the second term is the so called opposite chirality Higgs ($\tilde{\textrm{H}}$) couplings that connect the LH electron singlets to the RH electron doublets. Notice the latter coupling is absent in the SM and is a coupling strictly between KK modes. 

At the level of the first KK mode we thus write down the following mass terms in the flavor basis 
%%%%%%%%%%%%%%%%%%%%%%%%%
\begin{equation}\label{massflavorbasis}
{\cal L \supset} -\bar{\Psi}_{e,L}^{c,m} M_e \Psi_{e,R}^{c,m} =- \bar{\Psi}_{e,L}^{c,m} \begin{pmatrix}
\frac{v}{\sqrt{2}}\textrm{H} & M_{E_{\textrm{KK}}} \\
M_{L_{\textrm{KK}}} & \frac{v}{\sqrt{2}}\tilde{\textrm{H}}
\end{pmatrix} \Psi_{e,R}^{c,m},
\end{equation}
%%%%%%%%%%%%%%%%%%%%%%%%%
where the diagonal entries represent the mass contributions from EWSB while the off-diagonal entries represent the vector-like KK masses \eqref{EKKmass}. The Higgs contributions from EWSB in the flavor basis are given by the following overlap functions
%%%%%%%%%%%%%%%%%%%%%%%%%
\begin{align}
    \bar{e}^c_{L,n}\textrm{H}_{n,m} l^c_{e,R,m}=&\bar{e}^c_{L,n}\frac{g_5}{\sqrt{2}} \int_R^{R^\prime} \textrm{d}z\Big[-\vec{a}_n^\dagger C^\dagger_{z,c,n} C_{z,-c^\prime,m} \vec{b}_m^\prime \Big] \Big( \sqrt{\frac{2}{R}} \frac{z}{R^\prime} \Big)l^c_{e,R,m} \notag \\
    \bar{l}^c_{e,L,n} \tilde{\textrm{H}}_{n,m} e^c_{R,m} = & \bar{l}^c_{e,L,n} \frac{g_5}{\sqrt{2}} \int_R^{R^\prime} \textrm{d}z\Big[\vec{b}_n^{\dagger,\prime} S^\dagger_{z,c^\prime,n} S_{z,-c,m} \vec{a}_m \Big] \Big( \sqrt{\frac{2}{R}} \frac{z}{R^\prime} \Big) e^c_{R,m},
\end{align}
%%%%%%%%%%%%%%%%%%%%%%%%%
where the Higgs bulk profile is given by the function in parenthesis and note that H is a $6\times6$ matrix while $\tilde{\textrm{H}}$ is a $3\times 3$ matrix. Before diagonalizing this mass matrix and extracting the mass eigenstates, one must not forget go to the kinetic basis as discussed in Appendix \ref{electronsingletAppendix}. Indeed we still need to remove any kinetic mixing between the modes by the transformations
%%%%%%%%%%%%%%%%%%%%%%%%%
\begin{align}\label{yukawarot}
- \bar{\Psi}_{e,L}^{c,m} \begin{pmatrix}
\frac{v}{\sqrt{2}}\textrm{H} & M_{E_{\textrm{KK}}}\\
M_{L_{\textrm{KK}}} & \frac{v}{\sqrt{2}}\tilde{\textrm{H}}
\end{pmatrix} \Psi_{e,R}^{c,m} \rightarrow - \bar{\Psi}_{e,L}^{c,m} \begin{pmatrix}
V_L^e & 0\\
0 &  V_L^l
\end{pmatrix}^\dagger\begin{pmatrix}
\frac{v}{\sqrt{2}}\textrm{H} & M_{E_{\textrm{KK}}}\\
M_{L_{\textrm{KK}}} & \frac{v}{\sqrt{2}}\tilde{\textrm{H}}
\end{pmatrix} \begin{pmatrix}
 V_R^l & 0\\
0 & V_R^e
\end{pmatrix}\Psi_{e,R}^{c,m}\,.
\end{align}
The resulting mass matrix can be diagonalized by a bi-unitary transformation to obtain the mass basis via
%%%%%%%%%%%%%%%%%%%%%%%%%
\begin{align}\label{massrot}
    \Psi_{e,L}^{c,m} \rightarrow U_{R,e^c} \Psi_{e,L}^{c,m} \notag \\
    \Psi_{e,R}^{c,m} \rightarrow U_{L,e^c}  \Psi_{e,R}^{c,m}.
\end{align}
%%%%%%%%%%%%%%%%%%%%%%%%%
We can perform an analogous diagonalization for the neutrino sector, whose 5D Yukawas come from the bulk $\bf{6}$
%%%%%%%%%%%%%%%%%%%%%%%%%
\begin{align}
    \mathcal{L}\supset -\frac{g_5}{\sqrt{2}}\Big( \bar{l}^c_R H \nu^{c}_L + \bar{\nu}^{c}_R H l^c_L + \textrm{h.c.} \Big),
\end{align}
%%%%%%%%%%%%%%%%%%%%%%%%%
which results in similar rotations to get to the mass basis
%%%%%%%%%%%%%%%%%%%%%%%%%
\begin{align}
    \Psi_{\nu,L}^{c,m} \rightarrow U_{R,\nu^c} \Psi_{\nu,L}^{c,m} \notag \\
    \Psi_{\nu,R}^{c,m} \rightarrow U_{L,\nu^c}  \Psi_{\nu,R}^{c,m}\,.
\end{align}

\section{Fermion couplings and \texorpdfstring{$\mu\rightarrow e \gamma$}{mueg}}\label{AppendixC}

\subsection{Higgs couplings}

Armed with all the rotations needed to go to the mass basis, one can now find the couplings that will enter our computations for $\mu\rightarrow e \gamma$. First of all, the physical Higgs couplings $\Delta^h$ are obtained by performing the same series of rotations on the Yukawa matrix as in \eqref{yukawarot}, with no contributions from the KK masses, followed by the mass rotations from \eqref{massrot}, leading to
%%%%%%%%%%%%%%%%%%%%%%%%%
\begin{align} {\cal L} \supset &\ h\bar{\Psi}_{e,L}^{c,m} \Delta^h \Psi_{e,R}^{c,m} +\textrm{h.c.}= \notag \\
& \frac{h}{\sqrt{2}}\bar{\Psi}_{e,L}^{c,m} U_{R,e^c}^{\dagger}\begin{pmatrix}
V_L^e & 0\\
0 &   V_L^l
\end{pmatrix}^\dagger\begin{pmatrix}
\textrm{H} & 0\\
0 & \tilde{\textrm{H}}
\end{pmatrix} \begin{pmatrix}
V_R^l & 0\\
0 & V_R^e
\end{pmatrix}U_{L,e^c}\Psi_{e,R}^{c,m}+\textrm{h.c.}.
\end{align}
%%%%%%%%%%%%%%%%%%%%%%%%%%%
Therefore, the origin of the off-diagonal Higgs couplings lies in the the mass matrix being not aligned with the Yukawa couplings, which only contain EWSB mass effects and not the KK vector-like masses.

With the Yukawa couplings above one can calculate Higgs loop contribution to the $\mu\rightarrow e\gamma$ decay, which was already discussed previously in \cite{Agashe:2006iy}. The corresponding amplitude, with $q=p^\prime-p$ the incoming photon momentum and containing a sum over the internal leptons $m_{l,i}$ from which a photon is emitted, reads
%%%%%%%%%%%%%%%%%%%%%%%%%
\begin{align}
\mathcal{M_\mu}&=\int \frac{\textrm{d}^4k}{(2\pi)^4}\Big[ \bar{u}_{p^\prime} (-i)(\Delta_{ei}^hP_L+\Delta_{ei}^{h\dagger} P_R) \frac{i}{(\slashed{p}^{\prime}+\slashed{k}-m_{l,i})}\\\notag&(-i e \gamma_\mu) \frac{i}{(\slashed{p}+\slashed{k}-m_{l,i})} (-i) (\Delta_{i\mu}^hP_L+\Delta^{h\dagger}_{i\mu}P_R)  u_p\Big]\frac{i}{k^2-m_h^2}.
\end{align}
%%%%%%%%%%%%%%%%%%%%%%%%%
The contributions from this amplitude to the dipole coefficients $C_R$ and $C_L$ are
%%%%%%%%%%%%%%%%%%%%%%%%%
\begin{align}
    C_{R}(q^2=0) &= -\frac{e \Delta_{ei}^{h\dagger} m_\mu}{192 \pi^2 m_h^2(z_{i}-1)^4}\Big(6\Delta_{i\mu}^{h\dagger} m_h(-1+z_{i})\sqrt{z_{i}}(3-4z_{i}+z_{i}^2+2\log z_{i}) \notag \\ &+\Delta_{i\mu}^h m_\mu(2+3z_{i}-6z_{i}^2+z_{i}^3+6z_{i}\log z_{i})\Big) \notag \\
    C_{L}(q^2=0) &=- \frac{e \Delta_{ei}^h m_\mu}{192 \pi^2 m_h^2(z_{i}-1)^4}\Big(6\Delta_{i\mu}^h m_h(-1+z_{i})\sqrt{z_{i}}(3-4z_{i}+z_{i}^2+2\log z_{i}))\notag \\&+\Delta_{i\mu}^{h\dagger} m_\mu(2+3z_{i}-6z_{i}^2+z_{i}^3+6z_{i}\log z_{i}\Big),
\end{align}
%%%%%%%%%%%%%%%%%%%%%%%%%
with $z_{i}=(m_{l,i}/m_h)^2$ and where we sum over the SM leptons $i=1/e,2/\mu,3/\tau$ and first six KK modes $i=4,5,...,9$. These coefficients enter the decay for $\mu\rightarrow e \gamma $ through equation \eqref{eq:BRs}.

We can clearly distinguish two categories of contributions: the first term in both $C_R$ and $C_L$ corresponds to the diagram where the necessary chirality flip occurs in the internal line and is thus proportional to the mass of the internal fermion while in the second term the chirality flip occurs on the external muon leg and thus carries an extra factor of $m_\mu$. For a heavy KK internal lepton the first term is therefore enhanced by a factor $(m_{l,i}/m_\mu)$ with respect to the second term and the amplitude goes as $m_\mu/m_{l,i}$. For a light SM internal lepton the second term can clearly become dominant with the amplitude going as $m^2_\mu/m_h^2$. Thus, overall the decay is unsurprisingly dominated by heavy internal KK leptons.

\subsection{Z-boson couplings}

Let us now investigate the $Z$ boson couplings to the charged leptons for which the following 5D couplings are relevant
\begin{align}
    \mathcal{L}\supset &\frac{g}{\cos \theta_W} \Big( \bar{e}^c\gamma_\mu e^{c} + \bar{e}^{c\prime}\gamma_\mu e^{c\prime}  + \bar{l}^c \gamma_\mu l^{c}  + \bar{l}^{c\prime} \gamma_\mu l^{c\prime}
     \Big)Z_i^\mu.
\end{align}
As an example let us first extract the $Z$-couplings to the RH singlet electrons $\Delta^{Z^i}_R$ with the LH coupling $\Delta^{Z^i}_L$ fully analogous. The couplings in the mass basis, corrected by the various kinetic and mass transformations, are given by
%%%%%%%%%%%%%%%%%%%%%%%%%
\begin{align}
    {\cal L} \supset &\frac{g}{\cos \theta_W}\bar{\Psi}_{e,L}^{c,m} \gamma_\mu \Delta^{Z^i}_R \Psi_{e,L}^{c,m}Z^\mu_i= \notag \\
    &\frac{g}{\cos \theta_W}\bar{\Psi}_{e,L}^{c,m}
    U_R^{\dagger e} \begin{pmatrix}
V_L^e & 0\\
0 &  V_L^l
\end{pmatrix}^\dagger
    \gamma_\mu
    \begin{pmatrix}
    Z_{R}^{i} & 0 \\
    0 & \tilde{Z}_{R}^{i} \\
    \end{pmatrix}
     \begin{pmatrix}
 V_L^e & 0\\
0 &  V_L^l
\end{pmatrix} U_R^e
    \Psi_{e,L}^{c,m} Z^\mu_i,
\end{align}
%%%%%%%%%%%%%%%%%%%%%%%%%
where the index $i$ sums over the $Z$ boson KK modes and the $\mathord{\mathop{Z_R}\limits^{\scriptscriptstyle(\sim)}}$ matrices read
%%%%%%%%%%%%%%%%%%%%%%%%%
\begin{align}
    Z_{nm,R}^i=&\sin^2\theta_W \int_R^{R^{\prime}}\textrm{d}z f_{Z^{i}}(z)\Big[\vec{a}_n^\dagger C_{z,c,n}^\dagger C_{z,c,m} \vec{a}_m +\vec{a}_n^{\dagger} K_{e,n}^\dagger C_{z,c^\prime,n}^\dagger C_{z,c^\prime,m} K_{e,m}\vec{a}_m   \Big] \notag \\
    \tilde{Z}_{nm,R}^{i}=&(-\frac{1}{2}+\sin^2\theta_W)\int_R^{R^{\prime}}\textrm{d}z f_{Z^{i}}(z)\Big[\vec{b}_n^\dagger S_{z,c,n}^\dagger S_{z,c,m} \vec{b}_m +\vec{b}_n^\dagger K_{l,n}^\dagger S_{z,c^\prime,n}^\dagger S_{z,c^\prime,m} K_{l,m} \vec{b}_m  \Big]\,,
\end{align}
%%%%%%%%%%%%%%%%%%%%%%%%%
with the profile function for the SM $Z$ boson given in \eqref{eq:massbasisgauge}.
Notice that the coupling $\Delta^{Z^i}_R$ is hermitian as it should and we can similarly obtain the couplings to the LH electrons $\Delta^{Z^i}_L$. 

With the coupling $\Delta^{Z^i}_R$, $\Delta^{Z^i}_L$ in the mass basis computed, we can now compute the Z boson contribution to the $\mu\rightarrow e \gamma$ amplitude, where the photon is emitted from the internal lepton, leading to
%%%%%%%%%%%%%%%%%%%%%%%%%
\begin{align}
\mathcal{M_\mu}= (i\frac{g}{c_W})^2\int \frac{\textrm{d}^4k}{(2\pi)^4}\Big[\bar{u}_{p^\prime} \gamma_\alpha & (\Delta^{Z}_{L,e i}P_L+\Delta^{Z}_{R,e i} P_R) \frac{i}{(\slashed{p}^{\prime}+\slashed{k}-m_{l,i})}(-i e \gamma_\mu)\times \notag \\
& \frac{i}{(\slashed{p}+\slashed{k}-m_{l,i})}\gamma_\beta (\Delta^{Z}_{L,i\mu}P_L+\Delta^{Z}_{R,i\mu} P_R) u_p\Big] [i\Delta^{\alpha\beta}_Z(k)],
\end{align}
%%%%%%%%%%%%%%%%%%%%%%%%%
where we omit the index indicating the KK mode of the Z boson and with
\begin{align}
    \Delta^{\alpha\beta}_Z(k) =&\frac{ -g^{\nu\beta}+\frac{k^\nu k^\beta}{m_Z^2}}{k^2-m_Z^2},
\end{align}
the $Z$ boson propagator in unitary gauge.
The contributions from this amplitude to the dipole coefficients $C_R$ and $C_L$ become
%%%%%%%%%%%%%%%%%%%%%%%%%
\begin{align}
    C_{R}(q^2=0) &= - \frac{e g^2 \Delta^{Z}_{L,e i}m_\mu}{192 \pi^2\cos^2\theta_W^2m_Z  (z_i-1)^4}\Big(6\Delta^{Z}_{R,i\mu}m_Z(-1+z_i)\sqrt{z_i}(4-3z_i-z_i^3+6z_i\log(z_i)) \notag \\ &+\Delta^{Z}_{L,i\mu}m_\mu(8-38z_i+39z_i^2-14z_i^3+5z_i^4-18z_i^2\log(z_i)\Big) \notag \\
    C_{L}(q^2=0) &=- \frac{e g^2 \Delta^{Z}_{R,e i}m_\mu}{192 \pi^2\cos^2\theta_W^2m_Z  (z_{i}-1)^4}\Big(6\Delta^{Z}_{L,i\mu}m_Z(-1+z_i)\sqrt{z_i}(4-3z_i-z_i^3+6z_i\log(z_i)) \notag \\ &+\Delta^{Z}_{R,i\mu}m_\mu(8-38z_i+39z_i^2-14z_i^3+5z_i^4-18z_i^2\log(z_i)\Big), % \notag \\
\end{align}
%%%%%%%%%%%%%%%%%%%%%%%%%
with $z_{i}=(m_{l_i}/m_Z)^2$. Similarly as for the Higgs loop we identify two types of contributions, one with the chirality flip on the internal fermion that goes as $1/m_{l,i}$, while for the second contribution the chirality flip is on the external muon leg and it scales as $m_\mu/m_Z^2$.

\subsection{W-boson couplings}
The $W$ couplings can be extracted from the 5D couplings involving SU(2) doublets in the flavor basis
%%%%%%%%%%%%%%%%%%%%%%%%%
\begin{equation}
    \mathcal{L}\supset \frac{g}{\sqrt{2}} \Big( \bar{l}^c \gamma^\mu l^{c}  + \bar{l}^{c\prime} \gamma^\mu l^{c\prime}\Big)W^i_\mu.
\end{equation}
%%%%%%%%%%%%%%%%%%%%%%%%%
In the mass basis, the corresponding couplings $\Delta^{W^i}_L$ and $\Delta^{W^i}_R$ are again obtained after subsequent rotations
%%%%%%%%%%%%%%%%%%%%%%%%%
\begin{align}\label{Wcouplings}
   {\cal L} \supset &\frac{g}{\sqrt{2}}\bar{\Psi}_{e,R}^{c,m} \gamma^\mu \Delta^{W^i}_L \Psi_{\nu,R}^{c,m}W_\mu^i+\frac{g}{\sqrt{2}}\bar{\Psi}_{e,L}^{c,m} \gamma^\mu \Delta^{W^i}_R \Psi_{\nu,L}^{c,m}W_\mu^i+\textrm{h.c.} \notag \\
    =&\frac{g}{\sqrt{2}}\bar{\Psi}_{e,R}^{c,m}
    U_L^{\dagger e} \begin{pmatrix}
 V_R^l & 0\\
0 &  V_R^e
\end{pmatrix}^\dagger
    \gamma^\mu
    \begin{pmatrix}
    W^{i} & 0 \\
     0 & 0 \\
    \end{pmatrix}
     \begin{pmatrix}
 V_R^l & 0\\
0 &  V_R^\nu
\end{pmatrix} U_L^\nu
    \Psi_{\nu,R}^{c,m} W_\mu^i+\textrm{h.c.}\notag \\
    +&\frac{g}{\sqrt{2}}\bar{\Psi}_{e,L}^{c,m}
    U_R^{\dagger e} \begin{pmatrix}
V_L^e & 0\\
0 &   V_L^l
\end{pmatrix}^\dagger
    \gamma^\mu
    \begin{pmatrix}
    0 & 0 \\
     0 & \tilde{W}^{i} \\
    \end{pmatrix}
     \begin{pmatrix}
 V_L^\nu & 0\\
0 &   V_L^l
\end{pmatrix} U_R^\nu
    \Psi_{\nu,L}^{c,m} W_\mu^i+\textrm{h.c.},
\end{align}
%%%%%%%%%%%%%%%%%%%%%%%%%
where the index $i$ sums over the different $W$ boson KK modes and the overlap functions $\mathord{\mathop{W}\limits^{\scriptscriptstyle(\sim)}}$  read
%%%%%%%%%%%%%%%%%%%%%%%%%
\begin{align}
    W_{nm}^i=&\int_R^{R^{\prime}}\textrm{d}z f_{W^{i}}(z)\Big[\vec{b}_n^\dagger C_{z,c,n}^\dagger C_{z,c,m} \vec{b}_m +\vec{b}_n^{\dagger} K_{l,n}^\dagger C_{z,c^\prime,n}^\dagger C_{z,c^\prime,m} K_{l,m}\vec{b}_m   \Big] \notag \\
    \tilde{W}_{nm}^i=&\int_R^{R^{\prime}}\textrm{d}z f_{W^{i}}(z)\Big[\vec{b}_n^\dagger S_{z,c,n}^\dagger S_{z,c,m} \vec{b}_m +\vec{b}_n^{\dagger} K_{l,n}^\dagger S_{z,c^\prime,n}^\dagger S_{z,c^\prime,m} K_{l,m}\vec{b}_m   \Big].
\end{align}
%%%%%%%%%%%%%%%%%%%%%%%%%
From these couplings, $\Delta^{W^i}_L$ and $\Delta^{W^i}_R$, one can determine the $W$ loop contribution to the photon vertex. In the SM, this is the only mediator to the $\mu\rightarrow e \gamma$ decay and it is very small due to the unitarity of the PMNS matrix and the smallness of the neutrino masses. In SU(6) GHGUT, both of these effects are violated due to the presence of heavy KK neutrinos. The one-loop decay amplitude reads
%%%%%%%%%%%%%%%%%%%%%%%%%
\begin{align}
\mathcal{M_\delta}=&(i\frac{g}{\sqrt{2}})^2 \int \frac{\textrm{d}^4k}{(2\pi)^4}\Big[\bar{u}_{p^\prime} \gamma_\mu (\Delta^{W}_{L,e i} P_L+\Delta^{W}_{R,e i} P_R) \frac{i}{(\slashed{p}+\slashed{k}-m_i)} \gamma_\nu (\Delta^{W \dagger}_{L,i\mu} P_L +\Delta^{W \dagger}_{R,i\mu} P_R )u_p\Big] \times \notag \\
&[i\Delta^{\nu\beta}_W(k)][i\Delta^{\mu\alpha}_W(k-q)](-ie)\Gamma_{\alpha\beta\delta},
\end{align}
%%%%%%%%%%%%%%%%%%%%%%%%%
where we sum over the SM neutrinos and KK neutrinos with mass $m_i$ and omit the index indicating the KK mode of the $W$ boson. Moreover, $\Delta^{\nu\beta}_W(k)$ and $\Gamma_{\alpha\beta\gamma}$ are the gauge boson propagator in unitary gauge and the $W^{+}W^-\gamma$-vertex respectively:
%%%%%%%%%%%%%%%%%%%%%%%%%
\begin{align}
    \Delta^{\nu\beta}_W(k) =&\frac{ -g^{\nu\beta}+\frac{k^\nu k^\beta}{m_W^2}}{k^2-m_W^2}, \notag \\
    \Gamma_{\alpha\beta\gamma} = &(2k-q)_\gamma g_{\alpha\beta} + (-k-q)_\alpha g_{\beta \gamma} + (2q-k)_\beta g_{\gamma\alpha}.
\end{align}
%%%%%%%%%%%%%%%%%%%%%%%%%
In the SM the amplitude above is usually simplified by expanding the neutrino propagator for small neutrino masses, becoming
%%%%%%%%%%%%%%%%%%%%%%%%%
\begin{equation}
    \frac{\Delta^{W}_{L,e i} \Delta^{W\dagger}_{L,i \mu}}{(p+k)^2-m_i^2} = \frac{\Delta^{W}_{L,e i} \Delta^{W\dagger}_{L,i \mu}}{(p+k)^2} + \frac{\Delta^{W}_{L,e i} \Delta^{W\dagger}_{L,i \mu} m_i^2}{(p+k)^4}+... \,,
\end{equation}
%%%%%%%%%%%%%%%%%%%%%%%%%
where, due to the absence of KK modes, $i$ sums over the three flavors of neutrinos and $\Delta_L^{W}$ is unitary, which makes the first term vanish. The second term in the expansion then gives the leading contribution 
%%%%%%%%%%%%%%%%%%%%%%%%%
\begin{align}\label{eq:SMW}
    C_{1,R}(q^2=0) &= \sum_i \frac{e g^2 \Delta^{W}_{L,e i} \Delta^{W\dagger}_{L,i \mu} m_i^2 m_\mu^2}{128 m_W^4 \pi^2} \notag \\
    C_{1,L}(q^2=0) &= 0,
\end{align}
%%%%%%%%%%%%%%%%%%%%%%%%%
which due to the small neutrino masses remains well below experimental reach. In GHGUT the KK neutrino modes are heavy and one cannot use this expansion, using instead the full massive KK neutrino propagator. Moreover RH leptons also couple to the W boson resulting in the following amplitude
%%%%%%%%%%%%%%%%%%%%%%%%%
\begin{align}
    C_{R}(q^2=0) &= -\frac{e g^2 m_\mu\Delta^{W}_{L,e i}}{384 \pi^2 m_W^2  (z_{i}-1)^4}\Big( \Delta^{W\dagger}_{L,i \mu} m_\mu(10-43z_{i}+78z_{i}^2-49z_{i}^3+4z_{i}^4+18z_{i}^3\log(z_{i})) \notag \\
    & +6\Delta^{W\dagger}_{R,i \mu}  m_W(-1+z_i)\sqrt{z_i}(4-15z_i+12z^2-z_i^3-6z_i^2\log(z_i)) \Big)\notag \\
    C_{L}(q^2=0) &= -\frac{e g^2 m_\mu\Delta^{W}_{R,e i}}{384 \pi^2 m_W^2  (z_{i}-1)^4}\Big( \Delta^{W\dagger}_{R,i \mu} m_\mu(10-43z_{i}+78z_{i}^2-49z_{i}^3+4z_{i}^4+18z_{i}^3\log(z_{i})) \notag \\
    & +6\Delta^{W\dagger}_{L,i \mu}  m_W(-1+z_i)\sqrt{z_i}(4-15z_i+12z^2-z_i^3-6z_i^2\log(z_i)) \Big),
\end{align}
%%%%%%%%%%%%%%%%%%%%%%%%%
with $z_{i}=(m_i/m_W)^2$ and an internal sum over the (KK) neutrinos is implied. From this, one can recover the SM expression \eqref{eq:SMW} by expanding for small $z_{i}$ and turning off the RH couplings.

\subsection{S-leptoquark couplings}
Finally, let us discuss the couplings to the leptoquark $S$, important for the $\mu\rightarrow e \gamma$ decay, for which the following 5D yukawa couplings are relevant
\begin{align}
    \mathcal{L}\supset
    &-\frac{g_5}{\sqrt{2}}(\bar{u}_{R}^\alpha e_L^{c\prime}+\bar{q}_L^{\alpha,i}  l^{c\prime}_{R,i})S_{\alpha}+ \textrm{h.c.}.
\end{align}
Notice that in the SU(6) GHGUT model, there is a symmetry between the $e^c_L$ and $q_L$ fermions since they are embedded in identical SU(5) multiplets. Therefore, by this correspondence, the above leptoquark couplings are identical to the up-type quark Higgs Yukawa coupling and electron Higgs Yukawa coupling, respectively. The electron Yukawa couplings were already calculated in Section~\ref{sec:AppendixCMass} while the calculation of the up-type quark Yukawa couplings proceeds in a similar way. The leptoquark couplings in the mass basis, $\Delta^S_{R}$ and $\Delta^S_L$, defined via
\begin{align}
    \mathcal{L}\supset S\bar{\Psi}^m_{u,R}\Delta^S_{R}\Psi^{c,m}_{e,L}+S\bar{\Psi}^m_{u,L}\Delta^S_{L}\Psi^{c,m}_{e,R}+\textrm{h.c.}\,,
\end{align}
therefore differ from the up-type quark and electron Higgs Yukawa couplings solely by different mass rotations. As a consequence, non-negligible off-diagonal couplings amongst the SM fermions are induced. This is in contrast to Higgs mediated FCNCs amongst the SM fermions, which are very small due to the close alignment between the mass matrix and Yukawa coupling matrix. It may therefore be important to include the contribution from the SM fermions mediating the loop process although these are generally suppressed by the smaller chirality flip. The leptoquark contribution to $\mu \rightarrow e \gamma$ consists of two diagrams: one where the photon is emitted from the internal leptoquark and one where the photon is emitted from the internal quark line where the top quark and the light KK modes are most important. Their total contribution to the dipole coefficients reads
\begin{align}
    C_{R}(q^2=0) &= -\frac{e \Delta^{S\dagger}_{L,ei} m_\mu}{64 \pi^2 m_S^2  (z_{i}-1)^4} \Big( 2m_S\Delta^S_{R,i\mu}(-1+z_{i})\sqrt{z_{i}}(7-8z_{i}+z_{i}^2+4\log z_{i} +2z_{i}\log z_{i} )\notag \\
    &+\Delta^S_{L,i\mu}m_\mu(1+4z_{i}-5z_{i}^2+4z_{i}\log z_{i} +2z_{i}^2\log z_{i} ) \Big) \notag \\
    C_{L}(q^2=0) &= -\frac{e \Delta^{S\dagger}_{R,ei} m_\mu}{64 \pi^2 m_S^2  (z_{i}-1)^4} \Big( 2m_S\Delta^S_{L,i\mu}(-1+z_{i})\sqrt{z_{i}}(7-8z_{i}+z_{i}^2+4\log z_{i} +2z_{i}\log z_{i}) \notag \\
    &+\Delta^S_{R,i\mu}m_\mu(1+4z_{i}-5z_{i}^2+4z_{i}\log z_{i} +2z_{i}^2\log z_{i} ) \Big), 
\end{align}
with $z_i=(m_{T,i}/m_S)^2$, and a sum over the up-type quarks and their KK modes with mass $m_{T,i}$ is implied.

\bibliographystyle{unsrt}
\bibliography{GHGUTFLAVOR}
\end{document}